\documentclass[journal]{IEEEtran}

\usepackage{subfigure}
\usepackage{verbatim}

\usepackage{cite}

\ifCLASSINFOpdf

\usepackage[pdftex]{graphicx}
   \graphicspath{{../pdf/}{../jpeg/}}
   \DeclareGraphicsExtensions{.pdf,.jpeg,.png}
\else
\fi
\usepackage{amsmath}
\usepackage{algorithmic}

\usepackage{array}

\ifCLASSOPTIONcompsoc
  \usepackage[caption=false,font=normalsize,labelfont=sf,textfont=sf]{subfig}
\else
  \usepackage[caption=false,font=footnotesize]{subfig}
\fi

\ifCLASSOPTIONcaptionsoff
  \usepackage[nomarkers]{endfloat}
 \let\MYoriglatexcaption\caption
\renewcommand{\caption}[2][\relax]{\MYoriglatexcaption[#2]{#2}}
\fi
\begin{document}
%
\title{Parameterized Image Quality Score Distribution Prediction}
%
%
%

\author{Yixuan~Gao,
        Xiongkuo~Min,~\IEEEmembership{Member,~IEEE,}
        Wenhan~Zhu,
        Xiao-Ping~Zhang$^{*}$,~\IEEEmembership{Fellow,~IEEE,}
        and~Guangtao~Zhai$^{*}$,~\IEEEmembership{Senior Member,~IEEE}

\thanks{Yixuan~Gao, Xiongkuo~Min, Wenhan~Zhu and Guangtao~Zhai are with the Institute of Image Communication and Network Engineering, Shanghai Jiao Tong University, China (e-mail: gaoyixuan@sjtu.edu.cn; minxiongkuo@sjtu.edu.cn; zhuwenhan823@sjtu.edu.cn; zhaiguangtao@sjtu.edu.cn).}
\thanks{Xiao-Ping~Zhang is with the Department of Electrical and Computer Engineering, Ryerson University, Canada (e-mail: xzhang@ryerson.ca).}
\thanks{$^{*}$Corresponding authors: Guangtao~Zhai and Xiao-Ping~Zhang.}
\thanks{Part of this work was presented at the 2021 IEEE International Conference on Image Processing (ICIP) \cite{gao2021modeling}.}
}

\maketitle 

\begin{abstract}
Recently, image quality has been generally described by a mean opinion score (MOS).
 However, we observe that the quality scores of an image given by a group of subjects are very subjective and diverse. Thus it is not enough to use a MOS to describe the image quality.
 In this paper, we propose to describe image quality using a parameterized distribution rather than a MOS, and an objective method is also proposed to predict the image quality score distribution (IQSD). At first, the LIVE database is re-recorded.
Specifically, we have invited a large group of
subjects to evaluate the quality of all images in the LIVE database, and each image is evaluated by a large number of subjects (187 valid subjects), whose scores can form a reliable IQSD. By analyzing the obtained subjective quality scores, we find that the IQSD can be well modeled by an alpha stable model, and it can reflect much more information than a single MOS, such as the skewness of opinion score, the subject diversity and the maximum probability score for an image. 
Therefore, we propose to model the IQSD using the alpha stable model. 
Moreover, we propose a framework and an algorithm to predict the alpha stable model based IQSD, where 
quality features are extracted from each image based on structural information and statistical information, and support vector regressors are trained to predict the alpha stable model parameters. 
Experimental results verify the feasibility of using alpha stable model to describe the IQSD, and prove the effectiveness of objective alpha stable model based IQSD prediction method.

\end{abstract}


%
\IEEEpeerreviewmaketitle

\section{Introduction}
\label{sec:intro}

%
%
%
%

\IEEEPARstart{I}{mage} quality assessment (IQA) is one of the most challenging tasks in image processing \cite{2006Image,2006An,2004Image,LIVE,TID2013},
which can be divided into subjective IQA and objective IQA.
Subjective IQA means that a group of
subjects give opinion scores to an image through observation, and the image quality is generally described by the mean opinion score (MOS) of all subjective ratings. 
On the other hand, objective IQA predicts the MOS of an image by extracting and integrating image quality features via computational models, which is more efficient and widely applicable than subjective IQA. 
In particular,
objective IQA includes full reference (FR) methods \cite{2006A,Zhang2011FSIM}, reduced reference (RR)  methods \cite{2017Reduced,2009Reduced} and no reference (NR) methods \cite{Moorthy,2010A}.

FR IQA needs the reference image to evaluate the distorted image quality. RR IQA needs not only the distorted image, but also additional information of the reference image. NR IQA can predict the distorted image quality without using any information of the reference image.
In many cases, the reference image is difficult to get. 
Therefore, compared with FR and RR IQA, NR IQA is more widely used in practical applications, and its most important task is to extract quality features from an image. For example, researchers in \cite{Moorthy} proposed DIIVINE to extract features in wavelet domain. 
In \cite{Mittal}, Mittal \textit{et al.} proposed BRISQUE to extract features in spatial domain.
Michele \textit{et al.} have developed BLIINDS-II by using discrete cosine transform coefficients in \cite{Michele}. 
Image features in \cite{2015A} were extracted by integrating some statistical information of each image.
The above methods mainly extract the statistical information of an image. In fact, the structural information of an image is also very important. 
Min \textit{et al.} introduced BMPRI to extract structural features based on local binary pattern (LBP) in \cite{2018Blind}. The authors in \cite{8301594} utilized improved multiscale LBP for image feature extraction.
 Inspired by the above articles, this paper extracts quality-aware features of each image based on both structural and statistical information.

\begin{figure}[htb]
\centering
\subfigure[`Ocean'.]{
\begin{minipage}[t]{0.48\linewidth}
\centering
\includegraphics[width=4cm, height=3cm]{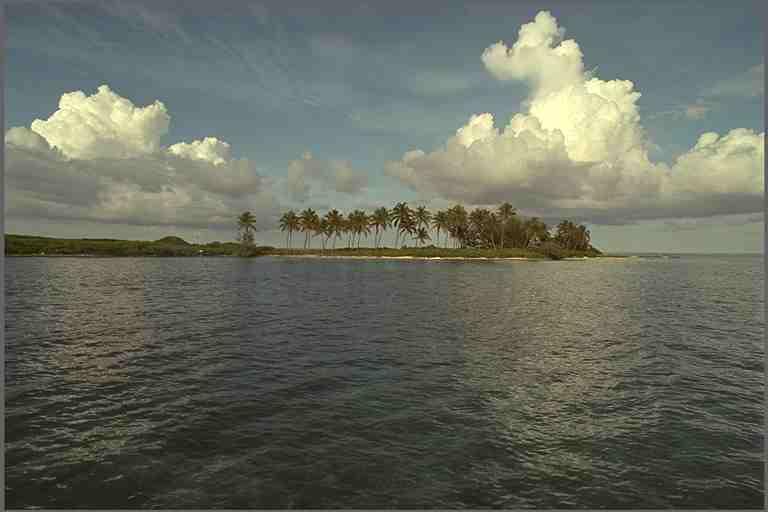}
\end{minipage}%
}%
\subfigure[`Buildings'.]{
\begin{minipage}[t]{0.48\linewidth}
\centering
\includegraphics[width=4cm, height=3cm]{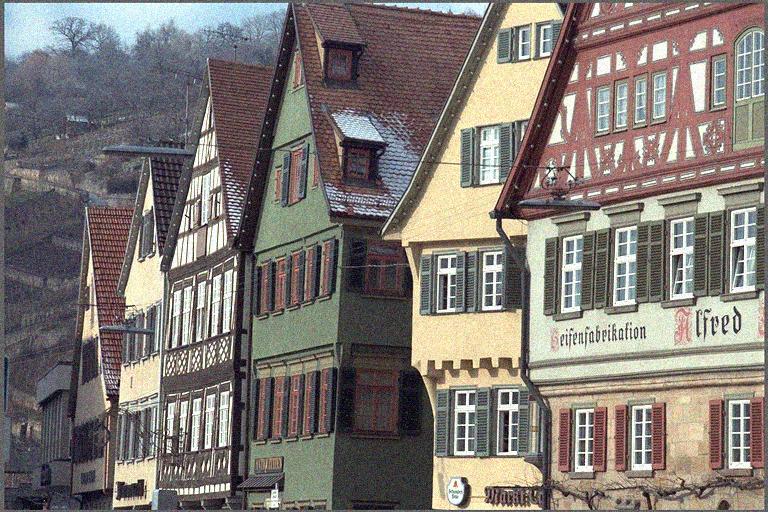}
\end{minipage}%
}%

\centering\subfigure[The quality score histogram of \newline `Ocean'.]{
\begin{minipage}[t]{0.48\linewidth}
\centering
\includegraphics[scale=0.23, trim=20 200 40 200, clip]{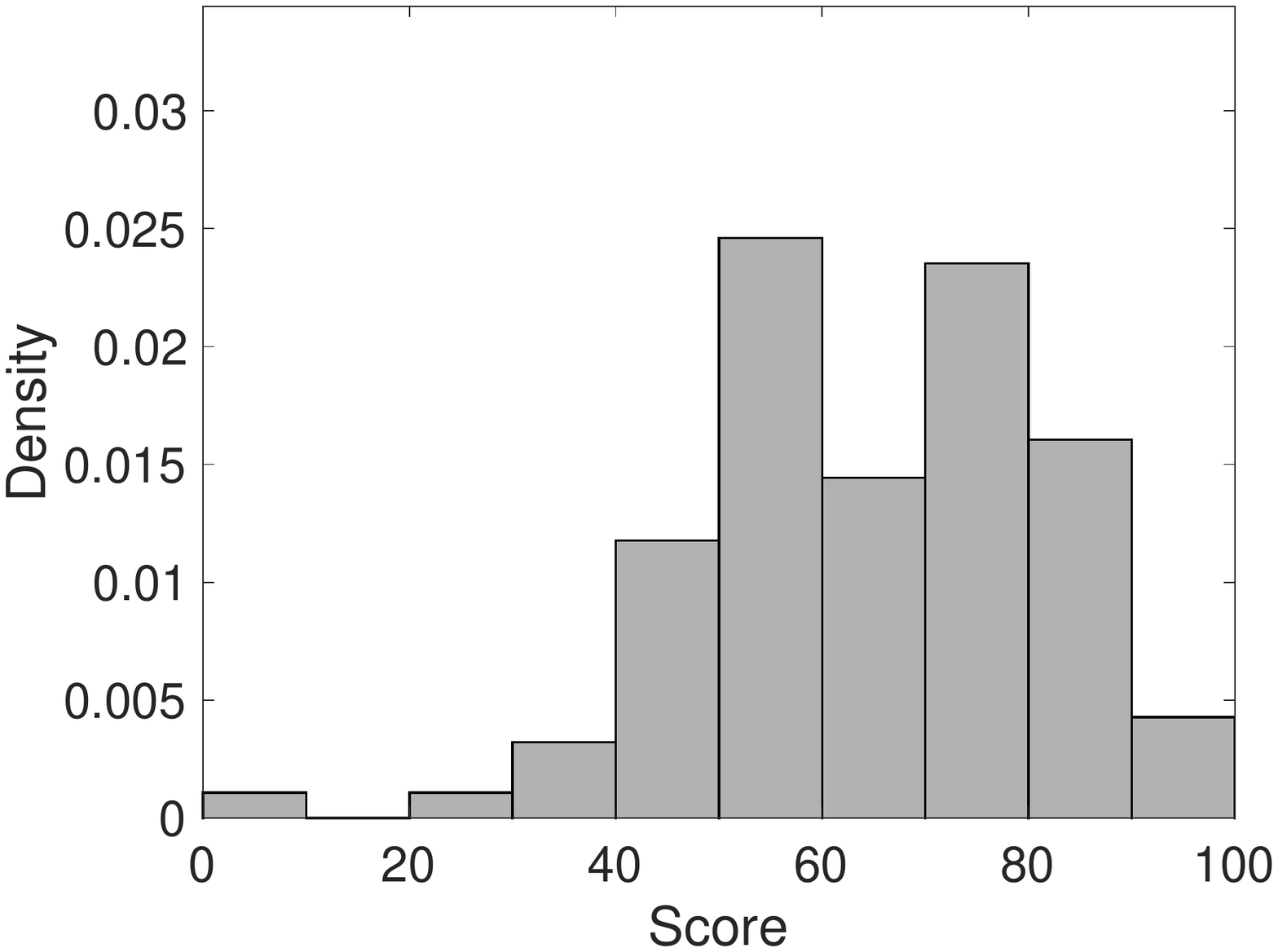}
\end{minipage}
}%
\centering\subfigure[The quality score histogram of \newline `Buildings'.]{
\begin{minipage}[t]{0.48\linewidth}
\centering
\includegraphics[scale=0.23, trim=20 200 40 200, clip]{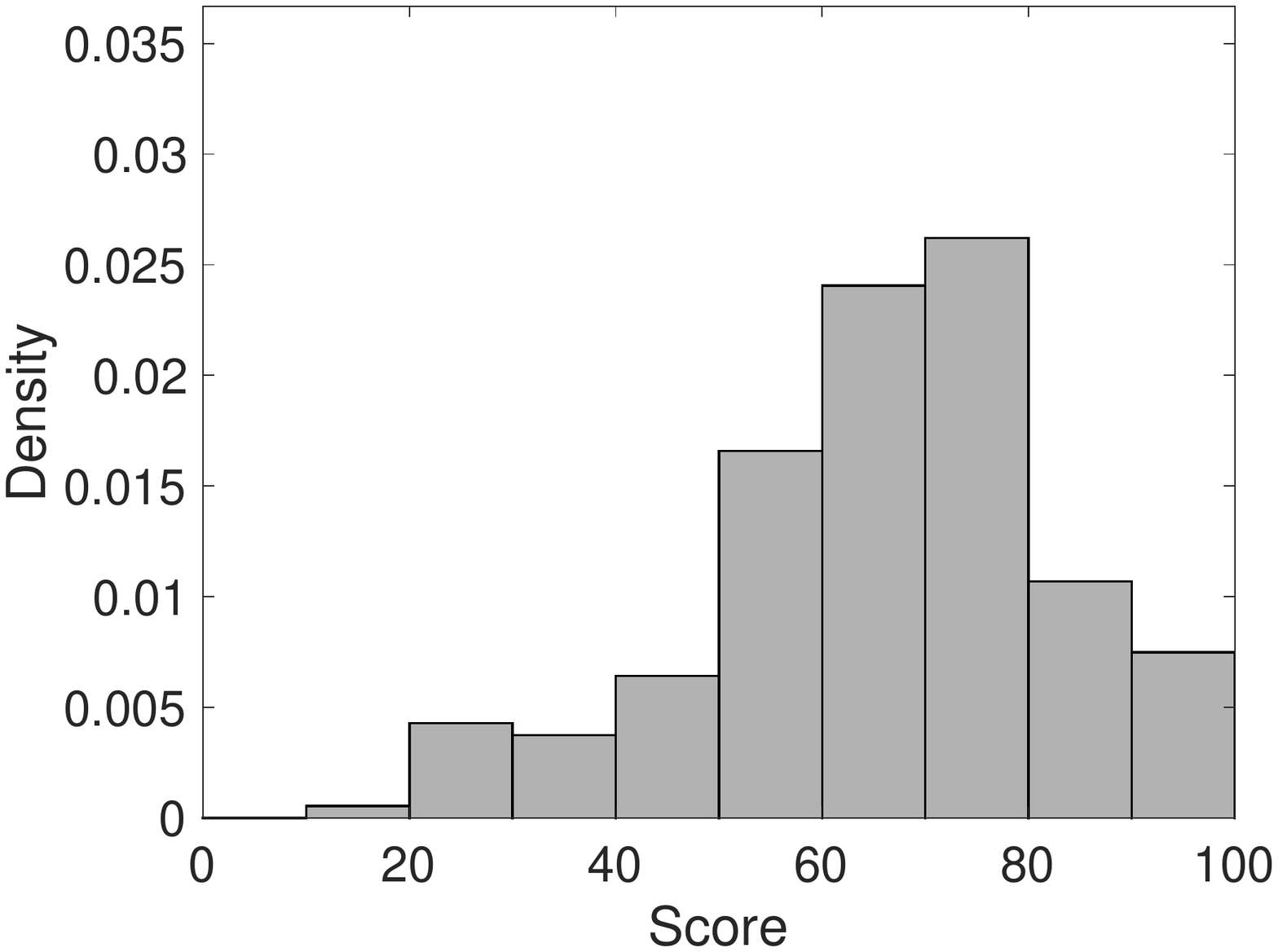}
\end{minipage}
}%
\centering
\caption{Two images with the same MOS and different quality score histograms.}
\label{Fig1}
\end{figure}

In the above articles, the proposed methods extract image quality features to predict the MOS of an image.
However, we should note that due to the different personal backgrounds of different subjects, such as experience, education and habit, different subjects may give different quality scores to the same image \cite{2007An}. Therefore, image quality is highly subjective and diverse, which is difficult to be described by a single MOS.
For example, in Fig.~\ref{Fig1}, two images have close MOS, which is about 66. However, they have different quality score histograms. 
It is difficult to see the difference between the two images just from the MOS, while it can be easily seen from Fig.~\ref{Fig1} (c) and (d) that the quality scores of `Buildings' are more concentrated than that of `Ocean', which means that subjects are more likely to agree on the scores of `Buildings'. 

Based on the above analysis, many researchers have broken through the limitation of MOS and tried to study different image quality descriptions \cite{6065690,2015QoE}.
For example, Tobias \textit{et al.} emphasized the importance of subjective score diversity in subjective quality of experience studies, and proposed to consider the standard deviation of opinion score (SOS) to evaluate subjective diversity in \cite{6065690}. Later in \cite{2015QoE}, Tobias \textit{et al.} also used quantiles to calculate subjective satisfaction/dissatisfaction.
Further, some researchers have tried to describe the image aesthetic quality by predicting the quality score histogram \cite{Zhang,Liu,jin2020deep}.
For example, the researchers predicted the aesthetic quality score histogram of each image in \cite{Zhang}. In \cite {Liu}, the authors predicted the image quality score histogram by assuming that the quality scores of each image follow the Gaussian distribution. 
In \cite {jin2020deep}, the authors proposed to use a deep drift-diffusion model to predict the image aesthetic score histogram. So far, the histogram of image aesthetic quality scores has been studied well by some researchers, and it can better reflect the subjective quality information of an image than MOS to some extent.
As for the image quality score histogram, no researcher has studied it before.
It is worth noting that the quality score histogram is discrete, which can not accurately represent the probability distribution of image quality score in a fixed quality range.
For example, what is the quality score distribution of `Buildings' like in Fig.~\ref{Fig1} between 60 and 70? 

To solve the above problem, we propose to parameterize the image quality score distribution (IQSD) by using a continuous distribution in this paper.
As a more universal form of continuous distribution, alpha stable model with four parameters can show some non-Gaussian statistical characteristics such as heavy tail, spike and skewness \cite{Mohammad2014Estimating}. Because of these special statistical characteristics, alpha stable model has been widely studied by many researchers and applied in many fields, such as signal processing \cite{2012Reverberation}, economy  \cite{2014Estimating}, etc. 
Considering its advantage, we also use the alpha stable model to describe the IQSD. 

Overall, our main contributions are summarized as follows. 
Firstly, we have invited a large group of subjects to give subjective quality ratings for all images in the LIVE database. 
By analyzing the obtained subjective quality scores, we observe that the alpha stable model can describe the IQSD well.
Then, we propose a framework and an algorithm to predict the alpha stable model based IQSD.
Specifically, we first extract image quality features based on both structural and statistical information, and then the alpha stable model parameters are predicted by four support vector regressors (SVRs).
Finally, the superiority and feasibility of using alpha stable model to describe the IQSD are verified by extensive  experiments. In details, experimental results show that the alpha stable model can describe the IQSD well, and the alpha stable model based IQSD can provide much more information than a single MOS, such as the skewness of opinion score, the subject diversity and the maximum probability score for an image.
Comparison experiments are carried out to verify the effectiveness of the objective alpha stable model based IQSD prediction method.

This paper is based on our previous work \cite{gao2021modeling}, but this paper has substantial differences from the previous work. The differences are summarized as follows. Firstly, images used in \cite{gao2021modeling} were only 100 images selected from the LIVE database, while this paper re-records all images in the LIVE database. Secondly, 
previous work \cite{gao2021modeling} extracted image quality features based on statistical information. 
In this paper, the feature extraction method of image quality is based on both structural and statistical information, which is completely different from the method used in \cite{gao2021modeling}.

The rest of the paper is organized as follows. In Section \ref{sec:assessment}, details of the subjective quality assessment study are provided. The proposed modeling and prediction framework are described in Section \ref{sec:2}. Section \ref{sec:3} gives the experimental validation and analysis of the alpha stable model based IQSD. Finally, we conclude the paper in Section \ref{Conclusion}.

\section{Subjective Quality Assessment}
\label{sec:assessment}

Since the existing databases only provide MOSs as the ground-truth of the image quality rather than all quality scores given by specific subjects, we have carried out a subjective quality assessment experiment to obtain the subjective quality scores for all images in the LIVE database \cite{LIVE}.


\subsection{Image Materials and Observers}
As an influential database in the field of image quality assessment, the LIVE database contains 779 distorted images generated from 29 reference images, with 4 distortion types including JPEG2000 (JP2K), JPEG, White noise (WN), Gaussian blur (Gblur) and Fast-fading rayleigh (FF). Each distortion type contains 4 or 5 distortion levels. 
The LIVE database uses differential MOS to represent the image quality.
In order to analyze the IQSD, we chose all images in the LIVE database as image materials, including 29 reference images and 779 distorted images. We invited 206 subjects to participate in the subjective quality assessment experiment, and all subjects rated all images included in the database.
The number of subjects per image is much larger than the 15 non-expert observers required by the Recommendation ITU-R BT.500 \cite{1}. 
Before the experiment, we have tested the visual conditions of all subjects. All 206 subjects have normal (corrected) visual acuity and color vision.
At the same time, we recorded the characteristics of all subjects, including age and occupation. The age range of subjects is large, ranging from 18 to 50 years old.
The occupational categories of subjects also span a large variety, including students, accounting, personnel, sales, self employment, etc., in which students account for the majority.

 \begin{figure*} [htb]
 \centering
    \includegraphics[scale=0.16]{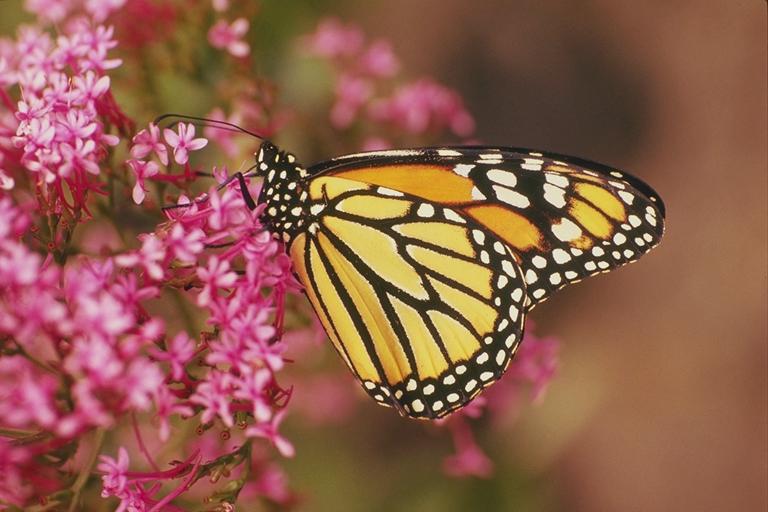}
    \includegraphics[scale=0.16]{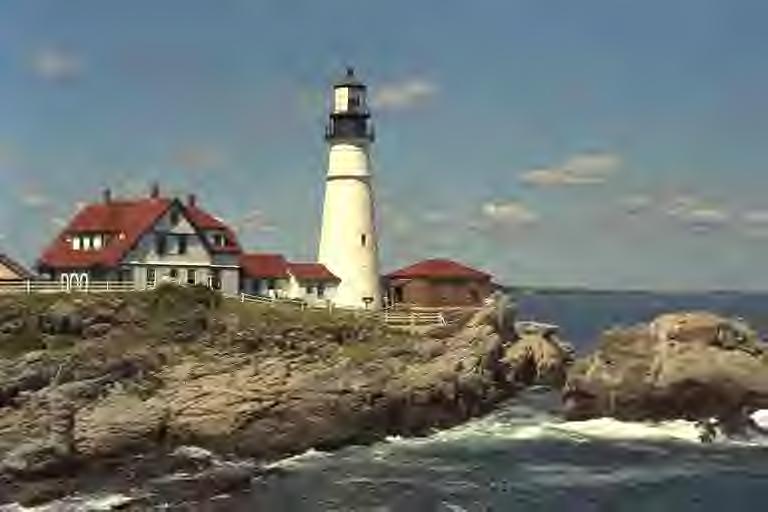}
    \includegraphics[scale=0.16]{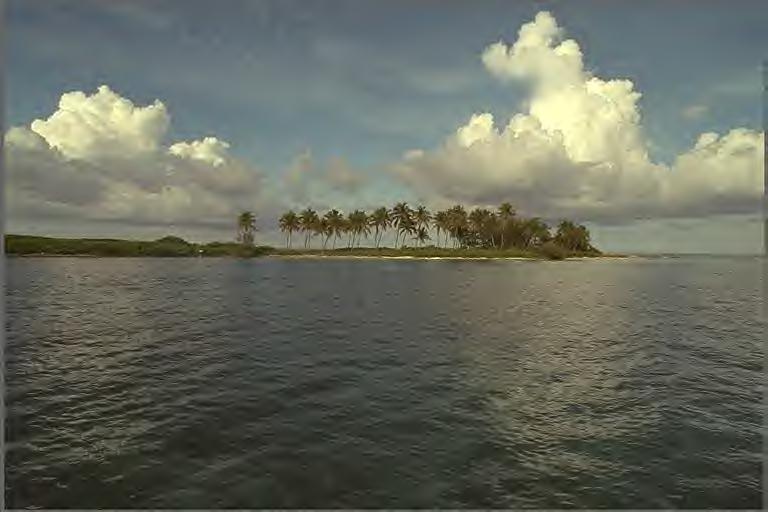}
    \includegraphics[scale=0.16]{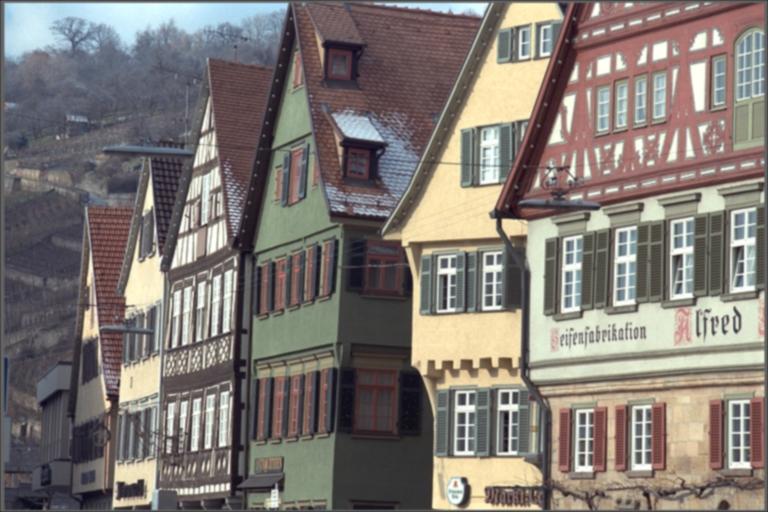}\\
\includegraphics[scale=0.23, trim=10 200 40 200, clip]{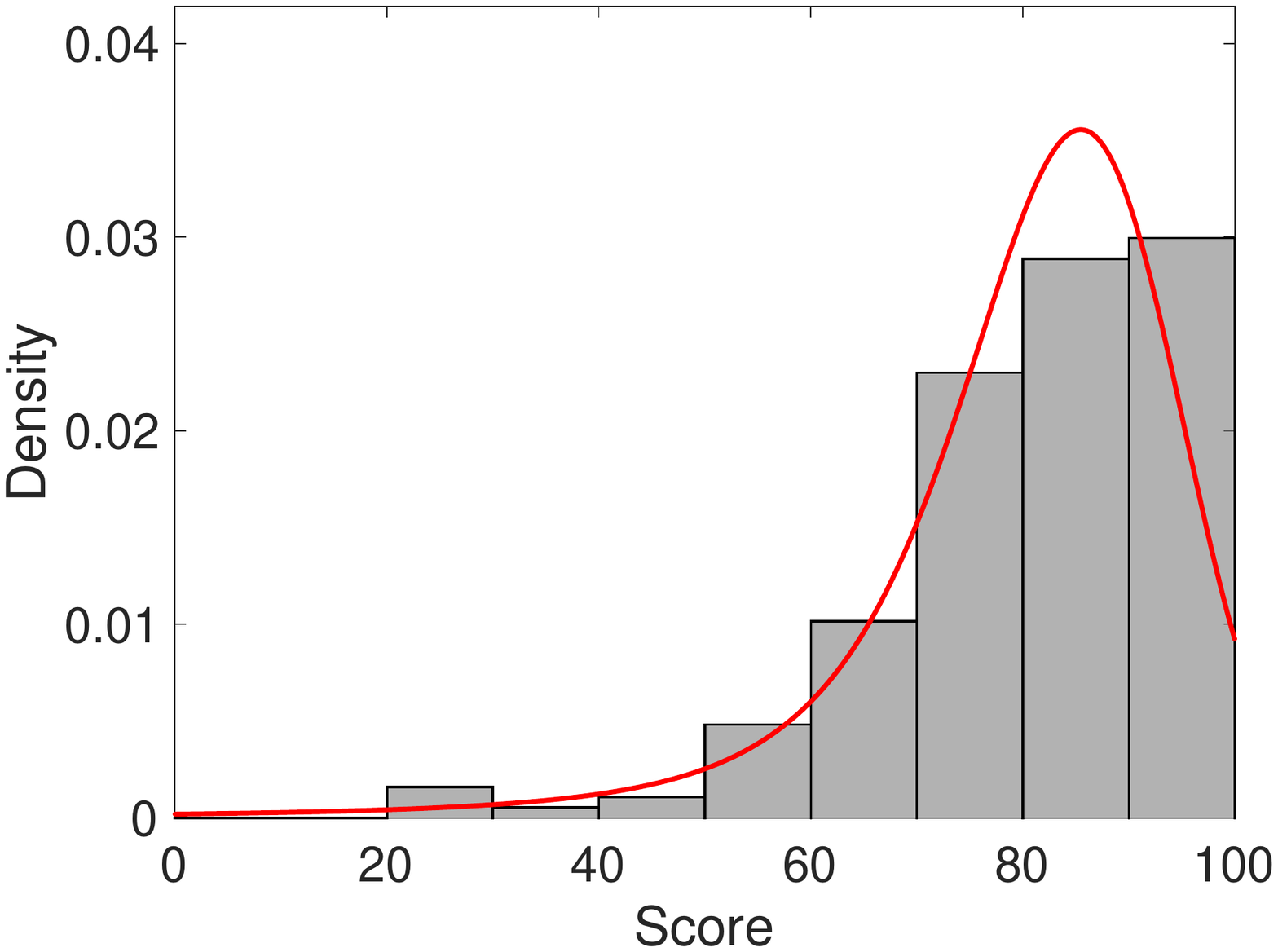}
    \includegraphics[scale=0.23, trim=20 200 40 200, clip]{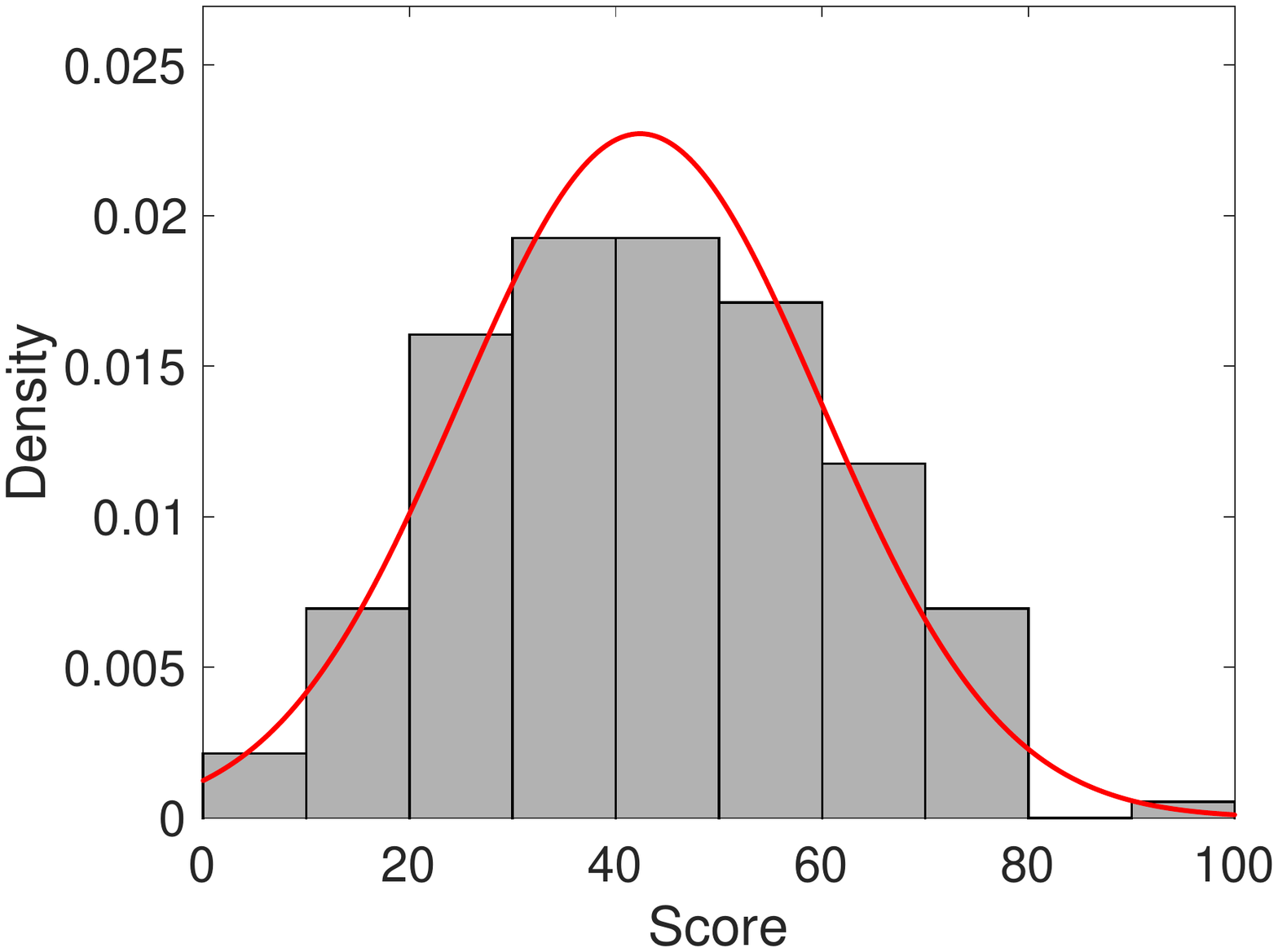}
    \includegraphics[scale=0.23, trim=20 200 40 200, clip]{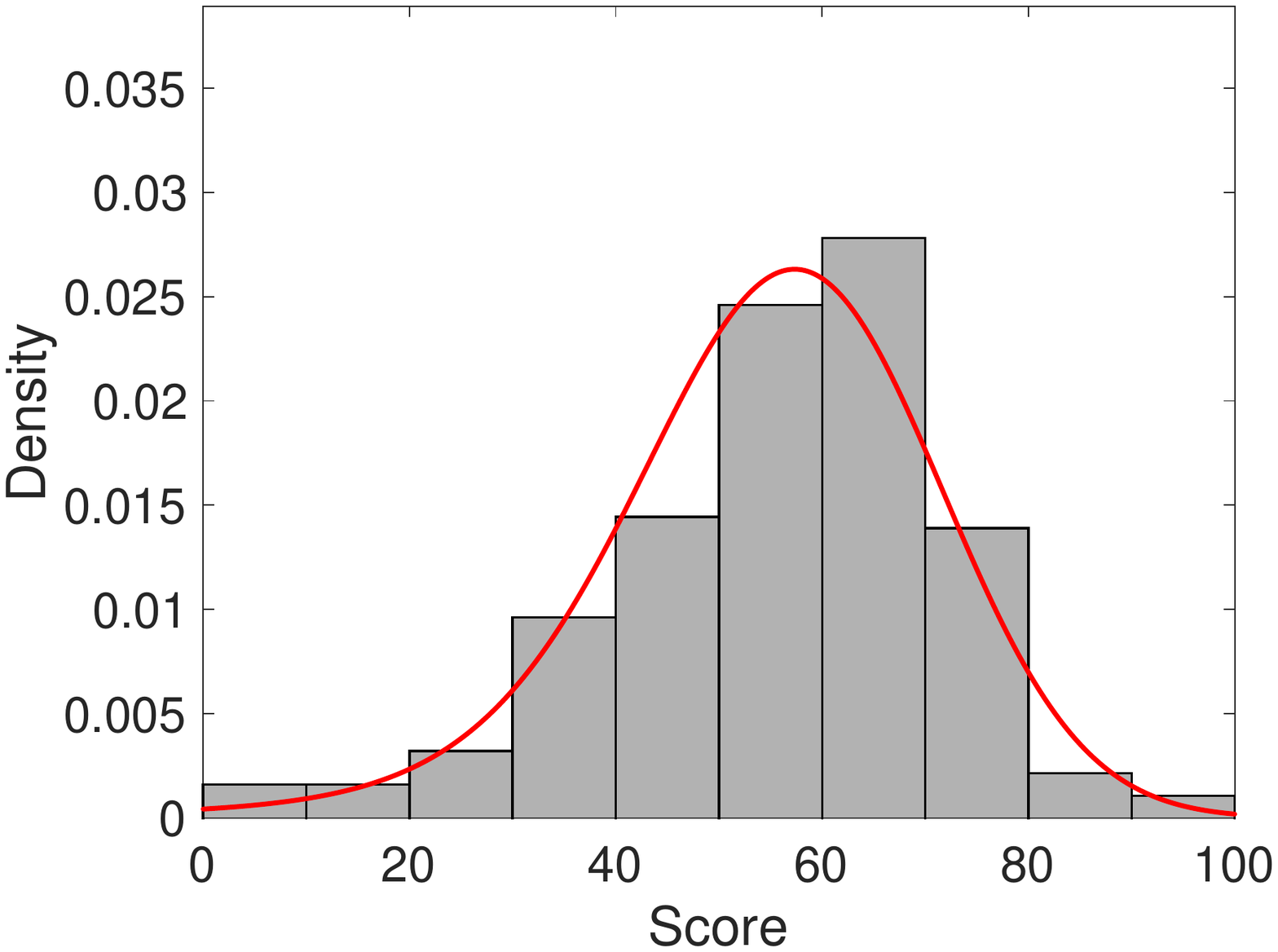}
    \includegraphics[scale=0.23, trim=20 200 40 200, clip]{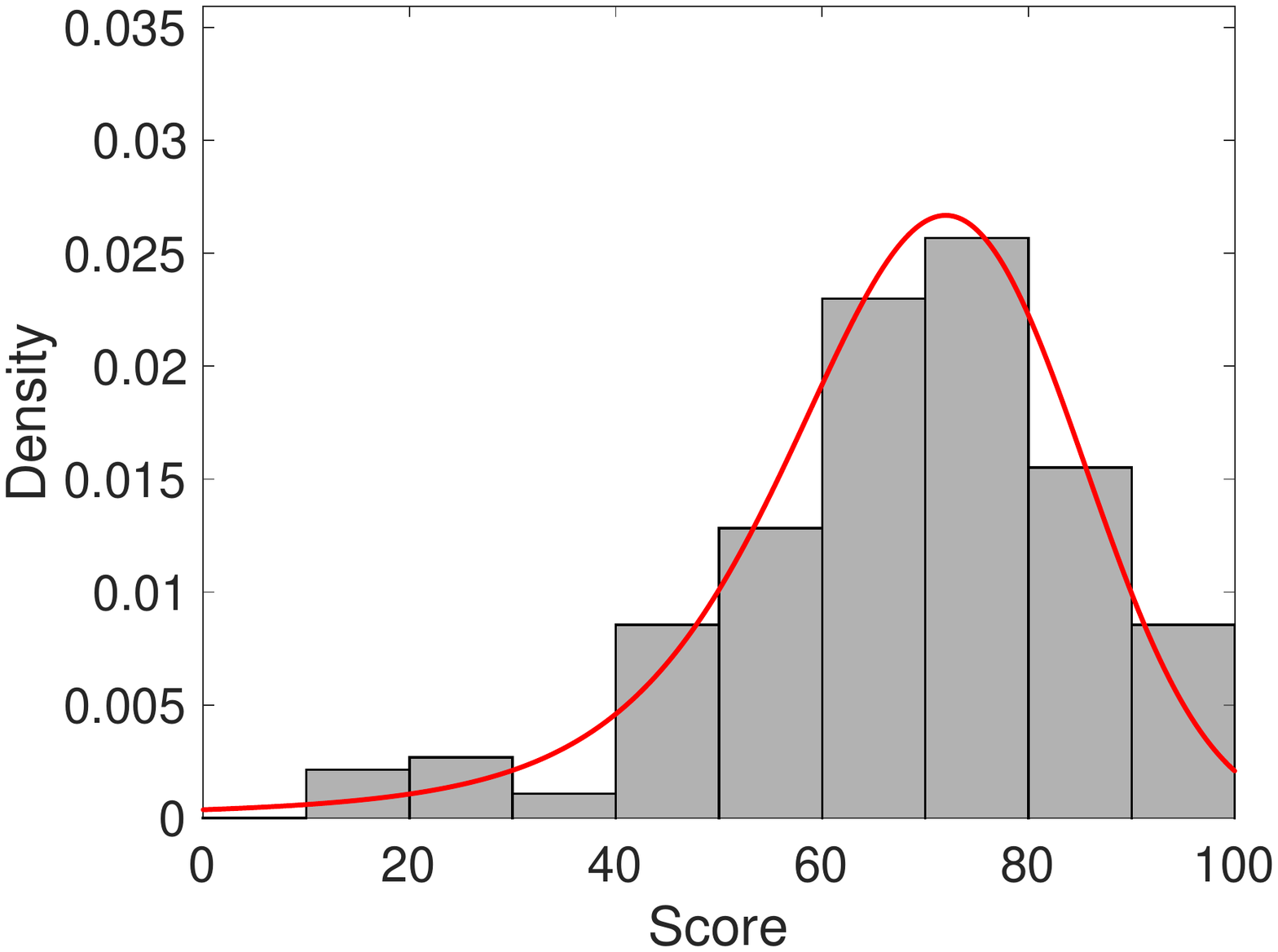}\\
       \includegraphics[scale=0.16]{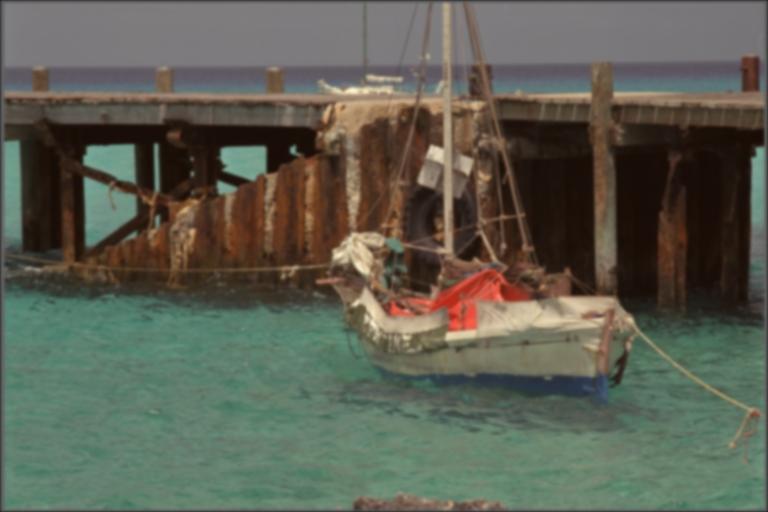}
    \includegraphics[scale=0.16]{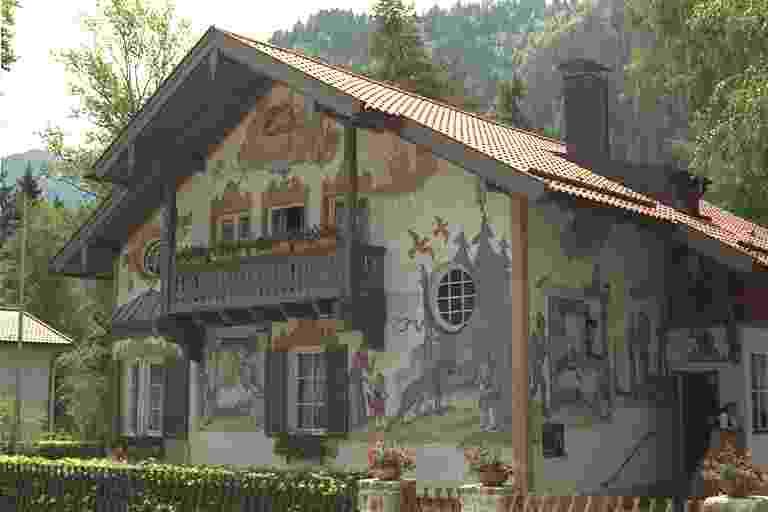}
    \includegraphics[scale=0.16]{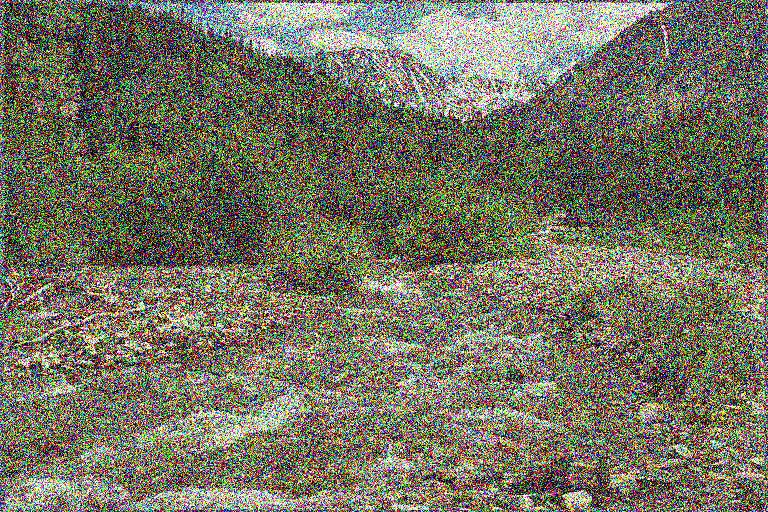}
    \includegraphics[scale=0.16]{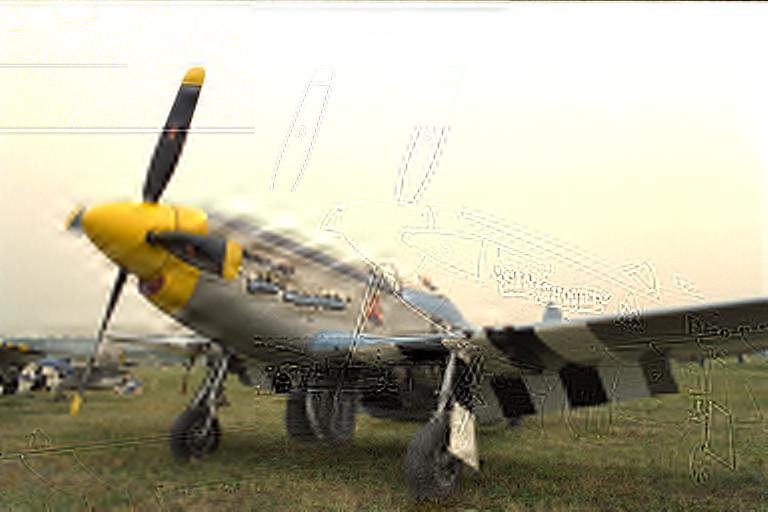}\\
\includegraphics[scale=0.23, trim=10 200 40 200, clip]{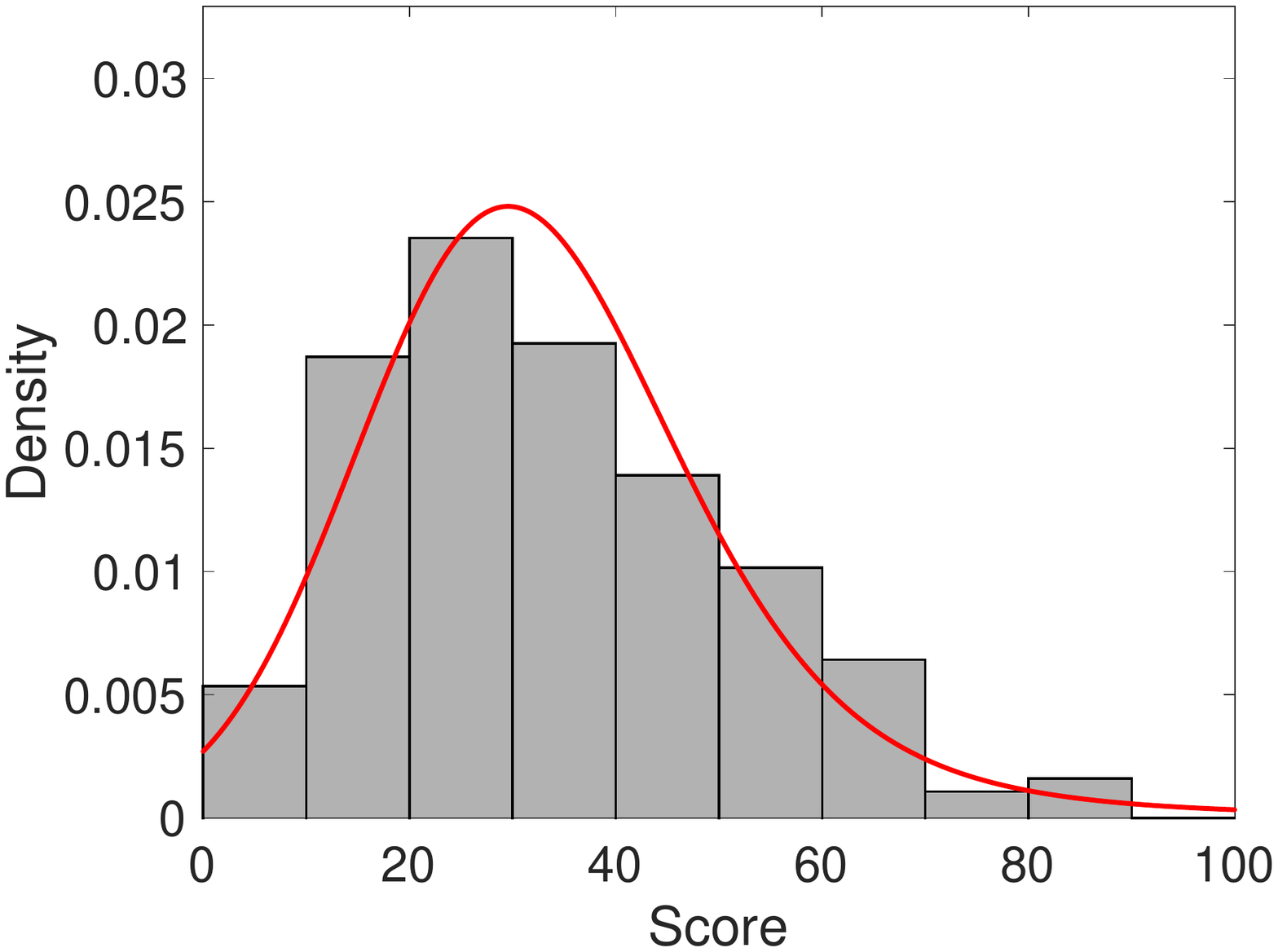}
    \includegraphics[scale=0.23, trim=20 200 40 200, clip]{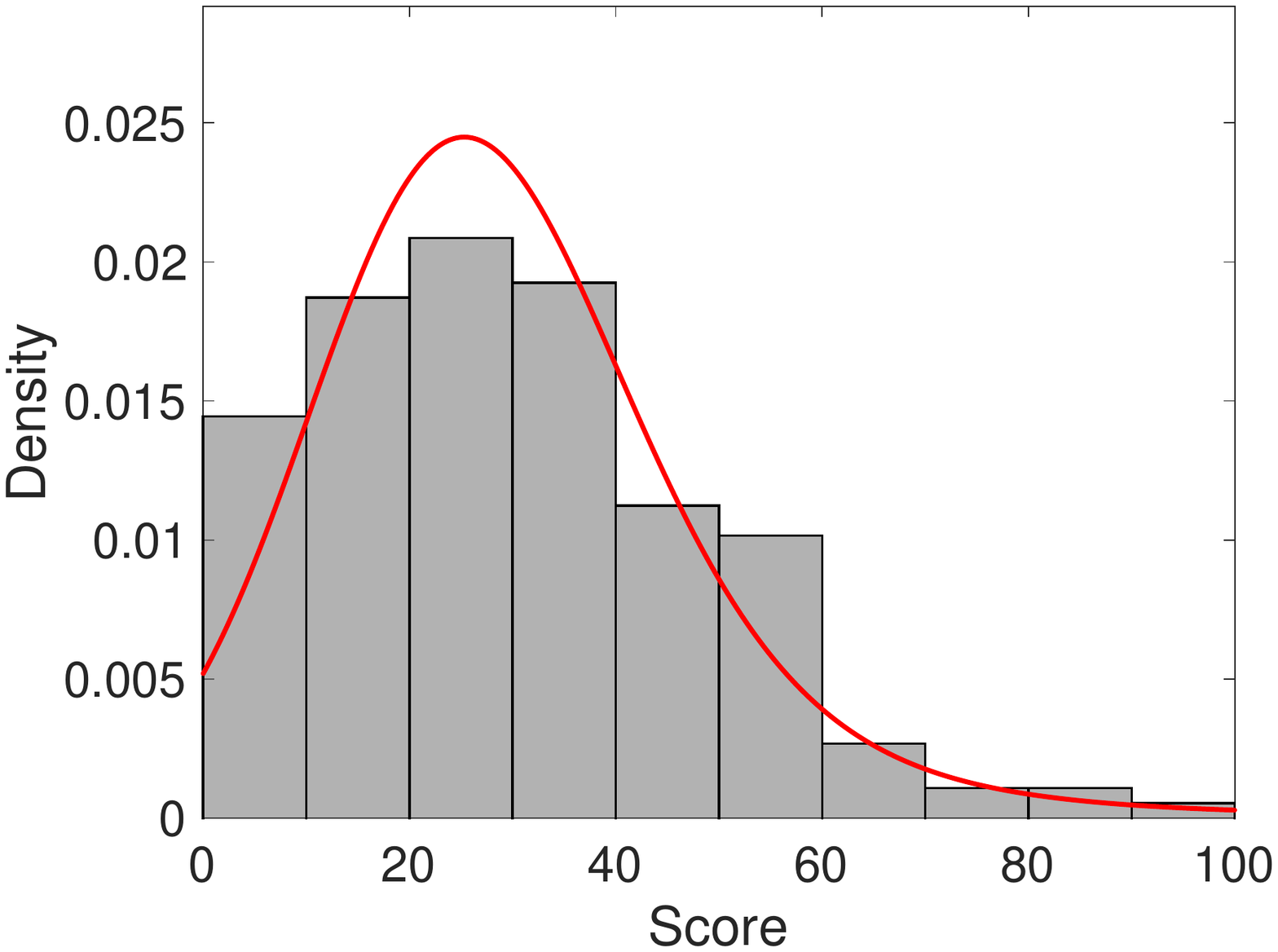}
    \includegraphics[scale=0.23, trim=20 200 40 200, clip]{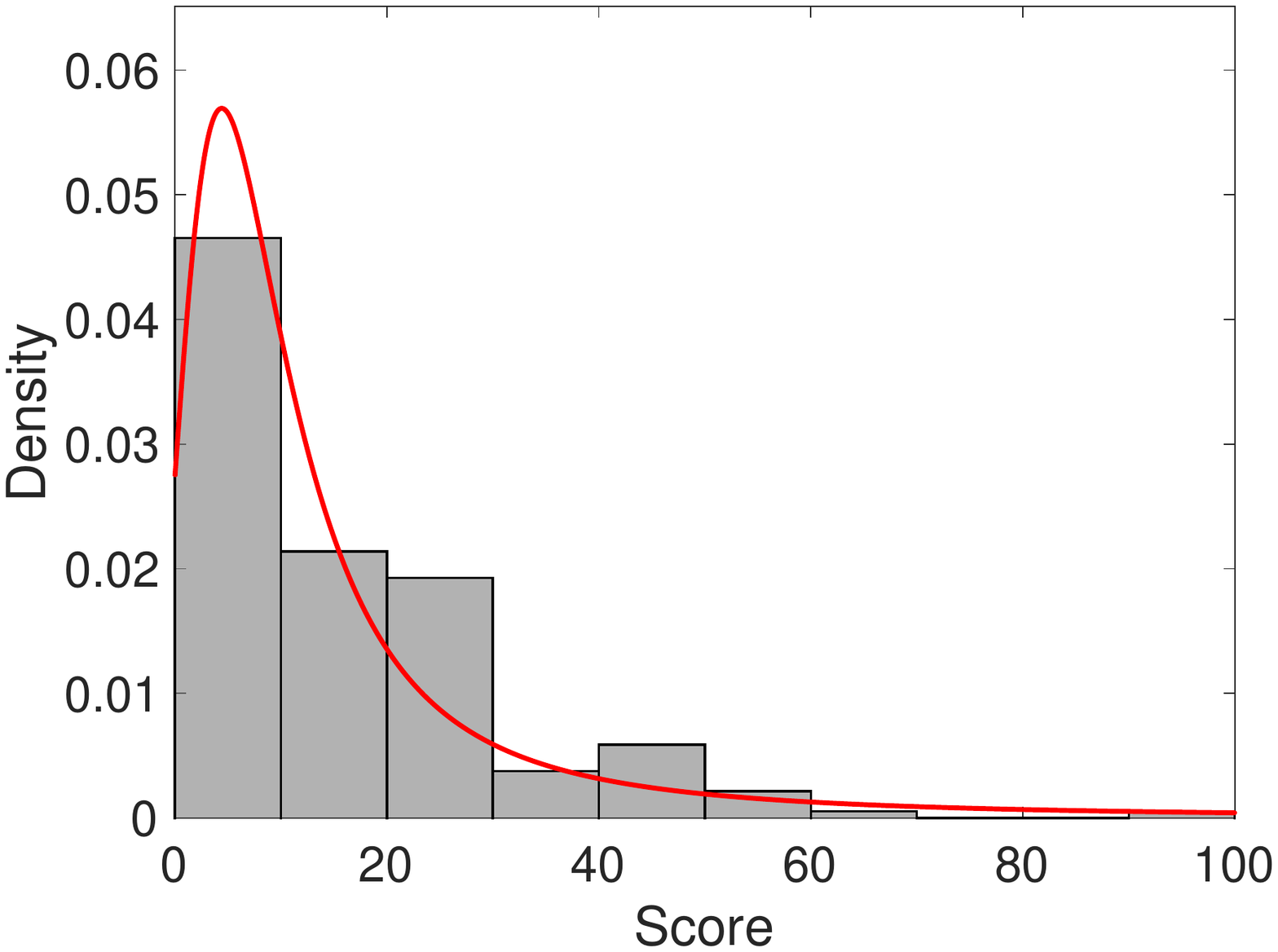}
    \includegraphics[scale=0.23, trim=20 200 40 200, clip]{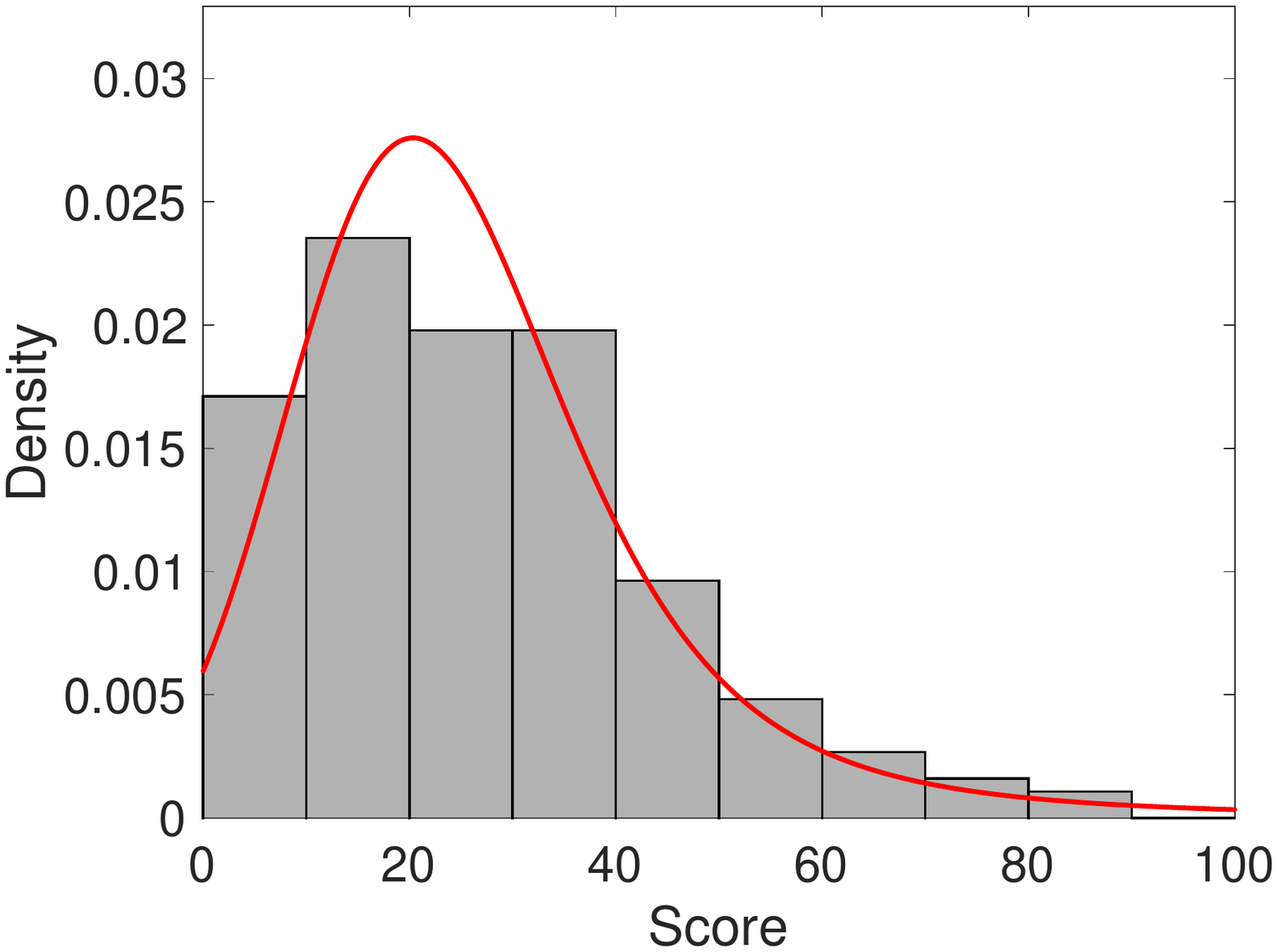}  
    \caption{Images and the corresponding quality score histograms with 10 bins. For histograms, the horizontal axis represents the image score, and the vertical axis represents the probability density. Red curves show the fitted IQSDs described by alpha stable model. }
\label{f0}
\end{figure*}

\subsection{Experiment Methodology}
Recommendation ITU-R BT.500 has described several subjective assessment methods of image quality, including the double stimulus improvement scale (DSIS) method, the double stimulus continuous quality scale (DSCQS) method, the single stimulus (SS) method, the stimulus-comparison method, the single stimulus continuous quality evaluation (SSCQE) method and the simulated double method stimulus for continuous evaluation (SDSCE) method. In this experiment, we used the SSCQE method to collect subjective quality scores. In details, 
we designed a web testing system to display images and collect image quality scores given by all subjects.
 Subjects were asked to score each image according to the overall impression about this image's quality, and marked the score on a continuous quality scale within the range of [0,100]. The quality scale was associated with the ITU-R quality and injury scale in the following ways: the score in the area between 0 and 20 means that the image quality is `Bad'; 
the score in the area between 20 and 40 means that the image quality is `Poor'; 
the score in the area between 40 and 60 means that the image quality is `Fair'; 
the score in the area between 60 and 80 means that the image quality is `Good'; 
the score in the area between 80 and 100 means that the image quality is `Excellent'. 

Recommendation ITU-R BT.500 suggests that before evaluating test images, training images should be displayed to subjects to show the range and types of impairments they will evaluate during the test.
Following Recommendation ITU-R BT.500, we selected 20 images with different contents and distortions as training images. The quality of training image covered the range from `Bad' to `Excellent' which would be seen by subjects in the test stage. In the web system, subjects needed to rate training images before the formal test stage, and all images (20 training images and 808 test images) were displayed to subjects in random order.

\subsection{Data Processing}
 After the experiment, we process the obtained quality scores of all images. First, we detect and discard the subjects with significant deviations from the mean values. Second, we calculate and remove the inconsistent subjects. After the above two steps, 19 subjects are excluded, and the number of valid subjects is 187.
Finally, the raw quality scores given by all these 187 effective subjects constitute the quality score histogram for each image. Fig.~\ref{f0} shows eight images and their corresponding quality score histograms. The first and third rows show images, and the figures below the images in the second and fourth rows show the corresponding quality score histograms.


 \begin{figure*} [h]
 \centering
\includegraphics[scale=0.55, trim=12 180 5 5, clip]{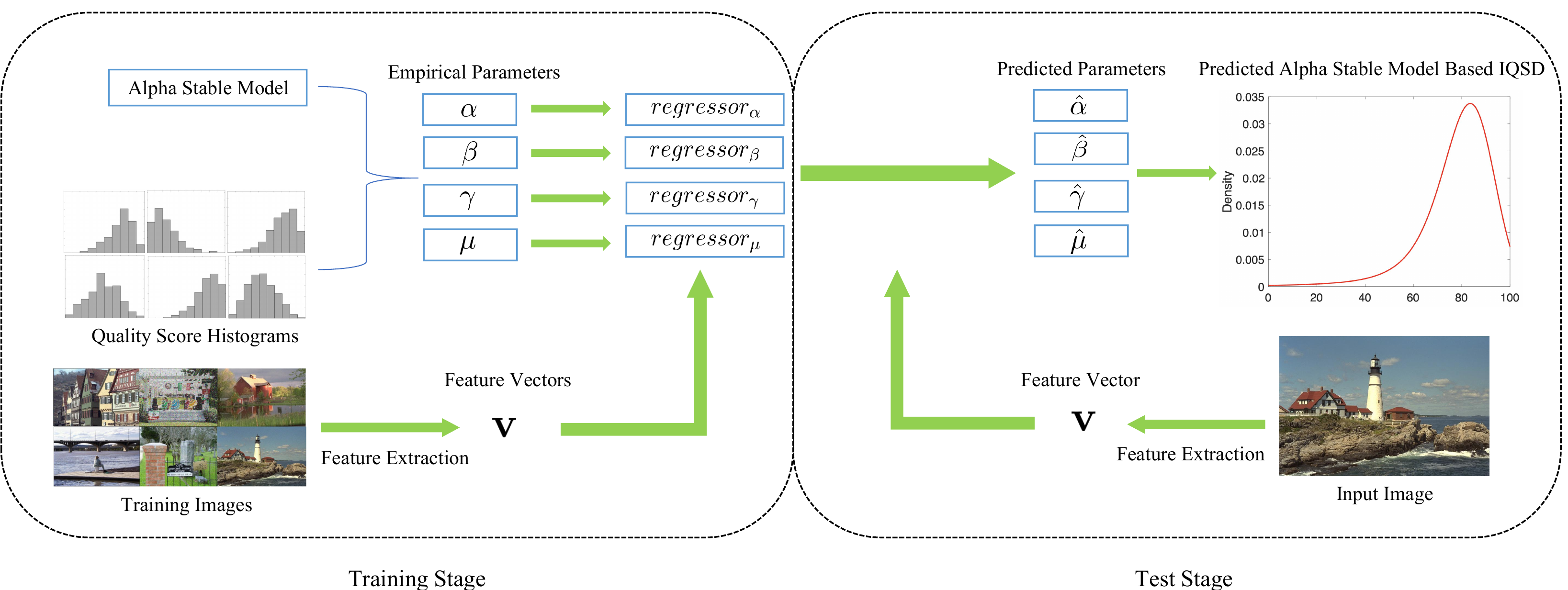}
  \caption{The diagram of the proposed prediction framework of the alpha stable model based IQSD.}
\label{diagram}
\end{figure*}


\begin{table}[h]
	\caption{The Fitting Results of Several Common Distributions. 
	  RMSE Describes the Mean RMSE Fitting Error for All Images. For Chi-Square Test, the Percentage of Images That Do Not Reject the Hypothesis That the Fitted Distribution and the Quality Score Histogram Come From the Same Distribution Is Given When the Significance Level Is 0.05. }
	\centering
\begin{tabular}{c|cccccccc}
\hline\hline
\textbf{Distribution} & \textbf{RMSE} &\textbf{Chi-Square Test}\\ \hline\hline

Gaussian & 0.0038&44.4\%\\
Half Normal & 0.0272&7.4\%\\
Exponential & 0.0093&8.9\%\\
Lognormal & 0.0071&2.0\%\\
Gamma & 0.0100&34.0\%\\
Generalized Pareto & 0.0082&12.3\%\\
Beta & 0.0148&41.7\%\\
Weibull &0.0137&42.3\%\\\hline\hline

\end{tabular}
	\label{goodness-of-fit}
\end{table}

\subsection{Data Analysis}
In order to find out which distribution can better describe these quality score histograms, we use the following common distributions to model the quality score histograms of all images: Gaussian distribution, half normal distribution, exponential distribution, lognormal distribution, gamma distribution, generalized Pareto distribution, beta distribution and Weibull distribution. The maximum likelihood estimation (MLE) is used to fit all distributions. In this paper, we use two measures to evaluate the fitting results. On the one hand, we use the root mean square error (RMSE) to quantify the difference between quality score histograms and the fitted distributions. We calculate the probability density histograms of the fitted distributions with 10 bins, which are the same as the number of bins of the quality score histograms. 
Thus, for each image, the RMSE between the quality score histogram and each fitted distribution can be written as 
\begin{equation}\label{e1}
RMSE=\sqrt{\frac{1}{n}\sum_{i=1}^{n}(h_{i}-d_{i})^2},
\end{equation}
where $h_{i}$ represents the probability density of the $i$-th bin of the image quality score histogram, $d_{i}$ represents the value of the $i$-th bin in the probability density histogram of the fitted distribution, and $n$ is the number of bins. For each kind of distribution, the RMSE in Table~\ref{goodness-of-fit} shows the mean value of RMSE fitting errors for all images.
On the other hand, 
the Chi-Square test is used as a goodness-of-fit test method to show whether the fitted distribution and quality score histogram come from the same distribution. 
When the significance level is 0.05, the percentage of images that do not reject the hypothesis that the fitted distribution and the quality score histogram come from the same distribution is also given in Table~\ref{goodness-of-fit}.

It can be seen from Table~\ref{goodness-of-fit} that the fitting performance of Gaussian distribution is the best among these common distributions.
 What's more, there are some quality score histograms follow Weibull distribution, beta distribution, exponential distribution, gamma distribution, generalized Pareto distribution and lognormal distribution. Although the fitting results of Gaussian distribution is the best, it can not describe the heavy tail and skewness of IQSD.
Therefore, none of the these distributions can describe image quality score histograms very well. 
 This inspires us to find a distribution that can show diversity of IQSD and has a better fitting performance.

\section{Alpha Stable Model based IQSD}
\label{sec:2}
In this part, we propose to use an alpha stable model to describe the IQSD. In addition, we introduce a framework to predict the alpha stable model based IQSD, where a feature extraction followed by feature regression method is adopted. The diagram of the proposed prediction framework of the alpha stable model based IQSD is shown in Fig.~\ref{diagram}. The details are described as follows.
\begin{figure*}[h]
\centering
\subfigure[]{
\begin{minipage}[t]{0.24\linewidth}
\centering
\includegraphics[scale=0.23, trim=15 200 40 200, clip]{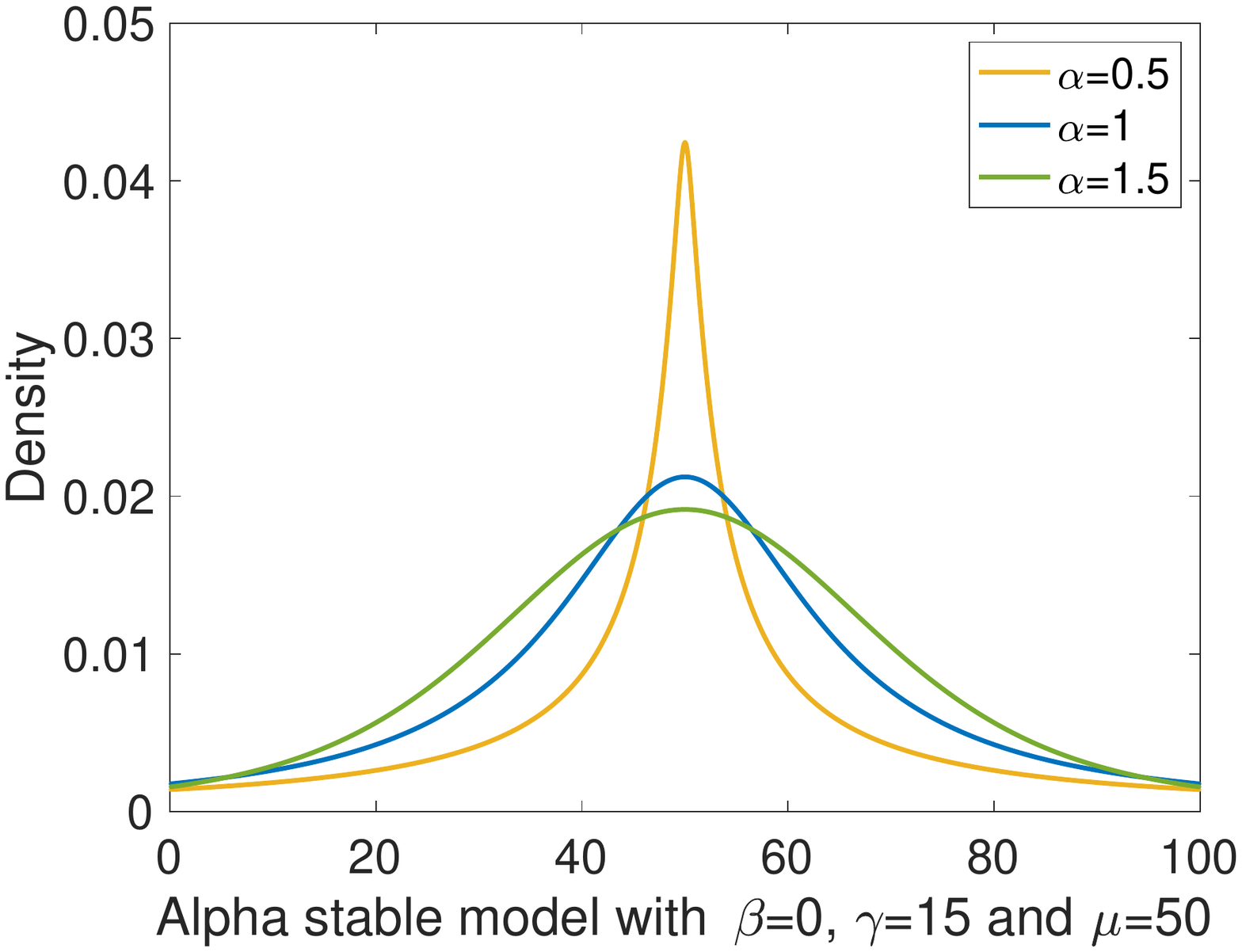}
\end{minipage}%
}%
\subfigure[]{
\begin{minipage}[t]{0.24\linewidth}
\centering
\includegraphics[scale=0.23, trim=15 200 40 200, clip]{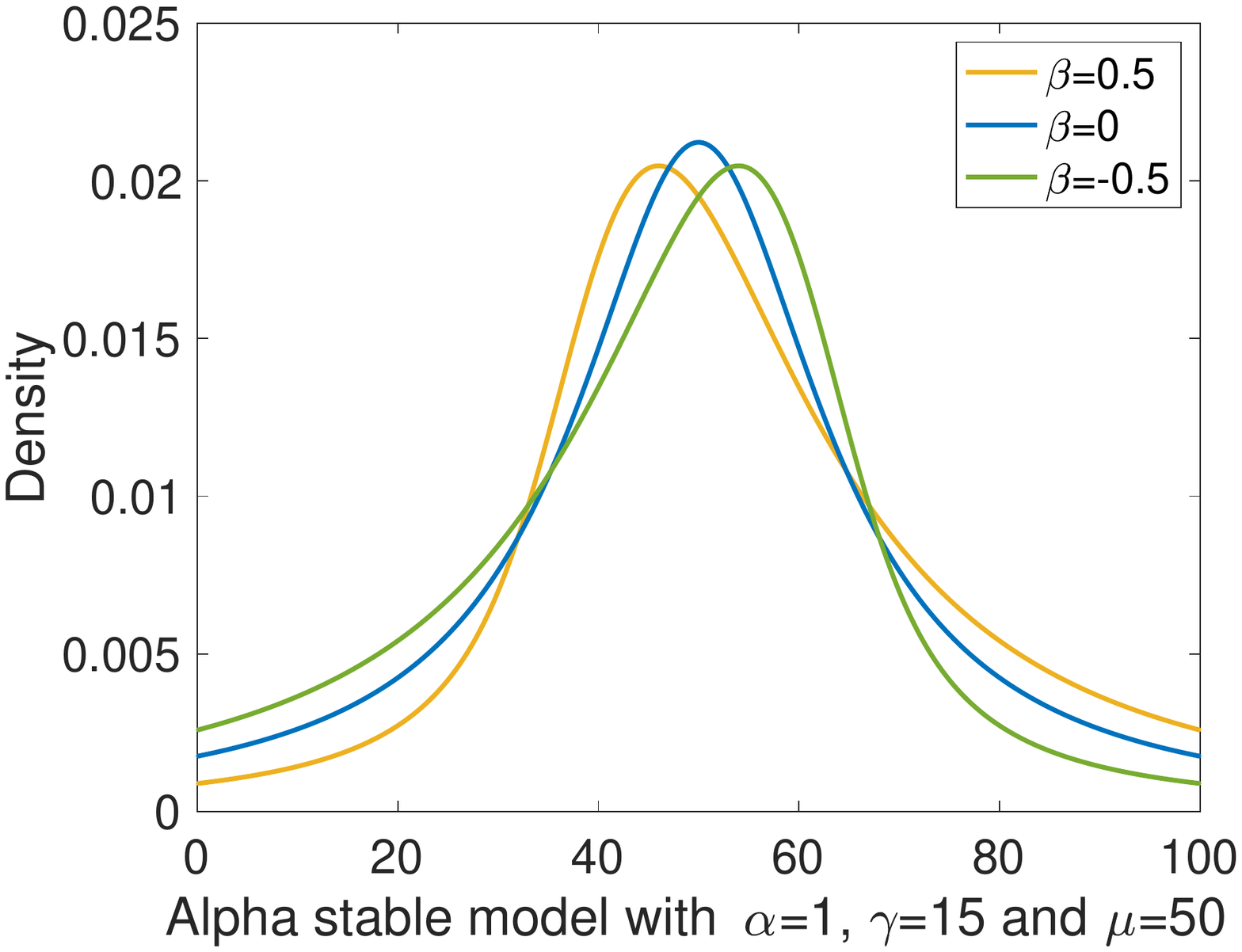}
\end{minipage}%
}%
\subfigure[]{
\begin{minipage}[t]{0.24\linewidth}
\centering
\includegraphics[scale=0.23, trim=15 200 40 200, clip]{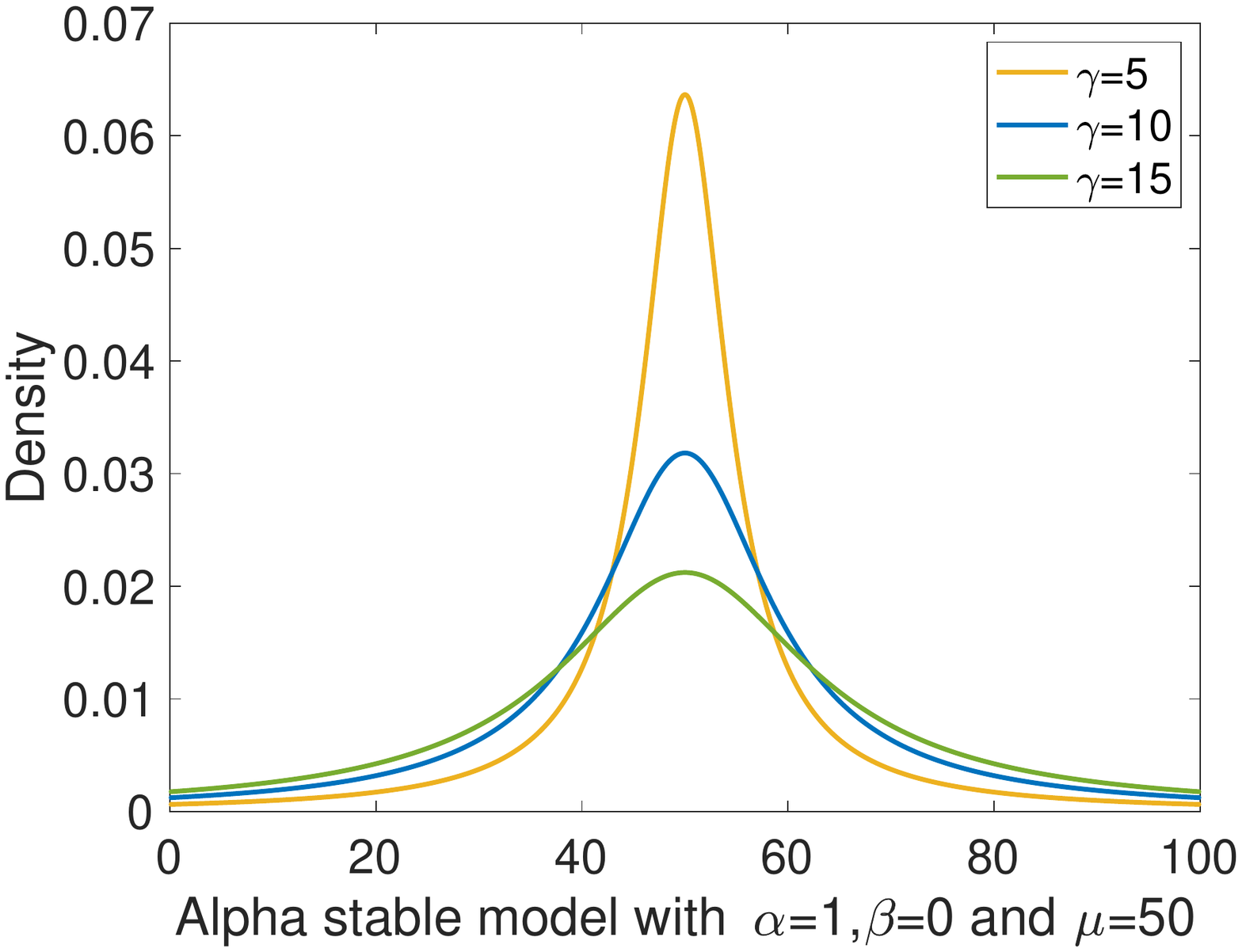}
\end{minipage}
}%
\subfigure[]{
\begin{minipage}[t]{0.24\linewidth}
\centering
\includegraphics[scale=0.23, trim=15 200 40 200, clip]{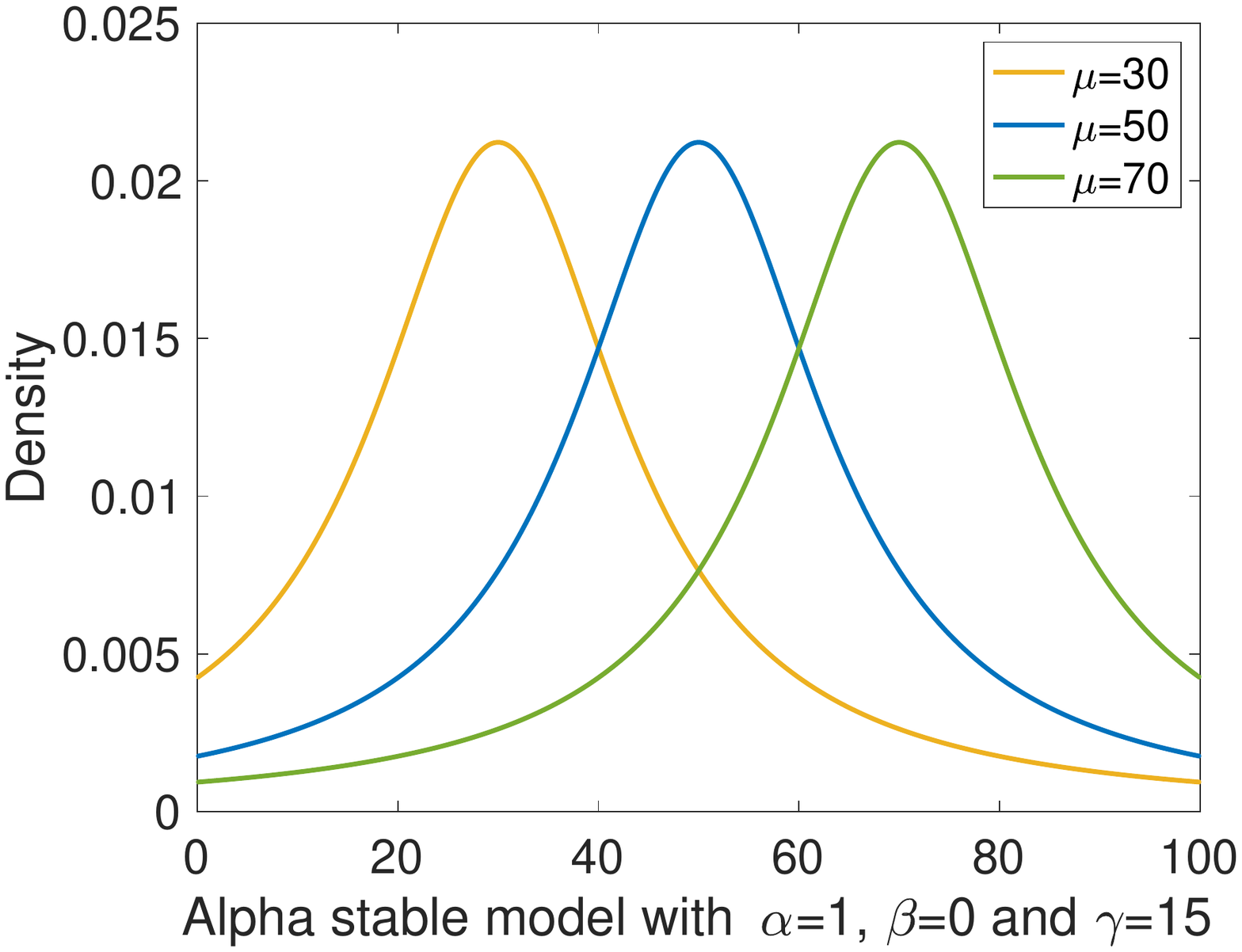}
\end{minipage}
}%
\centering
\caption{Alpha stable models. (a) shows alpha stable models with different parameter $\alpha$. (b) shows alpha stable models with different parameter $\beta$. (c) shows alpha stable models with different parameter $\gamma$. (d) shows alpha stable models with different parameter $\mu$.}
\label{f13}
\end{figure*}

\subsection{Alpha Stable Model}
\label{sssec:2.2.1}
Alpha stable model was proposed by French mathematician Levy in 1925, and developed by Feller \cite{1950An}, Samorodnitsky \cite{samorodnitsky1996stable} and others \cite{uchaikin1998chance}. 
It is a kind of probability distribution model with abundant forms, and Gaussian distribution is a special case of it. It has some characteristics that Gaussian distribution does not have, such as skewness, heavy tail and spike. Because of these important statistical characteristics, we propose to use the alpha stable model to parameterize the IQSD.
Since the probability density function (PDF) of the alpha stable model lacks analytical expression, 
this model can be completely described by its characteristic function
\begin{equation}\label{eq1}
\varphi(t;\alpha,\beta,\gamma,\mu)= \exp \left\{ j\mu t-\gamma|t|^{\alpha}\Phi(t;\alpha,\beta)\right\},
\end{equation}
where $\alpha,\beta,\gamma,\mu$ are four parameters for the alpha stable model, and
\begin{equation}\label{eq2}
\Phi(t;\alpha,\beta)=\left\{
                     \begin{array}{ll}
                      1-j {\rm sgn}(t)\beta\tan\left(\frac{\pi\alpha}{2}\right), & \hbox{$\alpha\ne 1$,} \\
                       1+j {\rm sgn}(t)\beta\frac{2}{\pi}\log|t|, & \hbox{$\alpha= 1$,}
                     \end{array}
                   \right.
\end{equation}
in which
\begin{equation}
{\rm sgn}(t)=\left\{
                     \begin{array}{ll}
                      1, & \hbox{$t> 0$,} \\
                      0, & \hbox{$t= 0$,} \\
                       -1, & \hbox{$t< 0$,}
                     \end{array}
                   \right.
\end{equation}
Specifically, 
$0<\alpha\leq2$ is the characteristic parameter, which reflects the `thickness' of the tail of the alpha stable model. The larger the $\alpha$ is, the smaller the tail is. The smaller the $\alpha$ is, the heavier the tail is. 
In particular, the alpha stable model degenerates to Gaussian distribution when $\alpha=2$.
$-1\leq\beta\leq1$ is the skewness parameter, which is used to measure the symmetry of alpha stable model. When $\beta=0$, the alpha stable model is symmetric. The greater the deviation of $\beta$ from 0, the more obvious the asymmetry of alpha stable model. If $\beta>0$, the model is inclined to the left, otherwise, it is inclined to the right.
$\gamma>0$ represents the scale parameter. The larger the $\gamma$ is, the more dispersed the model is. $-\infty<\mu<+\infty$ is the shift parameter. 
 Alpha stable models are shown in Fig. \ref{f13}.


According to the characteristic function $\varphi(t;\alpha,\beta,\gamma,\mu)$, the PDF of the alpha stable model, $F(x;\alpha,\beta,\gamma,\mu)$, can be obtained by inverse Fourier transform 

\begin{equation}\label{eq1}
F(x;\alpha,\beta,\gamma,\mu)= \frac{1}{2\pi}\int_{-\infty}^{
+\infty} \varphi(t;\alpha,\beta,\gamma,\mu)\exp\{-itx\} {\rm dt}.
\end{equation}
We propose to use the alpha stable model to describe the IQSD. Specifically, $x$ is the image quality score, and $\alpha,~\beta,~\gamma,~\mu$ are four parameters of the alpha stable model based IQSD which have the same meaning as the four parameters of the alpha stable model.



We deploy this alpha stable model to fit all IQSDs we obtained in the subjective quality assessment study. Then the empirical parameters of alpha stable model based IQSD for each image can be derived. In order to predict the alpha stable model based IQSD for a given image, a feature extraction method followed by four SVRs are used.


\subsection{Feature Extraction}
In this paper, image quality features are extracted from two inspects, including structural features and statistical features.
 \subsubsection{Structural Features}
 Instead of measuring the structural degradations from the distorted image directly, we measure the similarity between the image and its multiple pseudo reference images (MPRIs) \cite{2018Blind}. Specifically, 
  MPRIs are obtained by distorting the target image with different distortion types and degrees. In this paper, we use four most common distortion types, including JPEG compression (JPEG), JPEG2000 compression (JP2K), Gaussian blur (GBlur) and white Gaussian noise (WN), to distort the target image. Moreover, five different levels of each distortion type are considered. Then, the structural features are obtained by measuring the similarity between the target image and its MPRIs.

 Firstly, we use JPEG encoder to compress the target image $\textbf{I}$ into MPRIs $\textbf{I}_{c_{i}}$:
  \begin{equation}
\textbf{I}_{c_{i}}=JPEG(\textbf{I}, Q_{i}),~i=1,2,3,4,5,
\end{equation}
where $i$ means the $i$-th compression level of JPEG, $JPEG(\cdot,\cdot)$ donates the JPEG encoder, and $Q_{i}$ indicates the compression quality parameter. In this paper, we set $Q_{i}=2i-2$ for $i=1,2,3,4,5$.

Secondly, we use JP2K encoder to compress the target image $\textbf{I}$ into MPRIs $\textbf{I}_{k_{i}}$:
  \begin{equation}
\textbf{I}_{k_{i}}=JP2K(\textbf{I}, R_{i}),~i=1,2,3,4,5,
\end{equation}
where $i$ means the $i$-th compression level of JP2K, $JP2K(\cdot,\cdot)$ donates the JP2K encoder, and the compression ratio is $R_{i}$. In this paper, we set $R_{i}=25i$ for $i=1,2,3,4,5$.

Thirdly, we use Gaussian kernels to blur the target image $\textbf{I}$ into MPRIs $\textbf{I}_{b_{i}}$:
  \begin{equation}
\textbf{I}_{b_{i}}=\textbf{G}_{i}*\textbf{I},~i=1,2,3,4,5,
\end{equation}
where $i$ means the $i$-th blur level of GBlur, $*$ donates a convolution operator, and $\textbf{G}_{i}$ is a Gaussian kernel with the standard deviation $\sigma_{i}=0.5i$ for $i=1,2,3,4,5$.

Finally, the target image $\textbf{I}$ is added with the white Gaussian noise to generate MPRIs $\textbf{I}_{n_{i}}$:
  \begin{equation}
\textbf{I}_{n_{i}}=\textbf{I}+\mathcal{N}(0,\nu_{i}),~i=1,2,3,4,5,
\end{equation}
where $i$ means the $i$-th noise level of WN. $\mathcal{N}(0,\nu_{i})$ means the white Gaussian noise with 0 mean and $\nu_{i}$ variance, in which $\nu_{i}=0.1i$ for $i=1,2,3,4,5$. 
 
Now, for each image, we have got 20 MPRIs $\textbf{I}_{M}$ ($M\in\{c_{1},\dots,c_{5},k_{1},\dots,k_{5},b_{1},
 \dots,b_{5},n_{1},\dots,n_{5}\}$)
 by distorting the image $\textbf{I}$ with four different distortion types and five different distortion degrees. 
After that, the structural similarity between the image and its MPRIs is considered as quality features of the image. Specifically, we extract local binary pattern (LBP) features from the image and its MPRIs to get their LBP histograms, and then we measure the distance between the LBP histograms of the image and its MPRIs as the final structural features. The details of calculating LBP features are described as follows.
 
For a pixel $g_{c}$ in an image, its LBP value is calculated by the difference with its circularly symmetric neighboring pixels $g_{p}$:
 
  \begin{equation}
LBP_{P,R}=\sum_{p=0}^{P-1} u(g_{p}-g_{c}),
\end{equation}
 where $P$ is the number of neighboring pixels around the center pixel $g_{c}$, and $R$ is the distance between the neighbor pixel and the center pixel. In this paper, we set $P=4$ and $R=1$ for simplicity. 
 The unit step function $u(\cdot)$ can be written as
   \begin{equation}
u(x)=\left\{
                     \begin{array}{ll}
                      1, & \hbox{$x \geq 0$,} \\
                       0, & \hbox{$x < 0$.}
                     \end{array}
                   \right.
\end{equation}
 This means that if the gray level of the neighboring pixel is greater than or equal to the gray level of the center pixel, it is recorded as 1, otherwise it is recorded as 0. To sum up, the value of $LBP_{4,1}$ is an integer value in the range [0, 4], that is to say, for an image, there are up to five different LBP values. We calculate the LBP value of each pixel in an image, and derive the percentage of five LBP values to get the LBP histogram with five bins
\begin{equation}\label{H}
   H(k)=\sum_{m=1}^{M}\sum_{n=1}^{N}\frac{f(LBP_{4,1}(m,n),k)}{MN},~k=0,1,2,3,4,
\end{equation}
where 
 \begin{equation}
   f(x,y)=\left\{
                     \begin{array}{ll}
                      1, & \hbox{$x =y$,} \\
                       0, & \hbox{$x \neq y$,}
                     \end{array}
                   \right.
\end{equation}
 $M\times N$ is the size of the image, $k$ is the LBP value, and $LBP_{4,1}(m,n)$ represents the LBP value of pixel at point $(m,n)$ in an image, where $m\in\{1,2,\cdots,M\}$ and $n\in\{1,2,\cdots,N\}$.
 
 The LBP values can provide the structural information of an image \cite{1017623}, and distortion can change the structural information. Therefore, we attempt to measure the similarity between the image $\textbf{I}$ and its MPRIs $\textbf{I}_{M}$ to obtain the structural features according to their LBP histograms. Specifically, for a given image $\textbf{I}$ and its MPRIs $\textbf{I}_{M}$, the structural information can be extracted by Eq. (\ref{H}), and their LBP histograms are denoted as $\textbf{H}_{I}$ and $\textbf{H}_{M}$, respectively. 
 Then, the following metric is adopted to measure the similarity between $\textbf{H}_{I}$ and $\textbf{H}_{M}$:
  \begin{equation}
S_{M}=\sum _{k=0}^{K} \frac{(H_{I}(k)-H_{M}(k))^2}{H_{I}(k)+H_{M}(k)},
\end{equation}
where $K$ is the maximum LBP value and $S_{M}$ means the similarity between image $\textbf{I}$ and its MPRIs $\textbf{I}_{M}$.
The larger $S_{M}$ means the greater the deviation between $\textbf{H}_{I}$ and $\textbf{H}_{M}$. The smaller $S_{M}$ means the smaller the deviation between $\textbf{H}_{I}$ and $\textbf{H}_{M}$. If $S_{M}$ is 0, there is no deviation. 
 Finally, we can obtain twenty structural features $(S_{c_{1}},\dots,S_{c_{5}},S_{k_{1}}, \dots,S_{k_{5}}, S_{b_{1}},\dots,S_{b_{5}},S_{n_{1}},\dots,S_{n_{5}})$ for each image.

\subsubsection{Statistical Features}
In addition to extracting the structural features, we also calculate the statistical features for each image based on the classical natural scene statistics (NSS) model \cite{2009The,Mittal,2013Making}.
Specifically, we first compute the mean subtracted contrast normalized (MSCN) coefficients of each image 
\begin{equation}
   \hat I(m,n)=\frac{I(m,n)-\mu(m,n)}{\sigma(m,n)+1},
\end{equation}
where  
\begin{equation}
   \mu(m,n)=\sum_{k=-K}^{K}\sum_{l=-L}^{L}w_{k,l} I_{k,l}(m,n),
\end{equation}
\begin{equation}
   \sigma(m,n)=\sqrt{\sum_{k=-K}^{K}\sum_{l=-L}^{L}w_{k,l}( I_{k,l}(m,n)-\mu(m,n))^{2}},
\end{equation}
represent the local mean and local variance of the image, respectively. In which, $K=L=3$, and $w=\{w_{k,l}|k=-K,\cdots,K,l=-L,\cdots,L\}$ is a two-dimension (2-D) circularly-symmetric Gaussian weighting function. 
Then, a generalized Gaussian distribution (GGD) with zero mean is given by
\begin{equation}
   f(x;\nu,\sigma^2)=\frac{\nu}{2\omega\Gamma(1/\nu)} \exp \left( -\left( \frac{|x|}{\omega}\right)^\nu\right),
\end{equation}
in which 
\begin{equation}
  \omega=\sigma\sqrt{\frac{\Gamma(\frac{1}{\nu})}{\Gamma(\frac{3}{\nu})}},
\end{equation}
and
\begin{equation}
  \Gamma(a)=\int_{0}^{+\infty} t^{a-1}e^{-t}{\rm dt},~a>0,
\end{equation}
where $\nu$ is the shape parameter, and parameter $\sigma^2$ denotes the variance of the distribution.

According to \cite{2009The}, we use the GGD to fit the distribution of MSCN coefficients from two scales, including the original scale and the reduced resolution scale processed by low-pass filtering and a down sampling with the factor of 2, which gets two pairs of estimated parameters $(\nu,\sigma^2)$ and $(\nu_{*},\sigma_{*}^2)$.
These two pairs of parameters make up the statistical features for each image. 

In this part, we have extracted 24 features for each image, including 20 structural features and 4 statistical features. Next, we use SVRs to map these features to the four parameters of the alpha stable model based IQSD.

\subsection{Feature Regression for IQSD Prediction}\label{IQSD prediction}
SVR is widely used in various kinds of regression problems because of its simplicity and efficiency.
 In this paper, four SVRs are used to learn the mappings from the 
24 features extracted from each image to the four parameters of the alpha stable model based IQSD respectively. 
Specifically, we first fit the alpha stable model with IQSD, and get the four empirical parameters $(\alpha,\beta,\gamma~\rm{and}~\mu)$ for each image.
Then we catenate all features extracted from an image into a 24 dimensional feature vector
\begin{align}
\nonumber
  \textbf{v}=&[S_{c_{1}},\dots,S_{c_{5}},S_{k_{1}}, \dots,S_{k_{5}}, S_{b_{1}},\dots,S_{b_{5}},S_{n_{1}},\dots, \\
  &S_{n_{5}},\nu,\sigma^2,\nu_{*},\sigma_{*}^2].
\end{align}
Finally, using the images in the training set $\Phi$, 
we utilize the four empirical parameters $(\alpha,\beta,\gamma~\rm{and}~\mu)$ of the alpha stable model based IQSD and the feature vector $\textbf{v}$ to train four regressors.

The four regressors are obtained as
\begin{equation}
  regressor_{\alpha}={\rm SVR\_TRAIN_{\alpha}}( \textbf{v}_{i},\alpha_{i}),~i\in \Phi,
\end{equation}
\begin{equation}
  regressor_{\beta}={\rm SVR\_TRAIN_{\beta}}( \textbf{v}_{i},\beta_{i}),~i\in \Phi,
\end{equation}
\begin{equation}
  regressor_{\gamma}={\rm SVR\_TRAIN_{\gamma}}( \textbf{v}_{i},\gamma_{i}),~i\in \Phi,
\end{equation}
\begin{equation}
  regressor_{\mu}={\rm SVR\_TRAIN_{\mu}}( \textbf{v}_{i},\mu_{i}),~i\in \Phi,
\end{equation}
where $i$ is the image index, and $regressor_{\alpha}$, $regressor_{\beta}$, $regressor_{\gamma}$, $regressor_{\mu}$ represent the trained regressors for parameters $\alpha,\beta,\gamma,~\mu$, respectively. 

After training stage, for each test image, we can use four trained regressors and its feature vector $\textbf{v}$ to predict four parameters $(\hat\alpha,\hat\beta,\hat\gamma,\hat\mu)$ of alpha stable model based IQSD as follows
\begin{equation}
  \hat\alpha={\rm SVR\_PREDICT_{\alpha}}( \textbf{v},regressor_{\alpha}),
\end{equation}
\begin{equation}
  \hat\beta={\rm SVR\_PREDICT_{\beta}}( \textbf{v},regressor_{\beta}),
\end{equation}
\begin{equation}
  \hat\gamma={\rm SVR\_PREDICT_{\gamma}}( \textbf{v},regressor_{\gamma}),
\end{equation}
\begin{equation}
 \hat\mu={\rm SVR\_PREDICT_{\mu}}( \textbf{v}, regressor_{\mu}).
\end{equation}
In particular, LIBSVM \cite{Chang} is used to implement all SVRs, in which the radial basis function is chosen as the kernel function. 

The obtained four parameter regression models and the feature extraction model together constitute the prediction framework of the alpha stable model based IQSD. In this way, after feature extraction and parameter prediction, we can get the alpha stable model based IQSD for each image as follows
\begin{equation}\label{eq1}
F(x;\hat{\alpha},\hat \beta,\hat \gamma,\hat \mu)= \frac{1}{2\pi}\int_{-\infty}^{
+\infty} \hat {\varphi}(t;\hat{\alpha},\hat \beta,\hat \gamma,\hat \mu)\exp\{-itx\} {\rm dt},
\end{equation}
where 
\begin{equation}\label{eq1}
\hat\varphi(t;\hat{\alpha},\hat \beta,\hat \gamma,\hat \mu)= \exp \left\{ j\hat\mu t-\hat\gamma|t|^{\hat\alpha}\Phi(t;\hat\alpha,\hat \beta)\right\},
\end{equation}
in which 
\begin{equation}\label{eq2}
\Phi(t;\hat\alpha,\hat \beta)=\left\{
                     \begin{array}{ll}
                      1-j {\rm sgn}(t)\hat\beta\tan\left(\frac{\pi\hat\alpha}{2}\right), & \hbox{$\hat\alpha\ne 1$,} \\
                       1+j {\rm sgn}(t)\hat\beta\frac{2}{\pi}\log|t|, & \hbox{$\hat\alpha= 1$.}
                     \end{array}
                   \right.
\end{equation}

\section{Experiment and Analysis}
\label{sec:3}

In this section, experiments are carried out to verify the feasibility and effectiveness of our proposed alpha stable model based IQSD. The experiments consist of two parts. At first,
we analyze the alpha stable model based IQSD in details, and its superiority is proved. Secondly,
we conduct experiments to verify the effectiveness and feasibility of the proposed alpha stable model based IQSD prediction framework.

\begin{figure*}[h]
\centering
\subfigure[$\alpha$ vs. MOS]{
\begin{minipage}[t]{0.24\linewidth}
\centering
\includegraphics[scale=0.23, trim=15 200 40 200, clip]{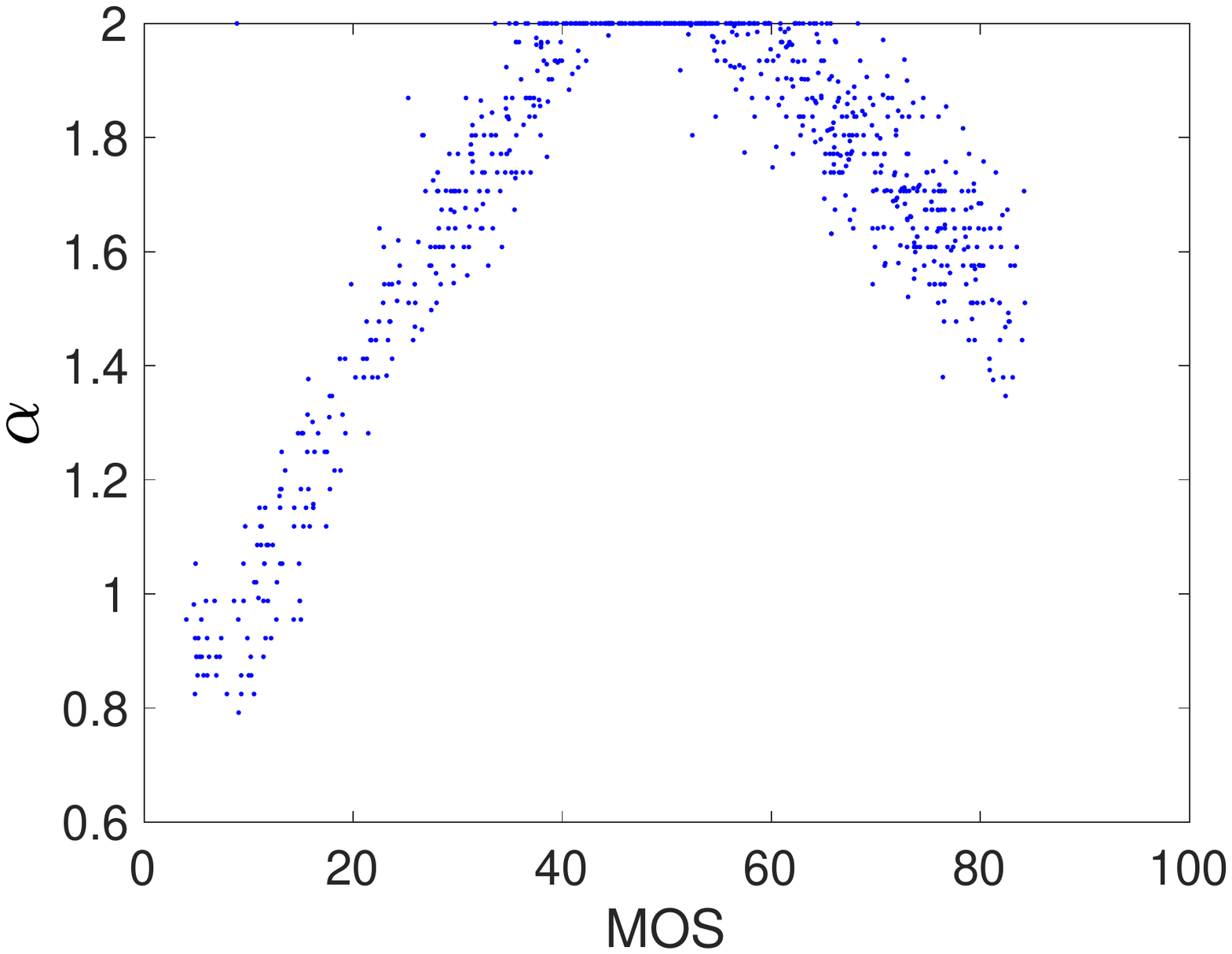}
\end{minipage}%
}%
\subfigure[$\beta$ vs. MOS]{
\begin{minipage}[t]{0.24\linewidth}
\centering
\includegraphics[scale=0.23, trim=15 200 40 200, clip]{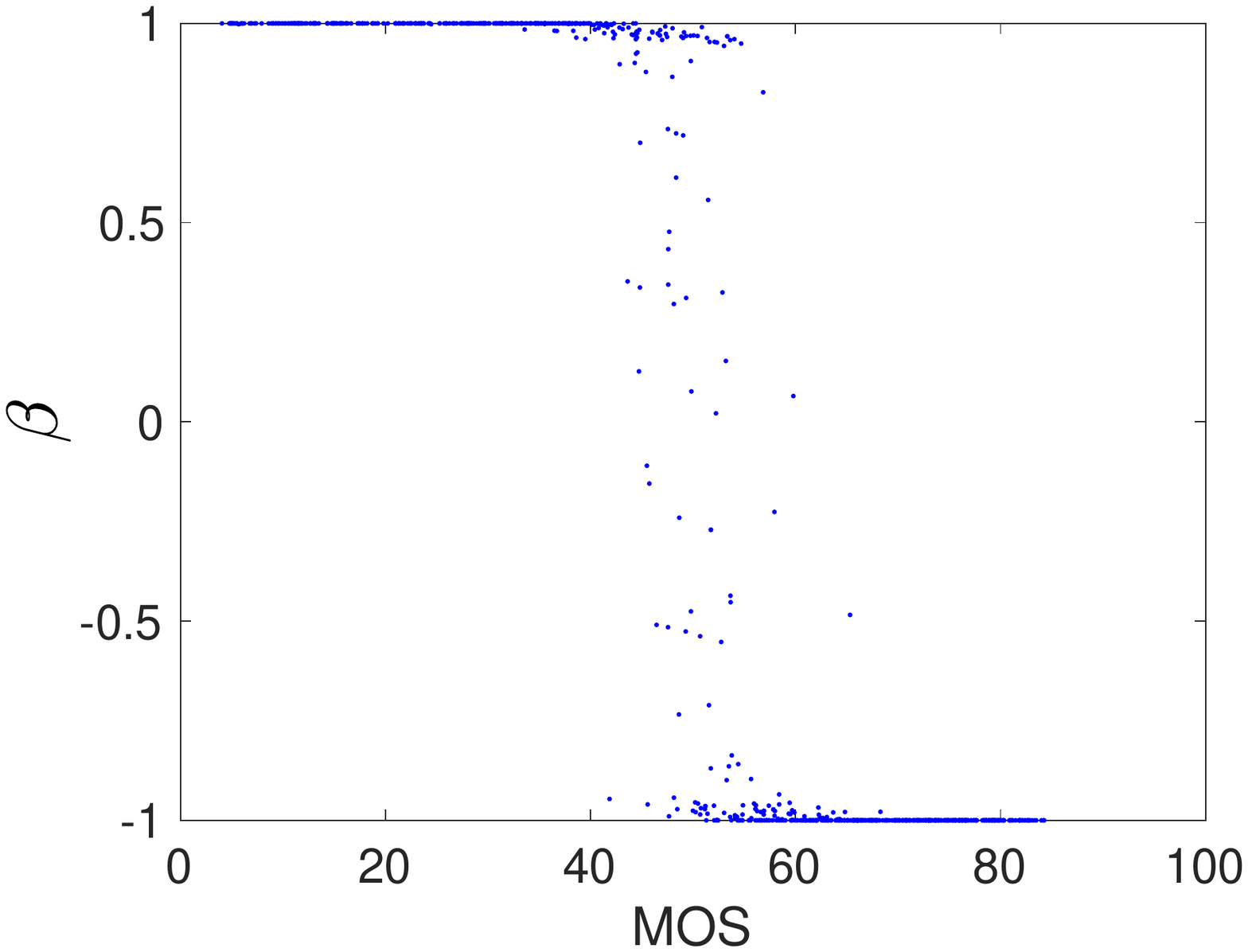}
\end{minipage}%
}%
\subfigure[$\gamma$ vs. MOS]{
\begin{minipage}[t]{0.24\linewidth}
\centering
\includegraphics[scale=0.23, trim=15 200 40 200, clip]{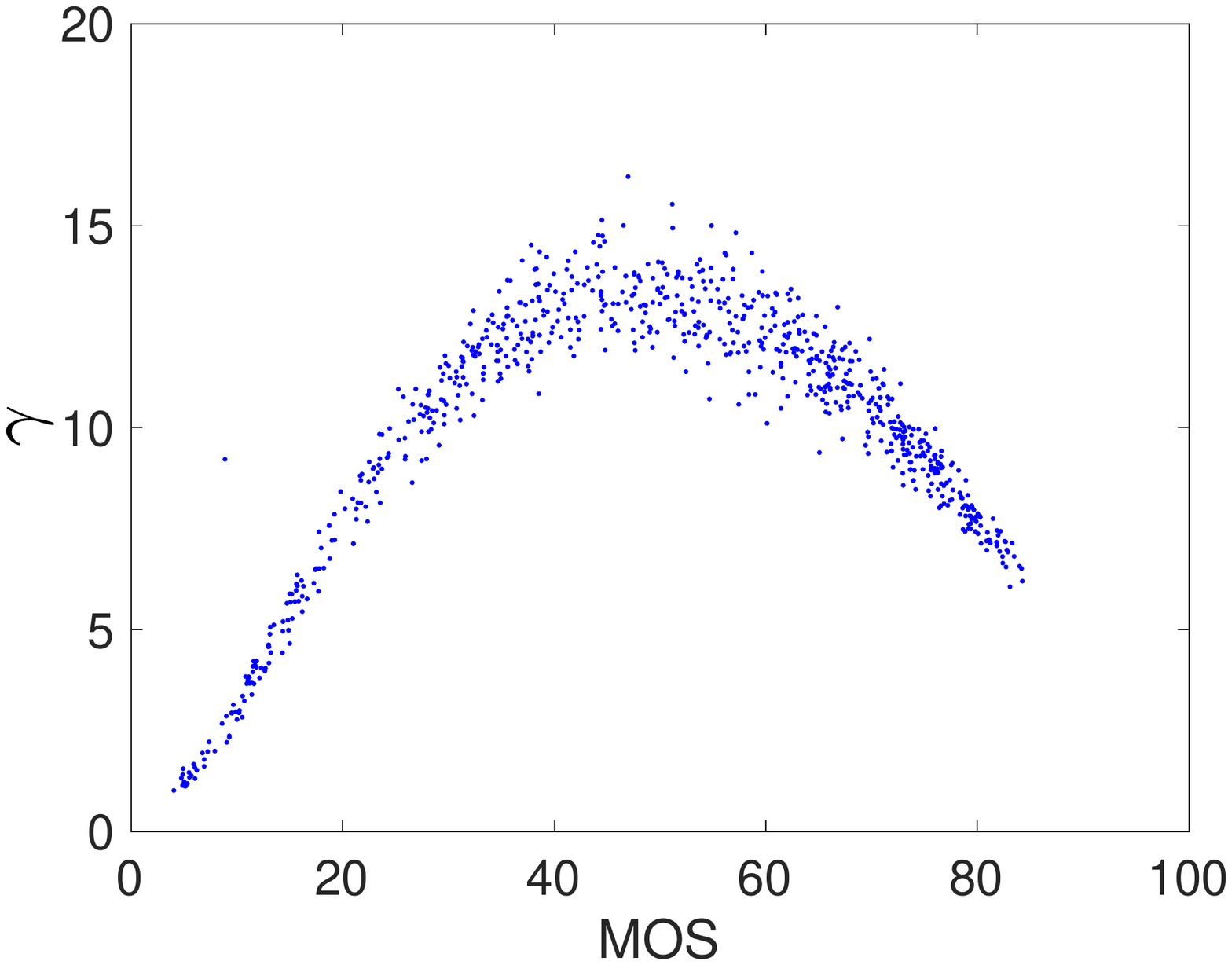}
\end{minipage}
}%
\subfigure[$\mu$ vs. MOS]{
\begin{minipage}[t]{0.24\linewidth}
\centering
\includegraphics[scale=0.23, trim=15 200 40 200, clip]{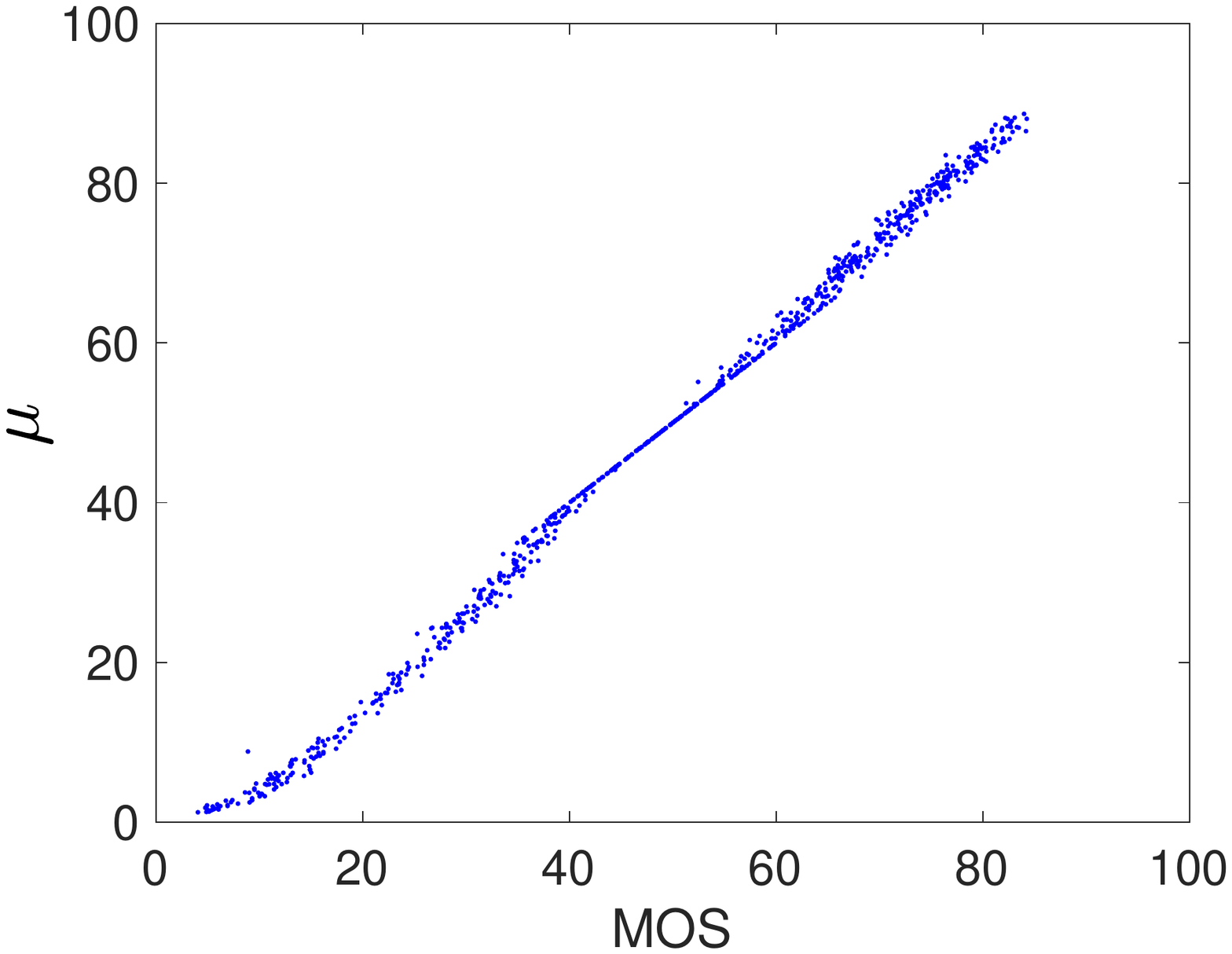}
\end{minipage}
}%
\centering
\caption{The relationship between four parameters of the alpha stable model based IQSD and MOS. }
\label{f4}
\end{figure*}

 \begin{figure*}[h]
\centering
\subfigure[$\alpha$ vs. SOS]{
\begin{minipage}[t]{0.24\linewidth}
\centering
\includegraphics[scale=0.23, trim=15 200 40 200, clip]{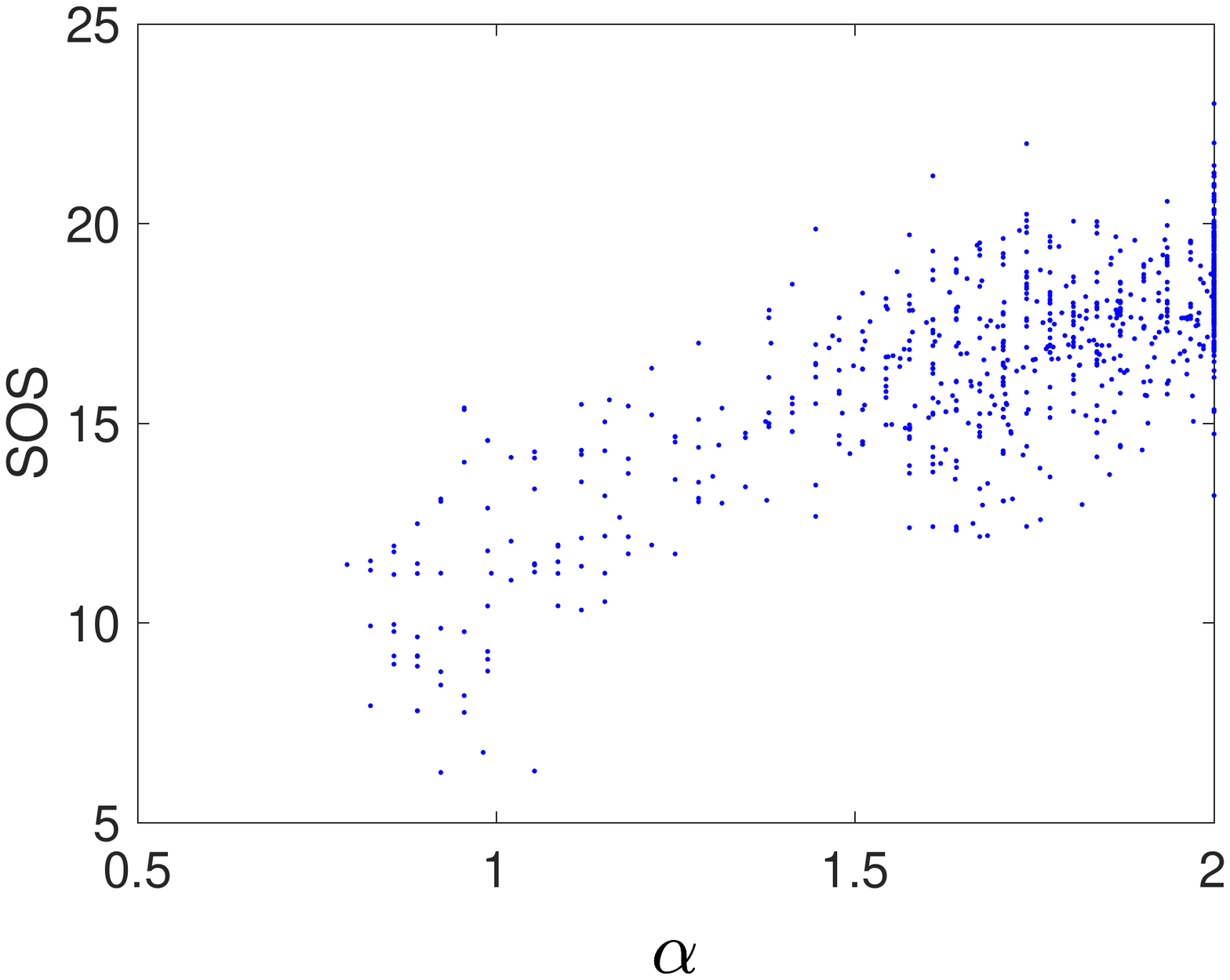}
\end{minipage}%
}%
\subfigure[$\beta$ vs. SOS]{
\begin{minipage}[t]{0.24\linewidth}
\centering
\includegraphics[scale=0.23, trim=15 200 40 200, clip]{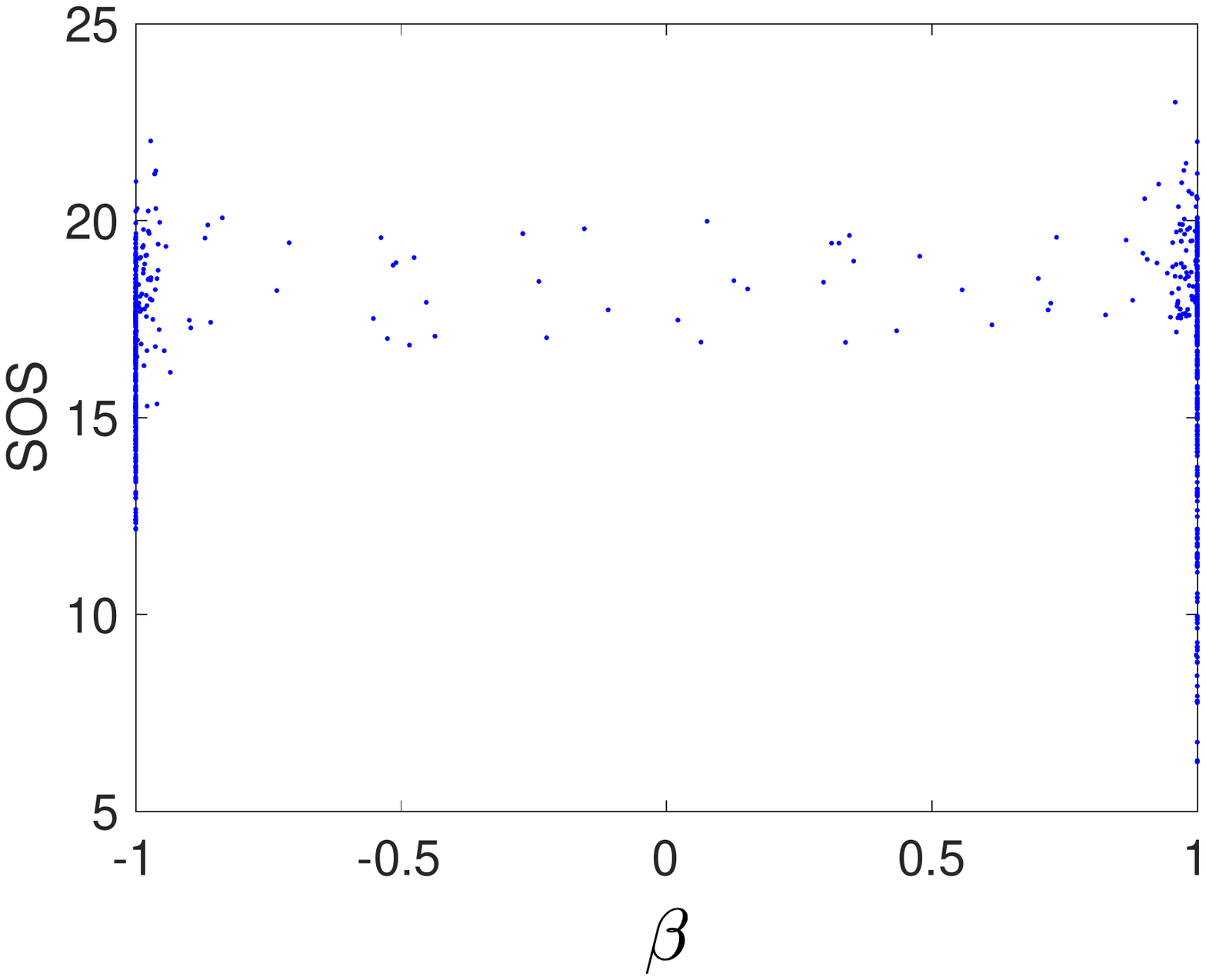}
\end{minipage}%
}%
\subfigure[$\gamma$ vs. SOS]{
\begin{minipage}[t]{0.24\linewidth}
\centering
\includegraphics[scale=0.23, trim=15 200 40 200, clip]{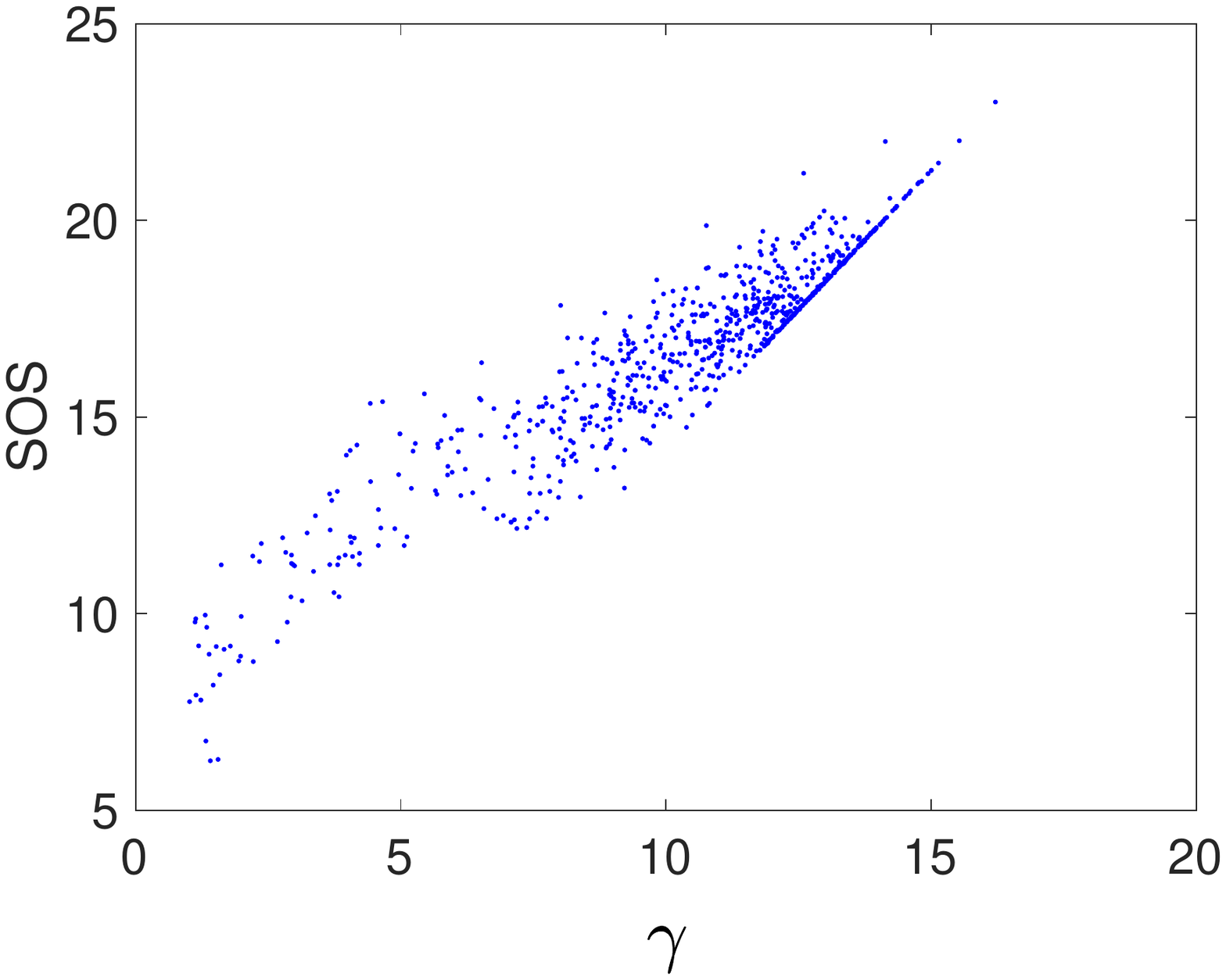}
\end{minipage}
}%
\subfigure[$\mu$ vs. SOS]{
\begin{minipage}[t]{0.24\linewidth}
\centering
\includegraphics[scale=0.23, trim=15 200 40 200, clip]{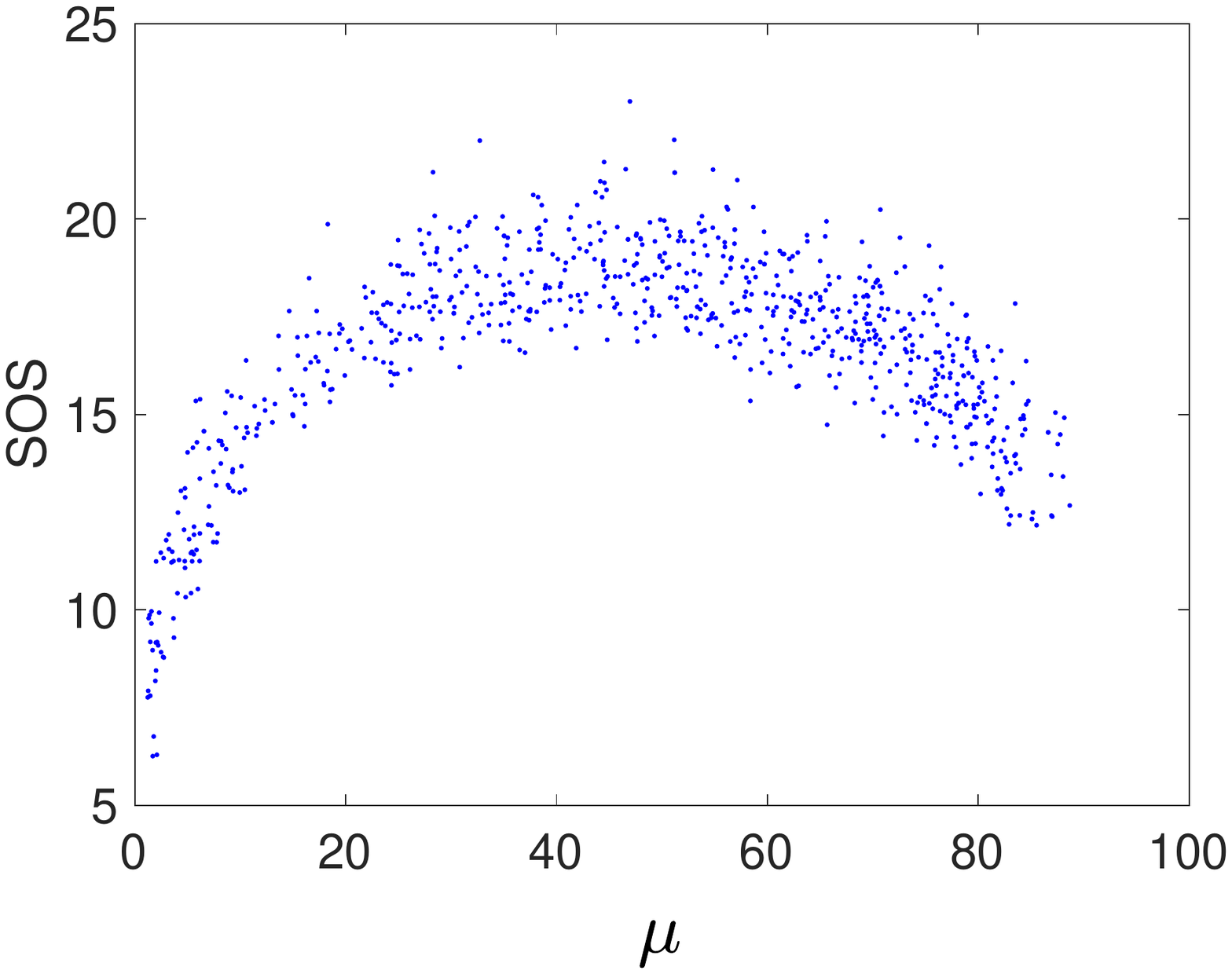}
\end{minipage}
}%
\centering
\caption{The relationship between four parameters of the alpha stable model based IQSD and SOS. }
\label{f14}
\end{figure*}

\begin{figure*}[h]
\centering
\subfigure[$\alpha$ vs. Skewness]{
\begin{minipage}[t]{0.24\linewidth}
\centering
\includegraphics[scale=0.23, trim=15 200 40 200, clip]{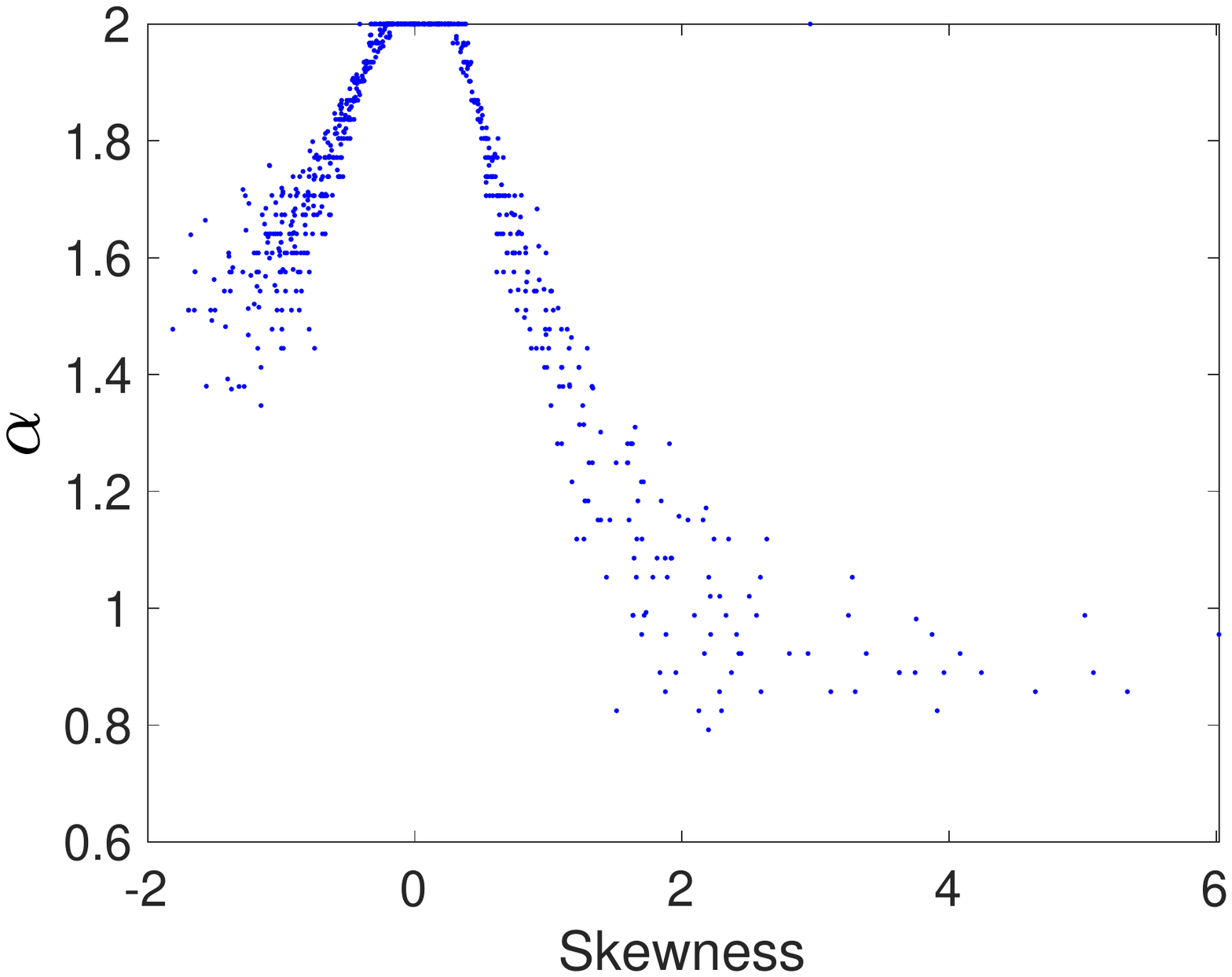}
\end{minipage}%
}%
\subfigure[$\beta$ vs. Skewness]{
\begin{minipage}[t]{0.24\linewidth}
\centering
\includegraphics[scale=0.23, trim=15 200 40 200, clip]{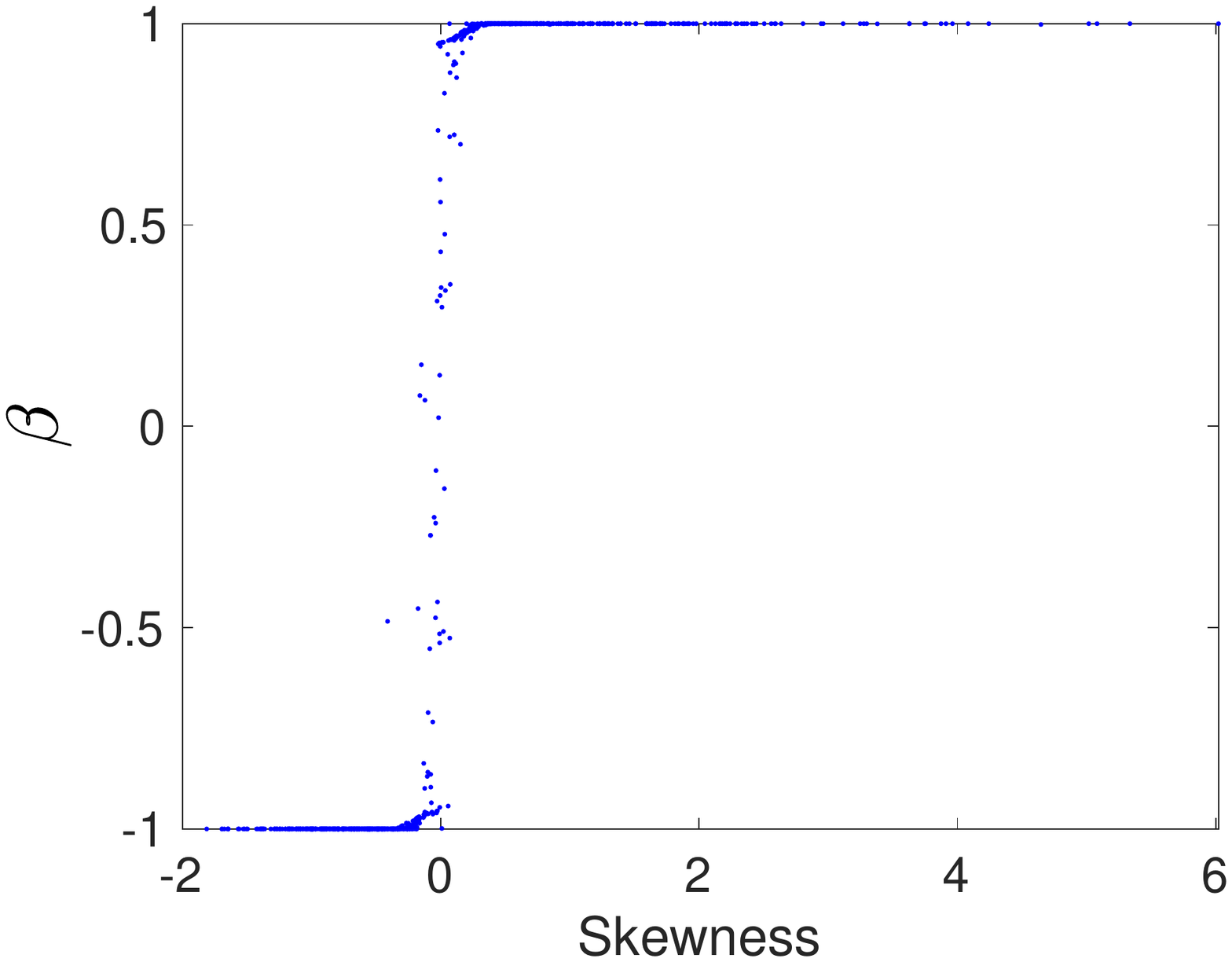}
\end{minipage}%
}%
\subfigure[$\gamma$ vs. Skewness]{
\begin{minipage}[t]{0.24\linewidth}
\centering
\includegraphics[scale=0.23, trim=15 200 40 200, clip]{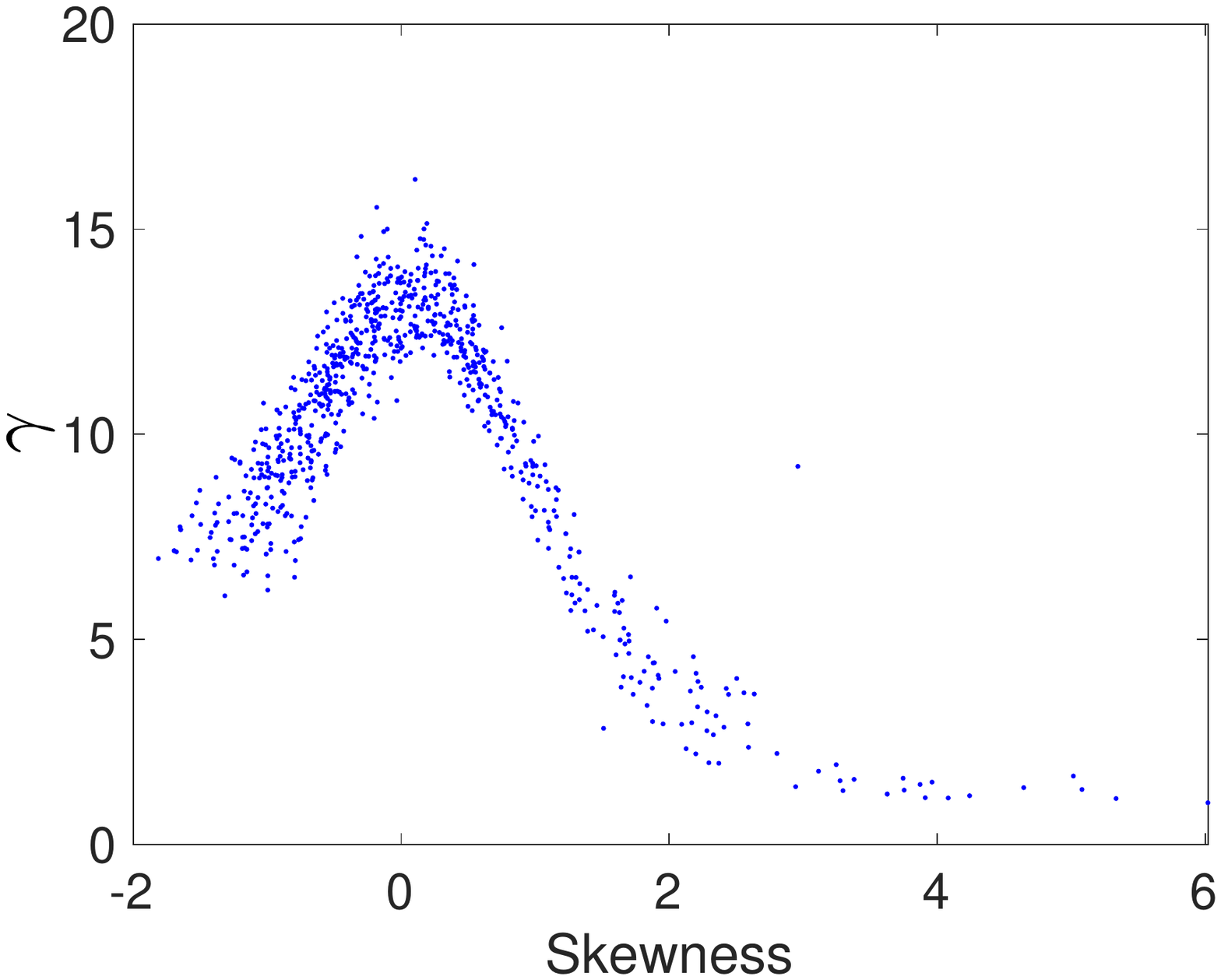}
\end{minipage}
}%
\subfigure[$\mu$ vs. Skewness]{
\begin{minipage}[t]{0.24\linewidth}
\centering
\includegraphics[scale=0.23, trim=15 200 40 200, clip]{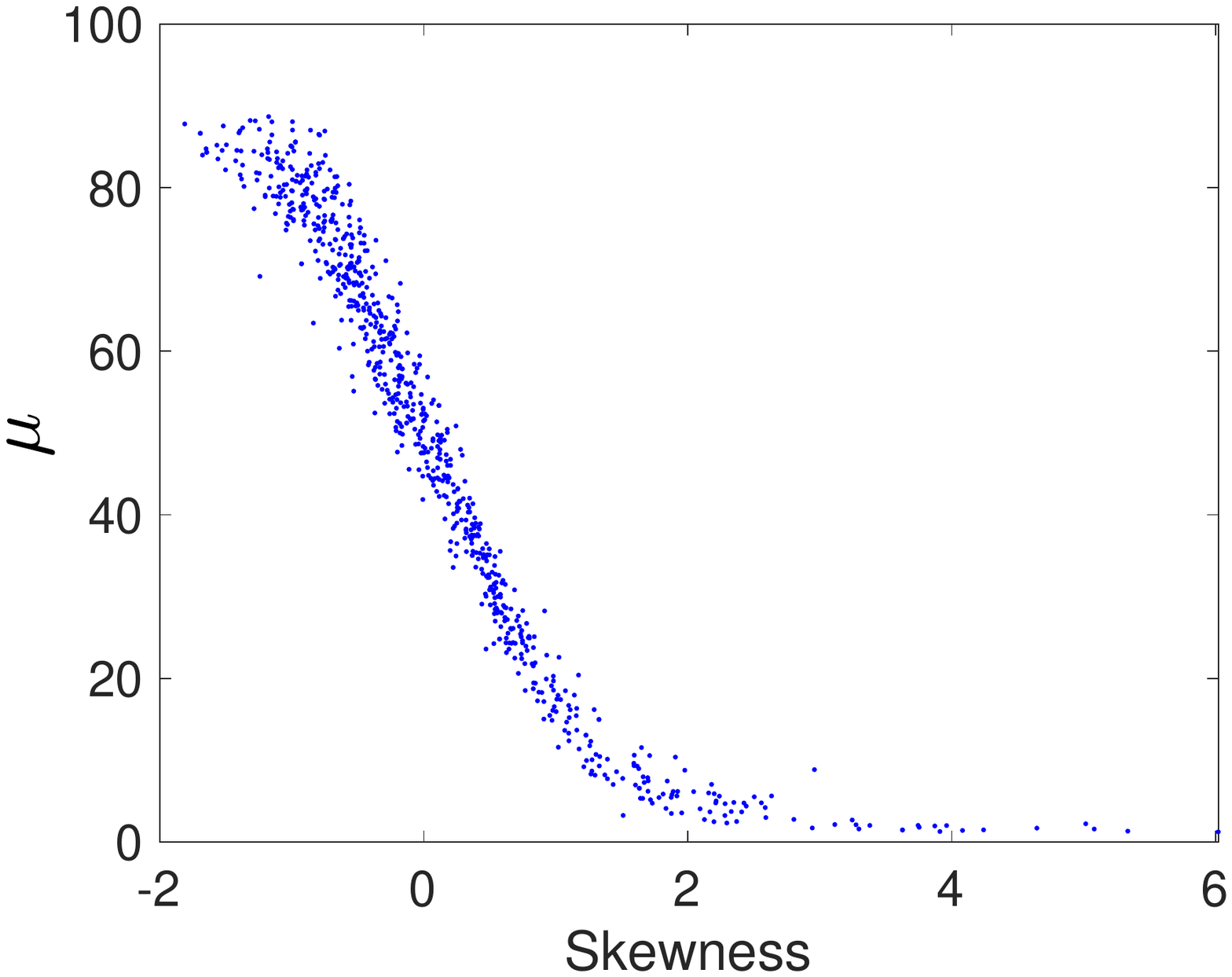}
\end{minipage}
}%
\centering
\caption{The relationship between four parameters of the alpha stable model based IQSD and the skewness of opinion score. }
\label{f16}
\end{figure*}



\begin{figure*}[h]
\centering
\subfigure[$\alpha$ vs. MOS for distortion types.]{
\begin{minipage}[t]{0.24\linewidth}
\centering
\includegraphics[scale=0.23, trim=15 200 40 200, clip]{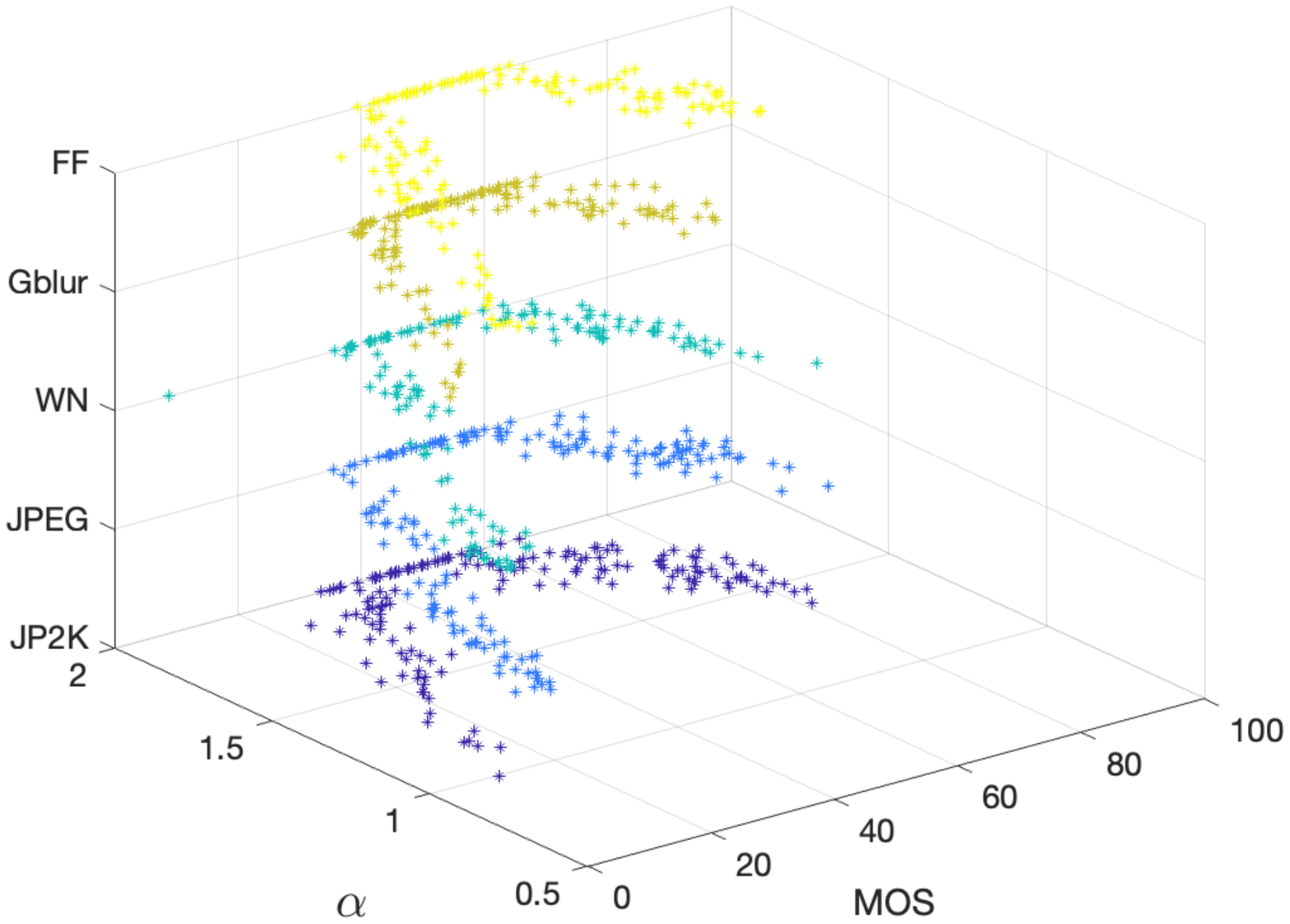}
\end{minipage}%
}%
\subfigure[$\beta$ vs. MOS for distortion types.]{
\begin{minipage}[t]{0.24\linewidth}
\centering
\includegraphics[scale=0.23, trim=15 200 40 200, clip]{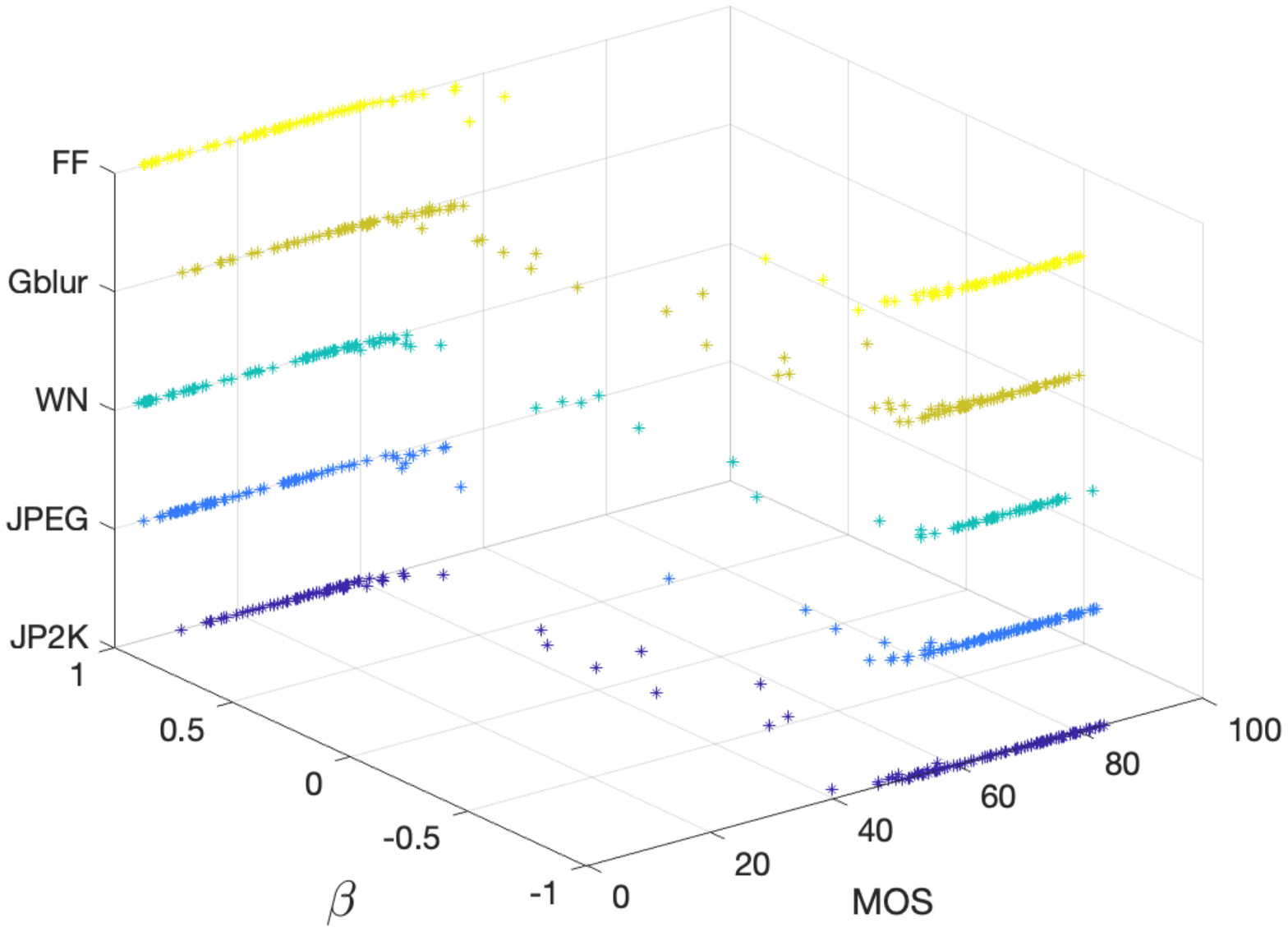}
\end{minipage}%
}%
\subfigure[$\gamma$ vs. MOS for distortion types.]{
\begin{minipage}[t]{0.24\linewidth}
\centering
\includegraphics[scale=0.23, trim=15 200 40 200, clip]{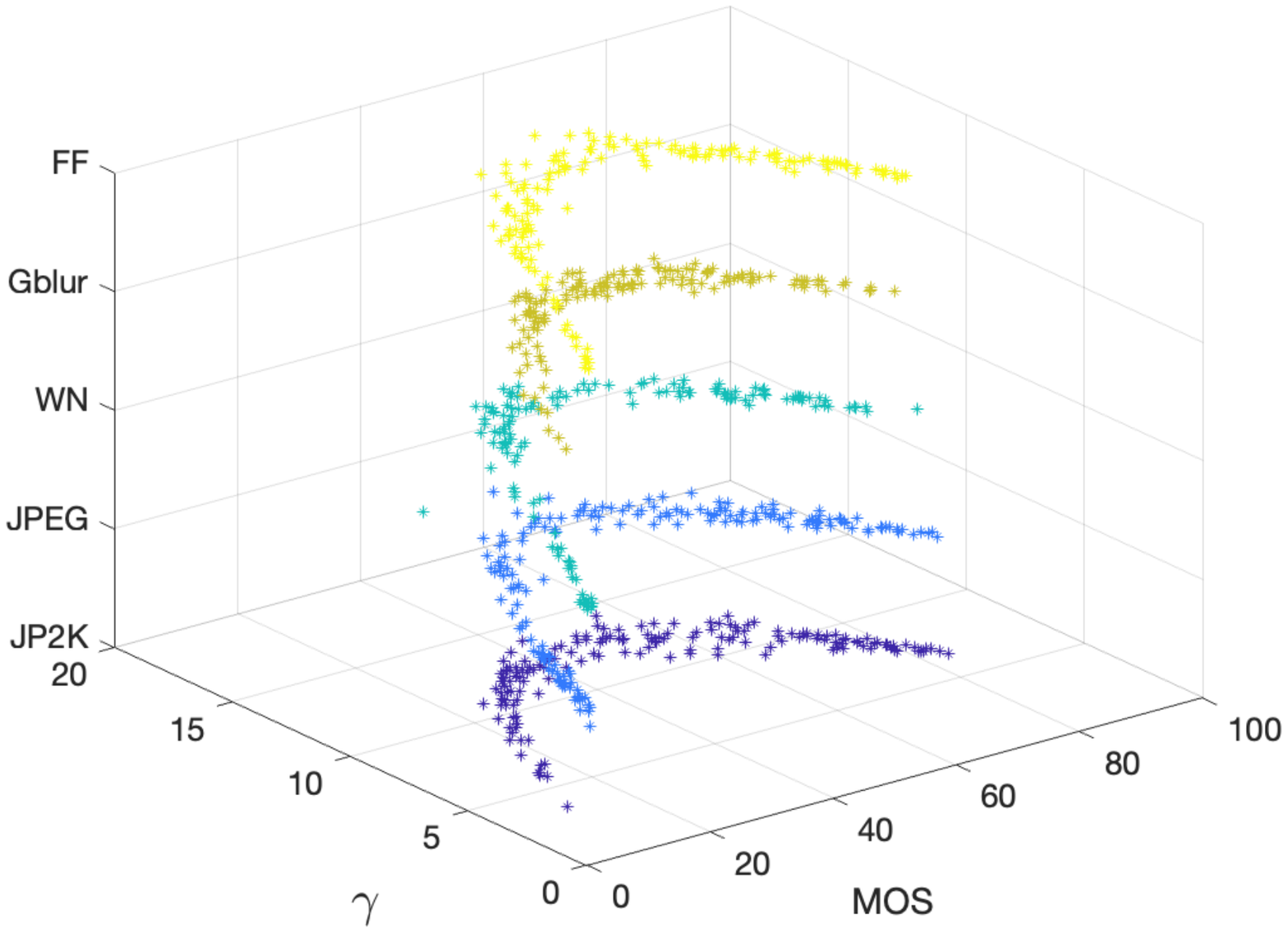}
\end{minipage}
}%
\subfigure[$\mu$ vs. MOS for distortion types.]{
\begin{minipage}[t]{0.24\linewidth}
\centering
\includegraphics[scale=0.23, trim=15 200 40 200, clip]{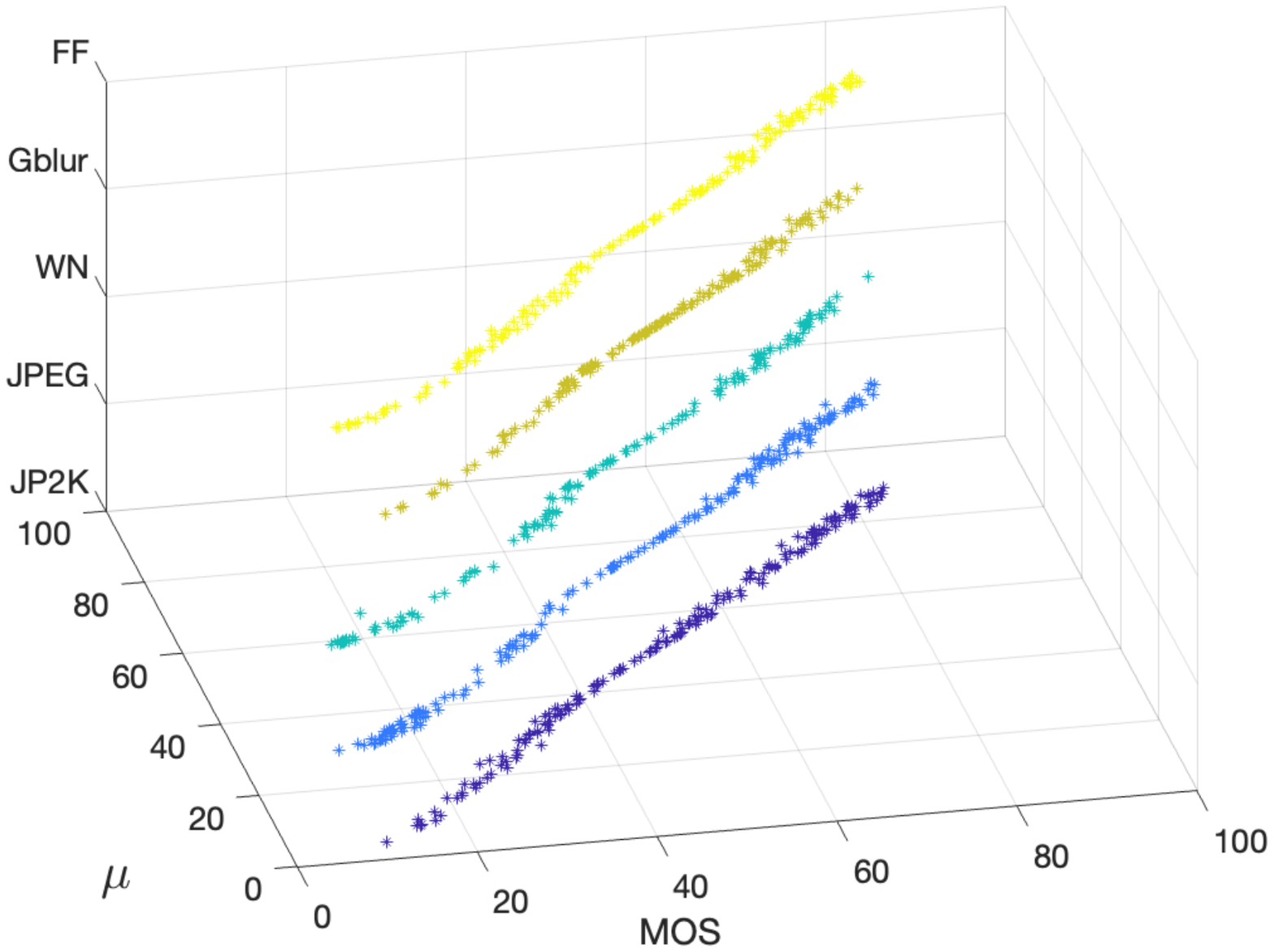}
\end{minipage}
}%

%
%
\subfigure[$\alpha$ vs. SOS for distortion types.]{
\begin{minipage}[t]{0.24\linewidth}
\centering
\includegraphics[scale=0.23, trim=15 200 40 200, clip]{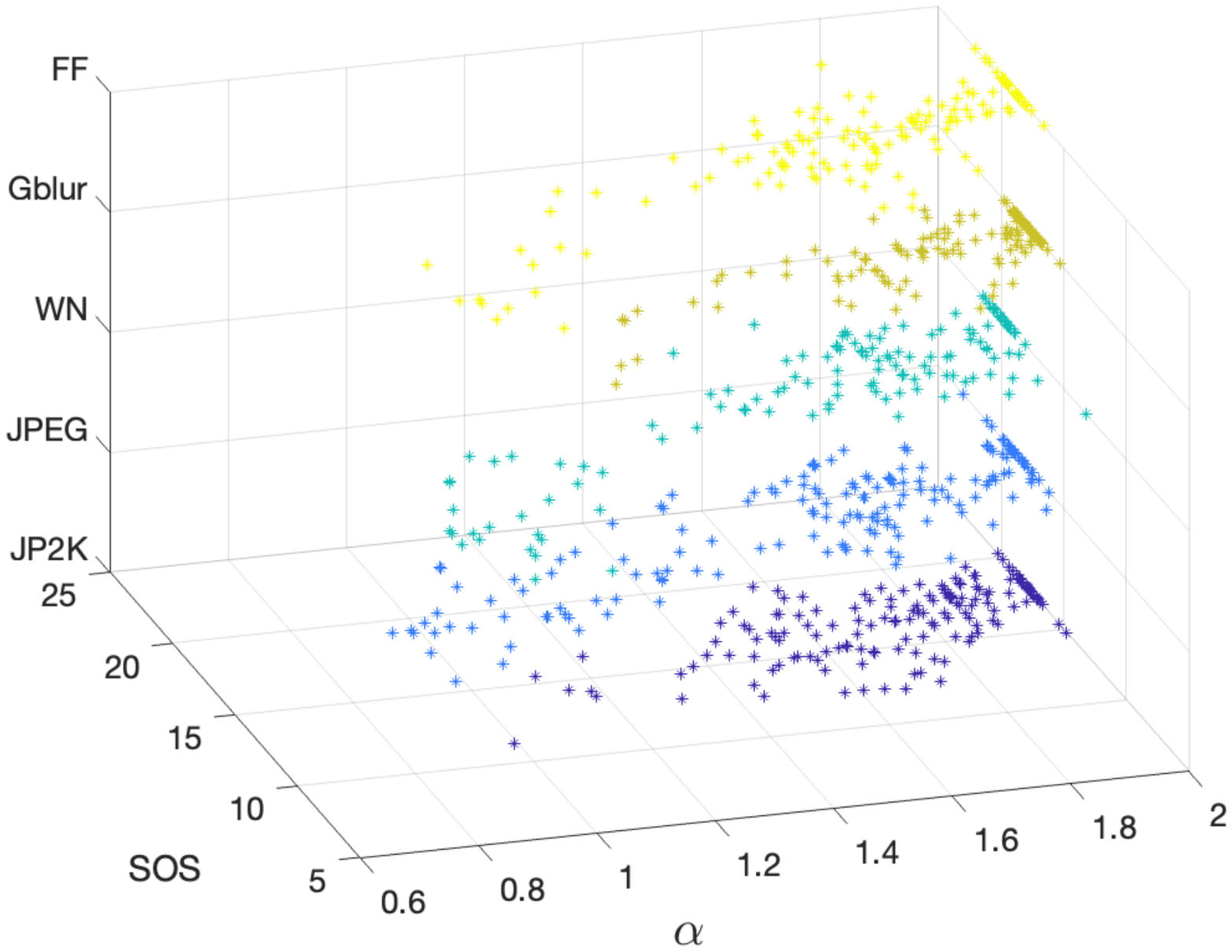}
\end{minipage}%
}%
\subfigure[$\beta$ vs. SOS for distortion types.]{
\begin{minipage}[t]{0.24\linewidth}
\centering
\includegraphics[scale=0.23, trim=15 200 40 200, clip]{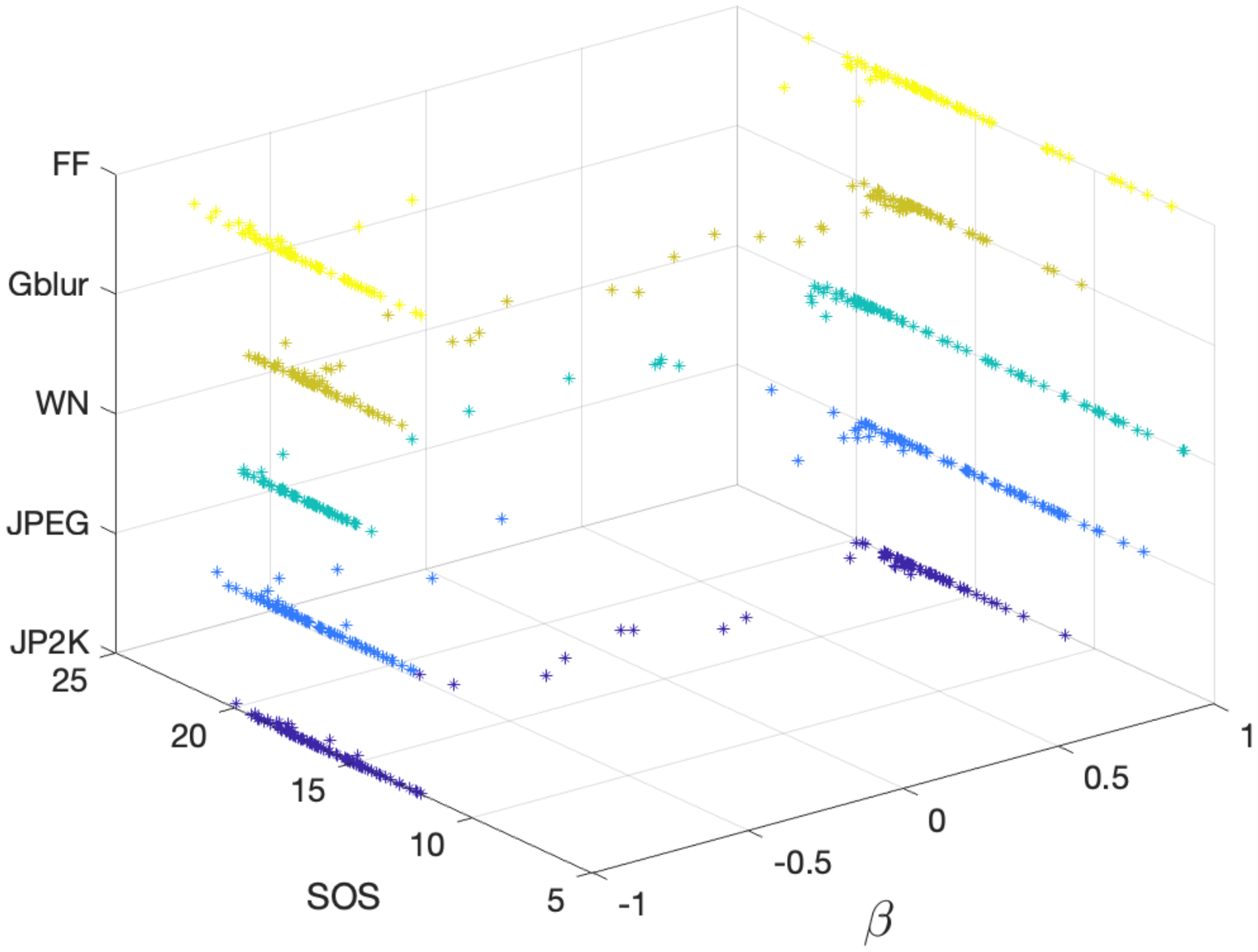}
\end{minipage}%
}%
\subfigure[$\gamma$ vs. SOS for distortion types.]{
\begin{minipage}[t]{0.24\linewidth}
\centering
\includegraphics[scale=0.23, trim=15 200 40 200, clip]{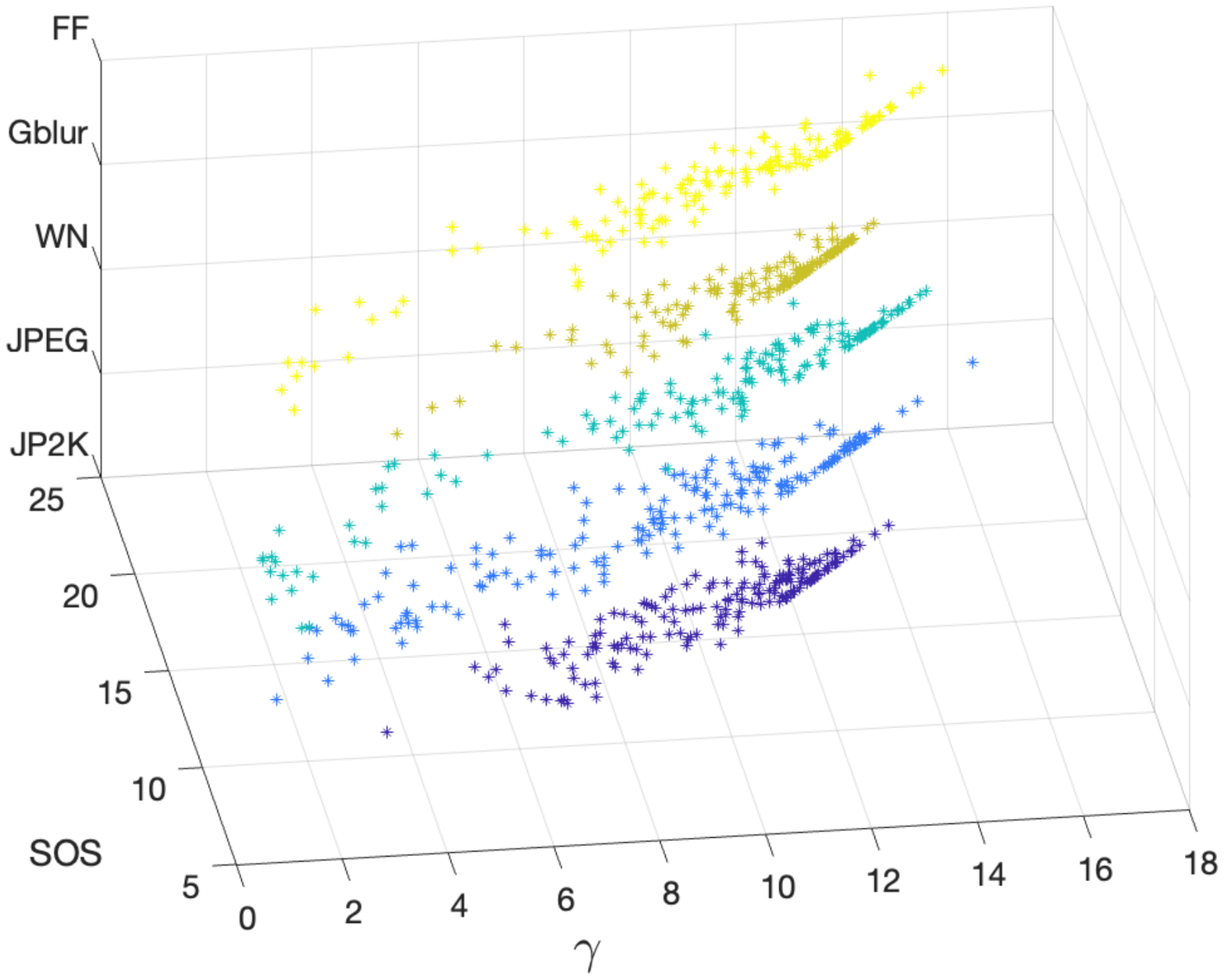}
\end{minipage}
}%
\subfigure[$\mu$ vs. SOS for distortion types.]{
\begin{minipage}[t]{0.24\linewidth}
\centering
\includegraphics[scale=0.23, trim=15 200 40 200, clip]{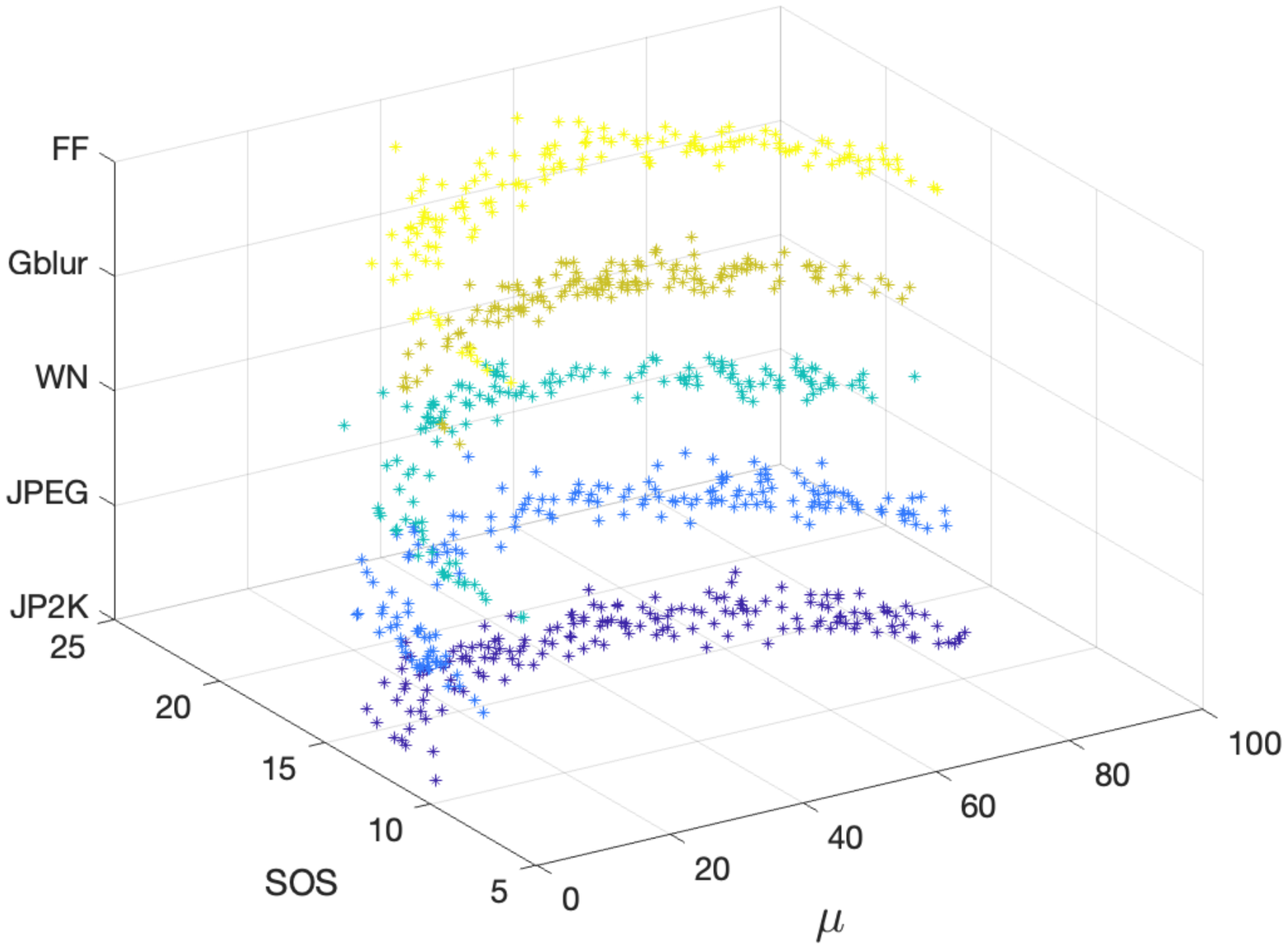}
\end{minipage}
}%

%
%
\subfigure[$\alpha$ vs. Skewness for distortion ~~~ \newline types.]{
\begin{minipage}[t]{0.24\linewidth}
\centering
\includegraphics[scale=0.23, trim=15 200 40 200, clip]{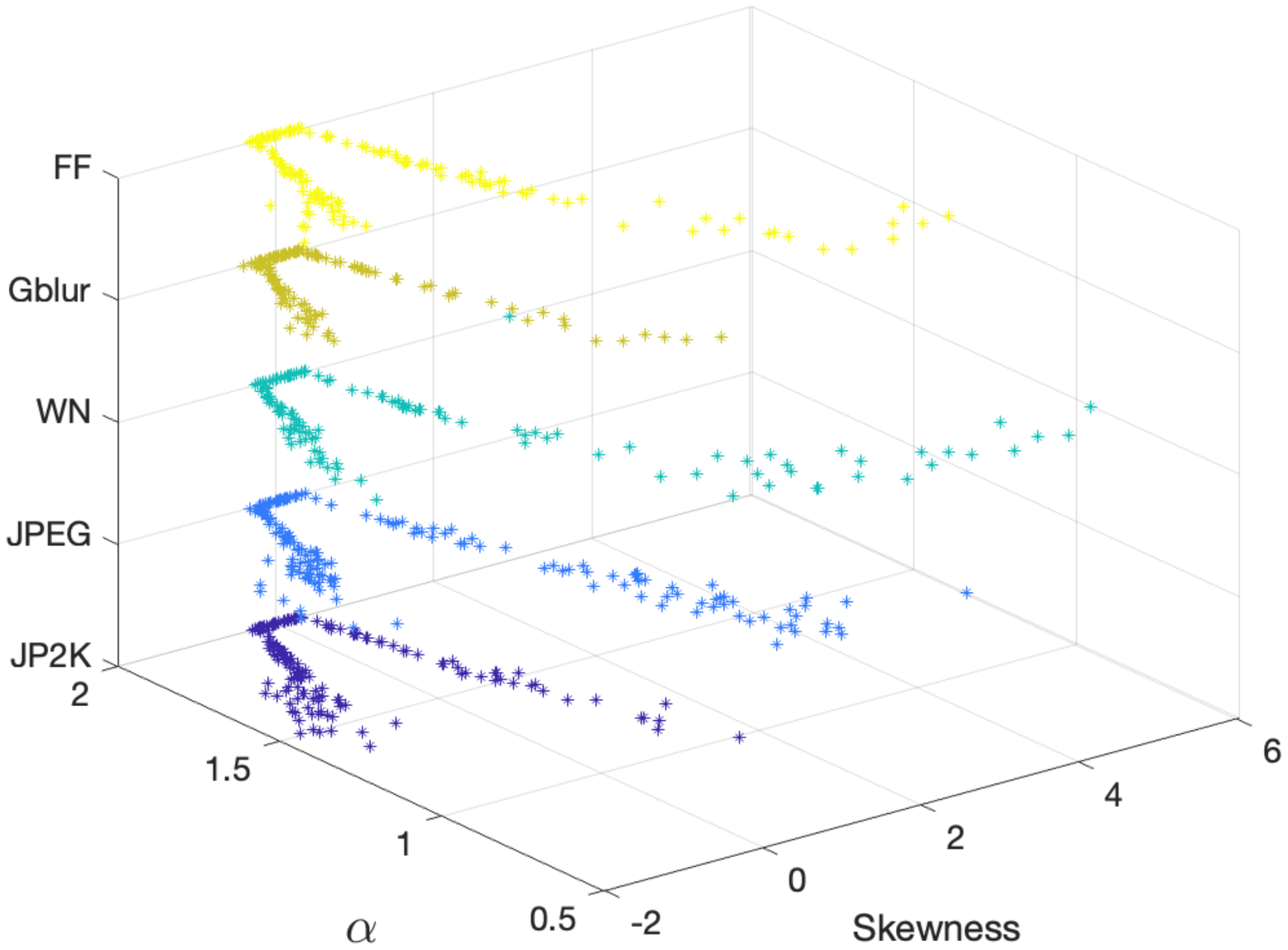}
\end{minipage}%
}%
\subfigure[$\beta$ vs. Skewness for distortion ~~~~~ \newline types.]{
\begin{minipage}[t]{0.24\linewidth}
\centering
\includegraphics[scale=0.23, trim=15 200 40 200, clip]{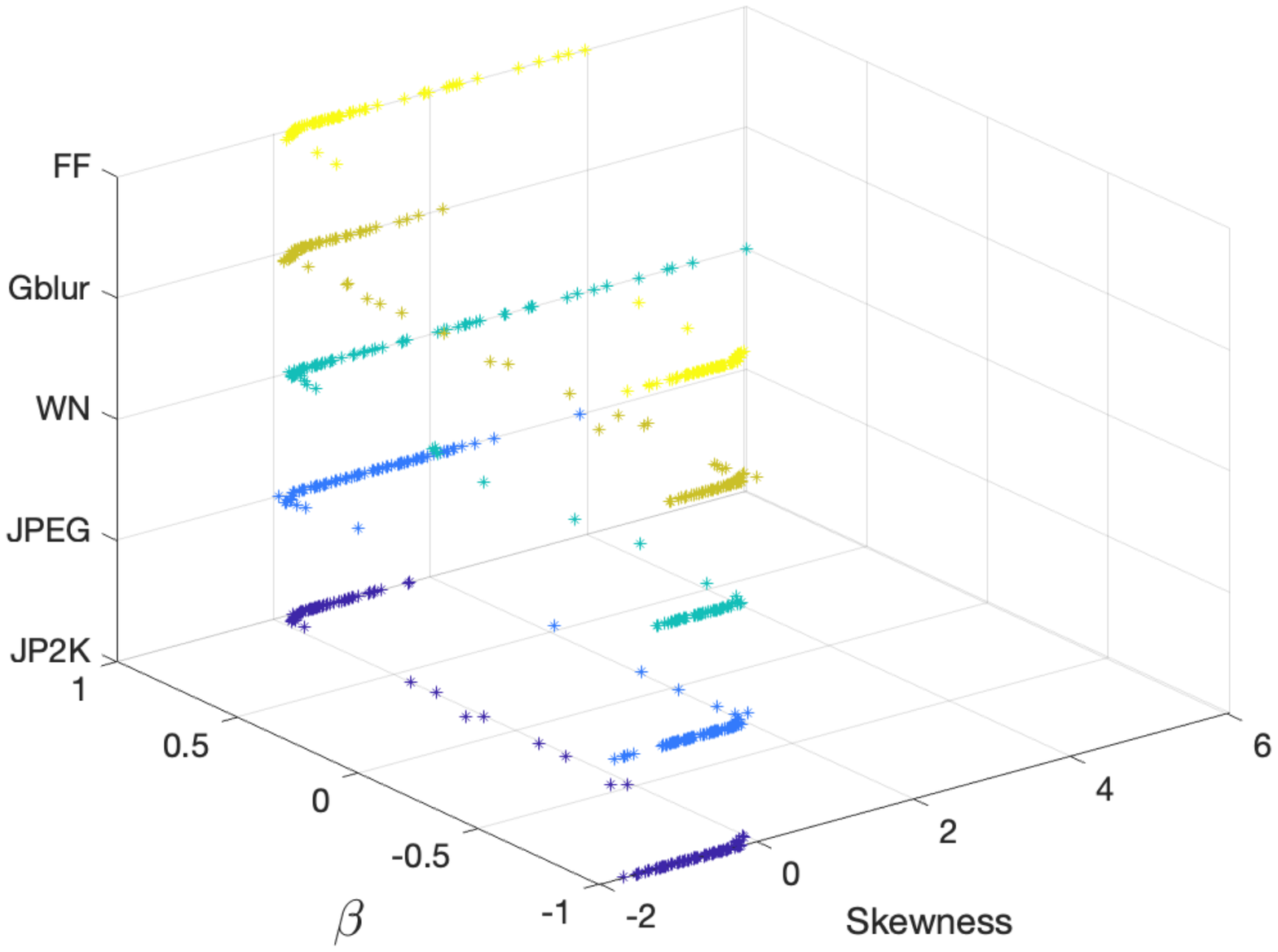}
\end{minipage}%
}%
\subfigure[$\gamma$ vs. Skewness for distortion ~~~~~ \newline types.]{
\begin{minipage}[t]{0.24\linewidth}
\centering
\includegraphics[scale=0.23, trim=15 200 40 200, clip]{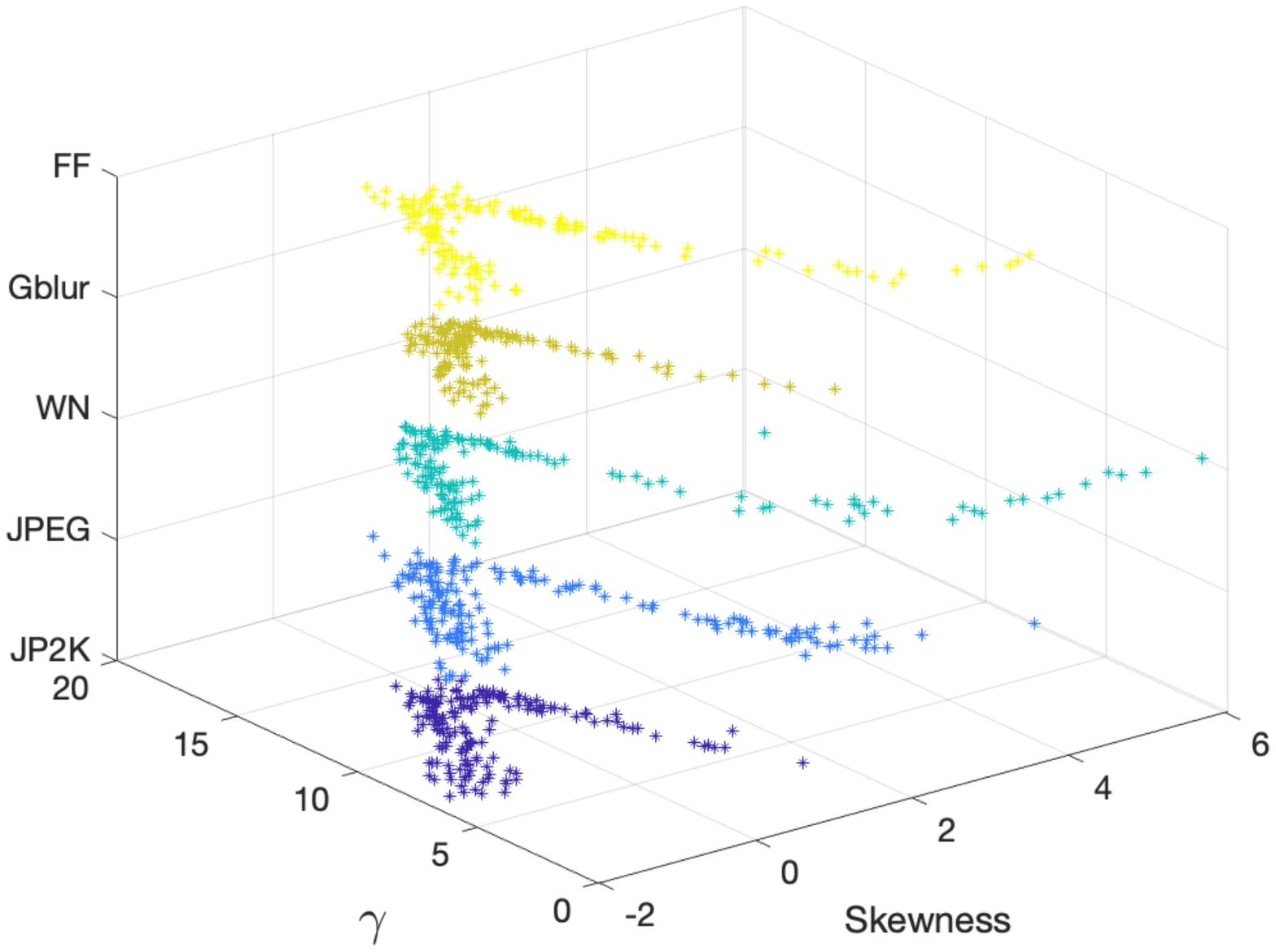}
\end{minipage}
}%
\subfigure[$\mu$ vs. Skewness for distortion ~~~~~ \newline types.]{
\begin{minipage}[t]{0.24\linewidth}
\centering
\includegraphics[scale=0.23, trim=15 200 40 200, clip]{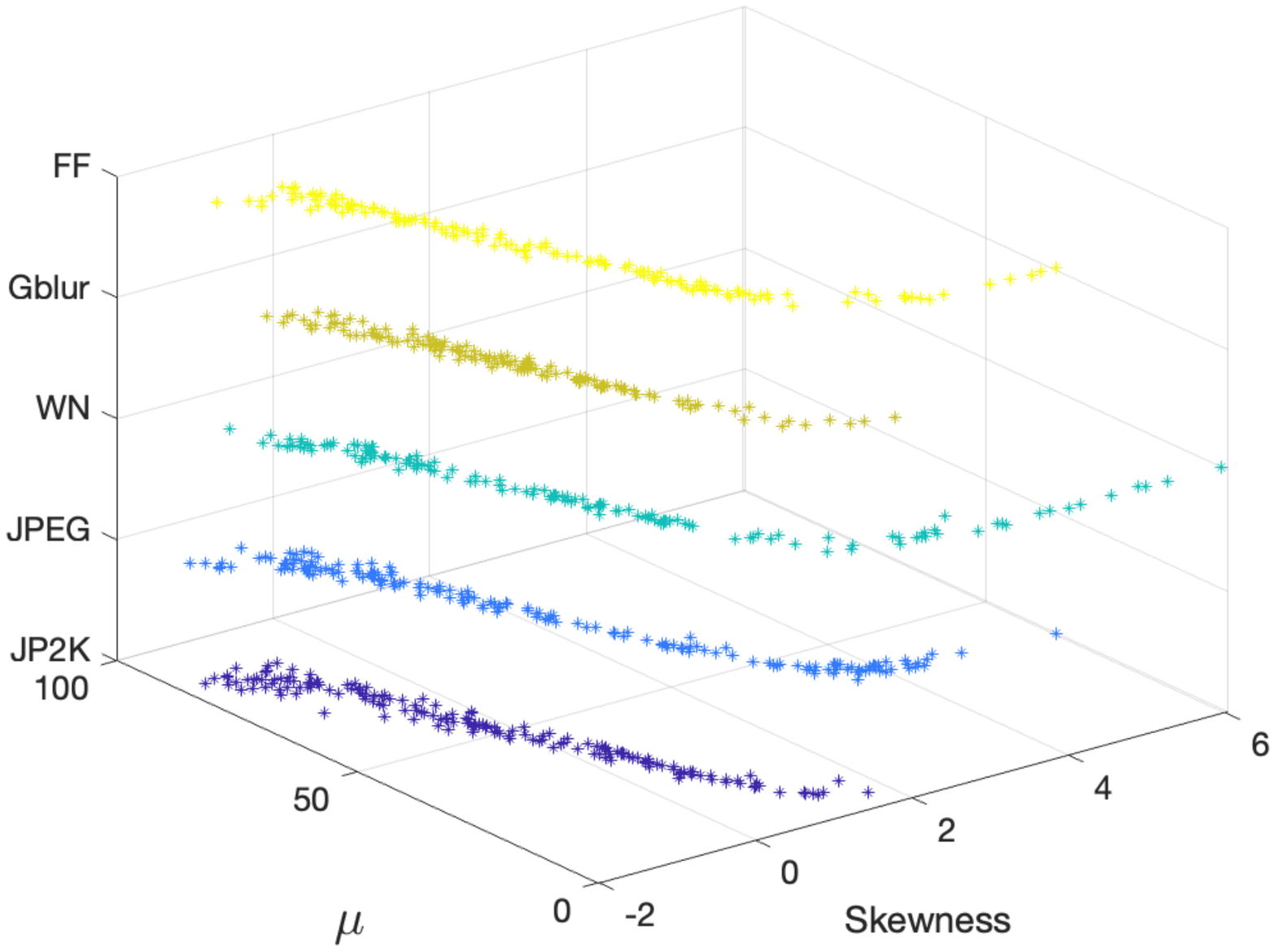}
\end{minipage}
}%
\centering
\caption{The relationships between four parameters of the alpha stable model based IQSD and MOS, SOS and the skewness of opinion score for images with different distortion types. }
\label{f17}
\end{figure*}


\subsection{Alpha Stable Model Based IQSD Analysis}
\label{subsection4.1}
 \subsubsection{Fitting results}
 
As the most popular parameter estimation method, MLE is used to fit the alpha stable model with IQSD for each image. The same methods and processes as described in Section \ref{sec:assessment} are used to measure the fitting results. 
 The mean RMSE is 0.0023,
 and the Chi-Square test result shows that IQSDs of about $57\%$ images follow the alpha stable model.
 Even if the alpha stable model cannot fully describe all IQSDs, it can be seen from Table \ref{goodness-of-fit} that the RMSE of the alpha stable model is smaller than that of other distributions, and the value of Chi-Square test is also larger than that of other distributions. Thus, the alpha stable model meets our needs to find a distribution that can minimize the difference with IQSDs and maximize the Chi-Square test result.
 Red curves in the second row of Fig.~\ref{f0} show the IQSDs described by alpha stable model for some images. By analyzing the collected subjective quality scores and the fitting results, it is easy to find that the alpha stable model can describe IQSDs well.
 
Besides, it is also meaningful to study the parameters of the alpha stable model based IQSD. The parameter $\alpha$ reflects the `thickness' of the tail of the distribution. 
The parameter $\beta$ is used to measure the asymmetry of the IQSD.
The scale parameter $\gamma$ measures the dispersion of the IQSD. 
For parameter $\mu$, it can be regarded as the quality score with the maximum probability of the IQSD. 
To further analyze the information that the alpha stable model based IQSD can provide,
we explore the relationships between four parameters of the alpha stable model based IQSD and MOS, SOS and the skewness of opinion score. 
 \subsubsection{Relationship with MOS}
\label{subsubsec:3.1.3}

The relationship between four parameters of the alpha stable model based IQSD and MOS is shown in Fig.~\ref{f4}. 
 In particular, Fig.~\ref{f4} (a) shows the relationship between parameter $\alpha$ and MOS. 
When the image quality is worse or better, the $\alpha$ value is smaller, which indicates that the tail of the IQSD is thicker.
 When the image quality is closer to `Fair', the $\alpha$ value is closer to 2, which means that the tail of the IQSD is thinner and closer to the tail of Gaussian distribution. 
Fig.~\ref{f4} (b) shows the relationship between parameter $\beta$ and MOS. It can be seen that the most alpha stable model based IQSDs are not symmetrical. 
In particular, when the image quality is lower than `Fair', $\beta$ is greater than 0 and the IQSD inclines to the left. When the image quality is higher than `Fair', $\beta$ is less than 0 and the IQSD inclines to the right.


Fig.~\ref{f4} (c) shows the relationship between parameter $\gamma$ and MOS. When the image quality is worse or better, the $\gamma$ value is smaller, which indicates that the IQSD is more concentrated.
 When the image quality is closer to `Fair', the $\gamma$ value is larger, which indicates that the IQSD is more dispersed. This is an interesting result because it shows that subjects seem to be more consistent when the image quality is good or bad, and their opinions on the medium image quality are more inconsistent.
It can also be seen from Fig.~\ref{f4} (d) that MOS has a linear relationship with parameter $\mu$. That is to say, the maximum probability score of the image increases with the improvement of image quality. The Spearman's rank correlation coefficient (SRCC) between MOS and parameter $\mu$ is 0.9990.
In addition, the vertical axis in Fig.~\ref{f4} (d) shows that the $\mu$ value ranges from 1 to 84, which indicates that the images presented to the subjects have a wide range of quality score.

 \subsubsection{Relationship with SOS}
 
 Fig.~\ref{f14} shows the relationship between four parameters of the alpha stable model based IQSD and SOS. Specifically, Fig.~\ref{f14} (a) shows the relationship between parameter $\alpha$ and SOS. Generally speaking, with the increase of $\alpha$, SOS is also increasing, which means that the subjective diversity is also increasing.
 Fig.~\ref{f14} (b) shows the relationship between parameter $\beta$ and SOS. It can be seen from this figure that there is no obvious relationship between the inclination of IQSD and subjective diversity.
Fig.~\ref{f14} (c) shows the relationship between parameter $\gamma$ and SOS. 
It is easy to see that parameter $\gamma$ is closely related to SOS, where the SRCC between SOS and parameter $\gamma$ is 0.9284, thus $\gamma$ is used to measure the subjects diversity in this paper. From the vertical axis in Fig.~\ref{f14} (c), we can also observe that the $\gamma$ value ranges from 1 to 16.2, which is non-ignorable. Therefore, it is necessary to consider the diversity of subjects when evaluating image quality.
Fig.~\ref{f14} (d) shows the square relationship between the parameter $\mu$ and SOS, which is consistent with the conclusion obtained in literature \cite{6065690}. When the image quality is close to `Fair', the subjective diversity of the image is very large. When the image quality is close to `Bad' or `Excellent', the subjective diversity of the image is very small.
 
  \subsubsection{Relationship with the skewness of opinion score}
 Fig.~\ref{f16} shows the relationship between four parameters of the alpha stable model based IQSD and the skewness of opinion score.
 Specifically, Fig.~\ref{f16} (a) shows the relationship between the parameter $\alpha$ and the skewness of opinion score, and Fig.~\ref{f16} (c) shows the relationship between the parameter $\gamma$ and the skewness of opinion score. 
 From these two figures, we can find that when parameters $\alpha$ and $\gamma$ reach the maximum, that is, the image quality approaches to `Fair', the skewness is close to 0. When parameters $\alpha$ and $\gamma$ decrease, there are two cases: the first case is that when the image quality becomes worse, the skewness decreases; the second case is that when the image quality gets better, the skewness increases.
 Fig.~\ref{f16} (b) shows the close relationship between the parameter $\beta$ and the skewness of opinion score. 
 When parameter $\beta$ is greater than 0, the skewness of opinion score is also greater than 0, which means that the majority of subjects think that the image quality is lower than MOS.
When $\beta$ is less than 0, the skewness of opinion score is also less than 0, which means that the majority of subjects think that the image quality is higher than MOS. 
   Therefore, parameter $\beta$ can explain the opinion of the majority of subjects relative to MOS.
 Fig.~\ref{f16} (d) shows the relationship between the parameter $\mu$ and the skewness of opinion score. It is easy to observe that the skewness of opinion score increases with the decrease of parameter $\mu$.

In conclusion, the four parameters of the alpha stable model based IQSD are closely related to MOS, SOS and the skewness of opinion score.
 Therefore, the alpha stable model based IQSD can provide much more information for the subjective quality of an image, such as the skewness of opinion score, the subject diversity and the maximum probability score.

 \subsubsection{Distortion types}
There are six types of images in the LIVE database, including JP2K, JPEG, WN, Gblur, FF distorted images and reference images.
To visualize the statistical relationships between four parameters of the alpha stable model based IQSD and MOS, SOS and the skewness of opinion score for images with different distortion types, we further draw scatter plots between them in Fig.~\ref{f17}. 
As shown in this figure, the relationships between four parameters of the alpha stable model based IQSD and MOS, SOS and the skewness of opinion score are similar between different distortion types. This means that the subjects are not sensitive to the distortion type of the image when assessing the image quality.

    \begin{table*}[h]\footnotesize
\centering
\caption {  Performance of the Proposed Feature Extraction Method and Other Competing Feature Extraction Methods in Predicting the Alpha Stable Model Based IQSD. The Best Performances Are In Bold.}
\begin{tabular}{c|cccccccccccc}
\hline\hline
 Criterion & \textbf{ BMPRI } &\textbf{ BPRI  }&\textbf{BLIINDS-II }&\textbf{ DIIVINE }&\textbf{BIQI }&\textbf{BRISQUE}&\textbf{CORNIA}&\textbf{NIQE}&\textbf{ResNet50}& \textbf{ Proposed}\\ \hline\hline
 
 JSD  $\downarrow $    &0.0088 & 0.0104 & 0.0084 & 0.0105 & 0.0111  & 0.0094 & 0.0304 & 0.0088 & 0.0220&\textbf{0.0081}\\
 RMSE$\downarrow $    &0.0069  & 0.0082 & 0.0070 & 0.0075 & 0.0078  & 0.0071 & 0.0127 & 0.0071& 0.0109&\textbf{0.0066}\\
Chebyshev$\downarrow $    & 0.0157 & 0.0188 & 0.0154 & 0.0169 & 0.0173 &0.0158 & 0.0243 &0.0156  &0.0228 &\textbf{0.0152}\\
Chi-Square$\downarrow $& 0.0204   &0.0212 & 0.0208  &0.0258  & 0.0210 & 0.0229 &0.0697  &0.0217 &0.0515 & \textbf{0.0200}\\

Cosine $\uparrow $ &0.8864 & 0.8584 &0.8848  & 0.8422 &0.8366&0.8593  &  0.5769 & 0.8752 &0.6571 &     \textbf{0.8902}      \\ \hline\hline
\end{tabular}
\label{tab:2}
\end{table*}

\begin{figure}[h]
  \includegraphics[scale=0.45, trim=0 200 0 200, clip]{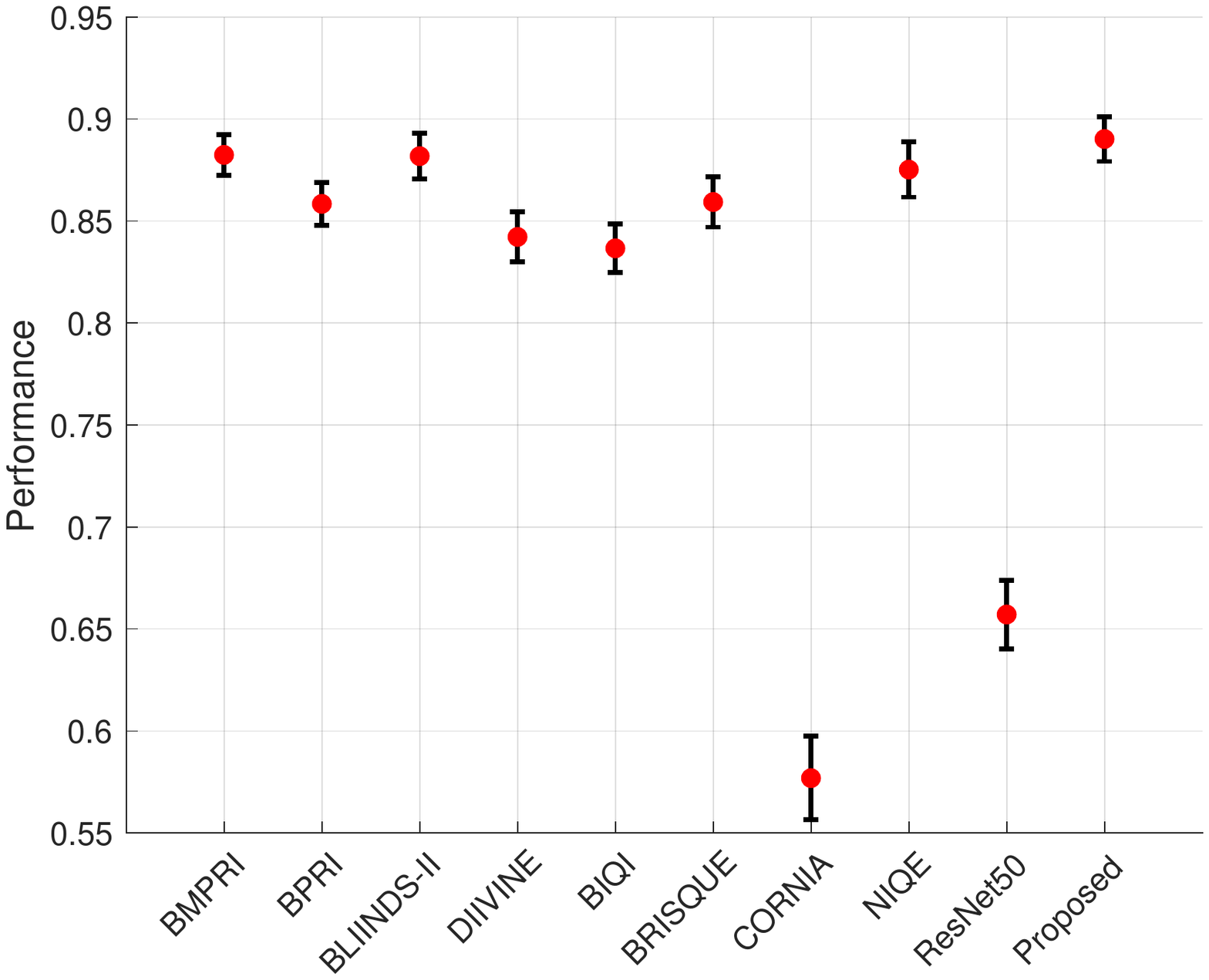}
  \caption{Mean and standard error bars of the cosine values from 1000 random train-test trials.}
  \label{error}
\end{figure}


  \begin{table*}[h]\footnotesize
	\centering
	\renewcommand\tabcolsep{5.0pt}
	\caption{Performance of the Proposed Feature Extraction Method and Other Competing Feature Extraction Methods in Predicting the Four Parameters of the Alpha Stable Model Based IQSD. The Best Performances Are In Bold.}
\begin{tabular}{c|ccccccccccccc}

\hline\hline
Parameter&Criterion& \textbf{ BMPRI } &\textbf{ BPRI  }&\textbf{BLIINDS-II }&\textbf{ DIIVINE }&\textbf{BIQI}&\textbf{BRISQUE}&\textbf{CORNIA}&\textbf{NIQE}&\textbf{ResNet50}& \textbf{ Proposed}\\ \hline\hline

&RMSE  $\downarrow $   & 0.1963 & 0.2072 & 0.1941 & 0.2201 &0.2413 & 0.2155 & 0.3074 & 0.2002 & 0.2966 &\textbf{0.1839}\\
&MAE  $\downarrow $    &  0.1329   & 0.1545   & 0.1374  & 0.1543   &0.1786&  0.1571  & 0.2369 &  0.1417  & 0.2149     & \textbf{0.1317}  \\
$\alpha$ &MAPE   $\downarrow $   & 0.0900   &  0.1043  &  0.0913  & 0.1073   & 0.1252   & 0.1076 & 0.1757  & 0.0972  & 0.1579      &\textbf{0.0872}   \\

&Correlation   $\uparrow $ &0.7820 & 0.7339 & 0.7734 & 0.7001 &0.6162  & 0.6990 & 0.1137 & 0.7546 & 0.2574 & \textbf{0.7942}\\
&Cosine $\uparrow $  & 0.9939  &  0.9927 & 0.9938    &  0.9919   &  0.9906     & 0.9945     &  0.9849 & 0.9934 &0.9855 & \textbf{0.9946}  \\\hline

&RMSE  $\downarrow $    &\textbf{0.5101} & 0.9196 & 0.7924 & 0.7708 &   0.7140 & 0.8122 & 1.3473 & 0.8935 & 1.0015 &0.6332 \\ 
&MAE  $\downarrow $    &   \textbf{0.4826} &    0.5231 &    0.5322    &     0.6717&     0.5676 &    0.6341   &     0.9521      &   0.5938   &0.8389  &  0.5154   \\
  $\beta$&MAPE  $\downarrow $    &    \textbf{0.5618}   &    0.5945   &    0.6207   &    0.7480  &   0.6318   &      0.7010  &   1.1753    &    0.6448   & 0.9068 &  0.5759  \\
    &Correlation $\uparrow $   &     0.7595 & 0.5398 & 0.7269 & 0.7617 &   0.6874     & 0.6991 &0.7793 & 0.5694 & 0.4398 & \textbf{0.7909} \\
    &Cosine  $\uparrow $  & 0.7673 & 0.7074  &   0.7156    &     0.6336   &   0.6852    &  0.6803    & 0.0893& 0.6244   &0.2507 &  \textbf{0.7793}  \\\hline

&RMSE $\downarrow $     &2.7559  & 3.1580 & 2.5866 & 3.1694 & 3.1407  & 3.1428 & 3.1923 & 5.8719 & 3.1771 & \textbf{1.9713} \\ 
&MAE  $\downarrow $    &      2.3445  &     2.6270 &    2.1288  &     2.2119   &     2.5996  &    2.0844 & 2.4160     &    2.3569 &     2.2171  &  \textbf{1.4707}   \\
$\gamma$&MAPE $\downarrow $     &    0.3539   &    0.4471  &     0.3167  &    0.3318  &    0.4309 &     0.3225  &  0.3706       &   0.5039 &      0.3318 &  \textbf{0.2501}\\
   
   &Correlation   $\uparrow $ &     0.5013 & 0.3906 & 0.5985 & 0.4917 & 0.2105    & 0.5696 & 0.5173 & 0.5437 & 0.4877 &\textbf{0.7799} \\ 
   &Cosine  $\uparrow $  & 0.9681  & 0.9575  &    0.9733  &   0.9695   &       0.9590    &   0.9743   &0.9690  & 0.9167 & 0.9690& \textbf{0.9829} \\\hline

&RMSE $\downarrow $     & 10.8554 & 12.3126 & 10.9991 & 12.0462 &  13.2073  & 12.4421 & 25.7155 & 10.9650 & 20.7216 & \textbf{10.7417} \\ 
&MAE  $\downarrow $    &    7.3496   &     8.5252    &    7.5038   &    8.5059   &    9.4765  &   8.4607     &  22.0320  &     7.3016 &    16.8451    &  \textbf{7.2546}\\
  $\mu$ &MAPE $\downarrow $     &    0.4651  &     0.5232  &     0.4352 &     0.6487   &    0.5799 &      0.5430 &  2.1397   &    0.5176  &    1.5873     &    \textbf{0.4265} \\
   &Correlation  $\uparrow $  &     0.9024 & 0.8743 & 0.9002 & 0.8834 & 0.8517 & 0.8721 & 0.1177 & 0.9018 & 0.6303 &\textbf{0.9044}\\
   &Cosine  $\uparrow $  & 0.9810  & 0.9756  &    0.9803  &   0.9761   &    0.9714  &   0.9804  & 0.8908& 0.9811&   0.9281 &\textbf{0.9815}  \\\hline\hline

\end{tabular}
	\label{t3}
\end{table*}

 \begin{table*}[h]
    \begin{center}
        \centering
         \caption{Performance Comparison of Predicting the Alpha Stable Model Based IQSD on Individual Distortions. The Best Performances Are In Bold.}\label{table2}
          \resizebox{1\hsize}{!}{
        \begin{tabular}{c|cccccccccccccc}
            \hline\hline
            Distortion type&Criterion & \textbf{ BMPRI } &\textbf{ BPRI  }&\textbf{BLIINDS-II }&\textbf{ DIIVINE }&\textbf{BIQI }&\textbf{BRISQUE}&\textbf{CORNIA}&\textbf{NIQE}&\textbf{ResNet50}& \textbf{ Proposed}  \\
          \hline\hline
            
            & JSD   $\downarrow $    &0.0116 & 0.0117 &0.0124 & 0.0210& 0.0131 & 0.0138 & 0.0263 & 0.0134&0.0298& \textbf{0.0112}\\
       & RMSE  $\downarrow $  & 0.0075 & 0.0076 & 0.0079 & 0.0099&0.0083 & 0.0084& 0.0114 & 0.0079&0.0123 & \textbf{0.0073} \\
JP2K &Chebyshev$\downarrow $&0.0163 & 0.0163 & \textbf{0.0164} & 0.0201 & 0.0177 & 0.0181 & 0.0209 & 0.0171 & 0.0230 & 0.0171\\
 &Chi-Square$\downarrow $&0.0290&	\textbf{0.0265}&	0.0269&	0.0471&	0.0287&	0.0332&0.0614	&	0.0308&0.0653&	0.0296\\
&Cosine $\uparrow $&0.8190 & 0.8299 & 0.8355 & 0.6968 & \textbf{0.8432} & 0.7831 & 0.6155 & 0.8005& 0.5662& 0.8206\\
           
           \hline

                       & JSD  $\downarrow $     & 0.0157& 0.0129& 0.0125& 0.0221 & 0.0107 & 0.0121& 0.0376 & 0.0123& 0.0347& \textbf{0.0104} \\
       & RMSE $\downarrow $   &0.0111 & 0.0085& 0.0084& 0.0111 & 0.0078 & 0.0085& 0.0145& 0.0083 & 0.0136 & \textbf{0.0075} \\
JPEG &Chebyshev$\downarrow $&0.0296&	0.0207&	0.0198&	0.0255&	0.0187&	0.0200&0.0278 &	0.0196&	0.0283&	\textbf{0.0185}\\
    &     Chi-Square$\downarrow $&\textbf{0.0227}  &  0.0301   & 0.0274  &  0.0495  &  0.0247 &   0.0269  & 0.0857   &  0.0282  &  0.0751  &  0.0254\\
         &Cosine $\uparrow $&\textbf{0.8942}&	0.8772&	0.8179&	0.6559&	0.8630&	0.8282	&  0.4773    &	0.8262&	0.5274&	0.8615\\
        \hline

                       & JSD   $\downarrow $    &0.0140& 0.0109 & 0.0106& 0.0126 & 0.0123 & 0.0095 & 0.0334 & 0.0104 & 0.0390 &  \textbf{0.0101}\\
      & RMSE  $\downarrow $  & 0.0101 & 0.0085 & 0.0082 & 0.0090&0.0090 & 0.0085& 0.0137 & 0.0081 & 0.0146 & \textbf{0.0079}\\
WN  &Chebyshev$\downarrow $&0.0269&	0.0199 &	0.0213 &	0.0245 &0.0229&0.0206 &  0.0278&	0.0202 &	0.0310 &\textbf{0.0197} &\\
           &Chi-Square$\downarrow $&0.0276 &   0.0235  &  0.0251   & 0.0313   & \textbf{0.0265}    &0.0237 & 0.0750   &   0.0262    &0.0827 &   0.0273\\
           &Cosine $\uparrow $&0.8915 &	0.8953 &	0.8818 &	0.8374 &\textbf{0.8970}&	0.8901 & 0.0750  &	0.8870 &	0.4587 &	0.8671 &\\
           \hline

         & JSD  $\downarrow $     &  0.0101 & 0.0108 & 0.0104& 0.0157& 0.0128& 0.0099 & 0.0194& 0.0113& 0.0234 & \textbf{0.0095}\\ 
            & RMSE  $\downarrow $  &  0.0066 & 0.0067 & 0.0067& 0.0084& 0.0077 & 0.0066 & 0.0095& 0.0068& 0.0107& \textbf{0.0065} \\ 
      Gblur  &Chebyshev$\downarrow $&\textbf{0.0134}&	0.0145&0.0135&	0.0164&0.0156&	0.0135&0.0174	&	0.0143& 0.0211&0.0137&\\
            & Chi-Square$\downarrow $& 0.0248  &  0.0277 &  \textbf{0.0231} &   0.0370  &  0.0349 &   0.0240  &      0.0464& 0.0278  &  0.0603  &  0.0260\\
             &Cosine $\uparrow $&0.8526&	0.8246&0.8567&	0.7798&0.8143&0.8619&0.0464	&	0.8495& 	0.6271&	\textbf{0.8568}\\
              \hline

               & JSD   $\downarrow $    &0.0146 & 0.0176 &0.0155 & 0.0211 &0.0177 & 0.0159 & 0.0336 & 0.0151 & 0.0330& \textbf{0.0144} \\
         & RMSE  $\downarrow $  &   0.0088 & 0.0094 & 0.0084& 0.0104 & 0.0100 & 0.0084 & 0.0136 & 0.0087 & 0.0132 &\textbf{0.0081} \\ 
     FF   &Chebyshev$\downarrow $&0.0202 & 	0.0211 & 	0.0204 & 	0.0226 & 	 0.0219 & 	0.0203 & 	0.0253 & 	0.0201 & 0.0261 & \textbf{0.0195} & \\
        &Chi-Square$\downarrow $&0.0371  &   0.0384 &    \textbf{0.0328} &    0.0469 &    0.0430  &   0.0348 &0.0795 &     0.0351    & 0.0725 &    0.0366 \\
        &Cosine $\uparrow $&0.7836 & 	0.7570 & 	0.7399 & 0.7054 & 	0.8040 & 0.7951 & 0.5137 &0.7788 & 0.5362 &\textbf{0.8170}\\

            \hline\hline
            
        \end{tabular}
        \label{t5}
        }
    \end{center}

\end{table*}

 \subsection{Prediction Framework Analysis}
 \label{Prediction Framework Analysis}
To verify the effectiveness of our prediction framework, we conduct experiments to analyze the feature extraction method used in this paper by comparing with some state-of-the-art NR quality methods. 
 \subsubsection{Experimental settings} 

   Since there is no method specially designed for IQSD prediction in previous literatures, we create some baseline IQSD prediction models based on our proposed IQSD prediction framework as competitors. Specifically, the same alpha stable model based IQSD prediction framework is still used, but we replace the features used in the proposed method with the features used in the state-of-the-art NR IQA methods, including BMPRI \cite{2018Blind}, BPRI \cite{Min}, BLIINDS-II \cite{Michele}, DIIVINE \cite{Moorthy}, BIQI \cite{5432998}, BRISQUE \cite{Mittal}, CORNIA \cite{2012Unsupervised}, NIQE \cite{2013Making} and a network feature extraction method ResNet50 \cite{he2016deep}. We should note that these baseline competitors are all created based on our novel alpha stable model based IQSD prediction framework, which is one of the core contributions of this paper.
   
   
   The experimental validation is conducted on the re-scored LIVE database which is obtained in subjective image quality assessment study described in Section \ref{sec:assessment}.
During the experiment, we randomly select $80\%$ images as the training set, and the remaining images as the test set. To reduce the bias caused by the random splitting of the training and test sets, we repeat this random split for 1000 times and report the mean performances.
    \subsubsection{Overall performance comparison} \label{Overall}
    In order to evaluate the prediction performance of the alpha stable model based IQSD prediction framework with different features, the predicted alpha stable model based IQSD is sampled to obtain the corresponding probability density histogram and compared with the raw quality score histogram.
    In particular, five criteria are used to evaluate the prediction performance, including Jensen-Shannon divergence (JSD), RMSE, Chebyshev distance, Chi-Square distance and cosine similarity, which are all commonly used metrics to measure the consistency of two histograms.
 The comparison results are shown in Table~\ref{tab:2}.

It can be seen from Table \ref{tab:2} that most feature extraction methods have little difference in the performance of predicting the alpha stable model based IQSD, that is, the proposed alpha stable model based IQSD prediction framework shows certain abilities of IQSD prediction no matter what features are used. 
In particular, the results show that the IQSD prediction framework based on the proposed feature extraction method has the best prediction performance. 

Note that we calculate the consistency of two distributions by repeating the train-test trial for 1000 times. Therefore, in addition to the mean performances reported in Table~\ref{tab:2}, we can also get the standard deviation from 1000 random train-test trials. In Fig. \ref{error}, standard deviation in terms of cosine similarity is shown for each feature extraction method. As can be seen from this figure, the IQSD prediction framework with the proposed feature extraction method has a small standard deviation, which means that it is relatively stable in predicting the alpha stable model based IQSD.

 \subsubsection{Performance on predicting distribution parameters} 
 The proposed alpha stable model based IQSD prediction framework consists of four parameter prediction models, whose performances are also tested. Specifically, the RMSE, mean absolute error (MAE), mean absolute percentage error (MAPE), correlation and cosine similarity are used as the measures to evaluate the prediction performance. The results are shown in Table~\ref{t3}. 
  It is easy to see that the proposed feature extraction method has the best prediction performance in predicting parameters $\alpha$, $\gamma$ and $\mu$.
For parameter $\beta$, the proposed feature extraction method is also competitive.

 \subsubsection{Performance on individual distortions} 
 \label{individual}
Further, we compare the prediction performance of these feature extraction methods on individual distortions. We conduct similar train-test trials as described in Section \ref{Prediction Framework Analysis}-$\textit{2)}$. The difference is that only the images with the same distortion type are used in each train-test trial. The average performance is shown in Table \ref{t5}. It is observed that the proposed feature extraction method can be comparable to the state-of-the-art methods when predicting the alpha stable model based IQSD on individual distortions. 

\subsubsection{Computational complexity} 
In order to analyze the computational complexity of the alpha stable model based IQSD prediction framework with different features, we test them on 100 images with a fixed resolution of 512$\times$768 on a computer with Apple M1 CPU and 16GB RAM, and report the mean running time (seconds/image) in Table~\ref{Computational complexity}. The running time includes both feature extraction and regression time. It can be observed from the table that the IQSD prediction framework with the proposed feature extraction method has low computational complexity.

    \begin{table}[t]
    \centering
\caption {Computational Complexity.}
\begin{tabular}{c|c}
\hline\hline
 Method& Time (seconds/image)\\ \hline\hline
 BMPRI &0.3376\\
BPRI&0.1551\\
BLIINDS-II& 1.0342      \\ 
DIIVINE&0.5762\\
BIQI &0.0130\\
BRISQUE&0.0336\\
CORNIA&  2.2235     \\ 
NIQE&0.0312\\
ResNet50&1.5579\\
Proposed& 0.2338\\ \hline\hline
\end{tabular}
\label{Computational complexity}
\end{table}

\subsubsection{Ablation analysis} 
The feature extraction method proposed in this paper is to extract two types of features for each image, including structural features and statistical features. Therefore, it is meaningful to study the impact of these two types of features on the prediction performance.

On the one hand, we compare the performance of each type of features and their combination in predicting the alpha stable model based IQSD. The results are shown in Table~\ref{tab:5}. From this table, we can draw the following conclusions. Firstly, the prediction performance of statistical features alone is poor. Secondly, structural features alone lead to medium prediction performance. Thirdly, the combination of the two types of features obtains the best performance.
    \begin{table}[t]
    \centering
\caption {Performance of Each Type of Features and Their Combination in Predicting the Alpha Stable Model Based IQSD. `+' Represents the Combination of Two Types of Features. The Best Performances Are In Bold.}
\begin{tabular}{c|ccc}
\hline\hline
 Criterion & \textbf{Structural} &\textbf{Statistical}&\textbf{Structural+Statistical}\\ \hline\hline
 JSD  $\downarrow $    &0.0098 & 0.0130 & \textbf{0.0081} \\
 RMSE$\downarrow $    &0.0074  & 0.0085 & \textbf{0.0066} \\
Chebyshev$\downarrow $    & 0.0170 & 0.0182 & \textbf{0.0152}        \\ 
Chi-Square$\downarrow $& 0.0241   &0.0318 & \textbf{0.0200}  \\
Cosine $\uparrow $ &0.8679 & 0.8053 &\textbf{0.8902}   \\ \hline\hline
\end{tabular}
\label{tab:5}
\end{table}

On the other hand, we also compare the performance of each type of features and their combination in predicting the four parameters of the alpha stable model based IQSD in Table~\ref{t6}. From this table, we can draw the conclusion that 
the prediction performance of the combination of the two types of features is also the best. 

In summary, the structural and statistical features we extracted can be well used for alpha stable model based IQSD prediction.

  \begin{table}[t]\footnotesize
	\centering
	\renewcommand\tabcolsep{5.0pt}
	\caption{Performance of Each Type of Features and Their Combination in Predicting the Four Parameters of the Alpha Stable Model Based IQSD. `+' Represents the Combination of Two Types of Features. The Best Performances Are in Bold.}
\begin{tabular}{c|cccc}

\hline\hline
& & & &\textbf{Structural}\\
Parameter&Criterion& \textbf{ Structural } &\textbf{ Statistical }&\textbf{+}\\
& & & &\textbf{Statistical}\\ \hline\hline

&RMSE  $\downarrow $   & 0.2080    & 0.2393   & \textbf{0.1839}\\
&MAE  $\downarrow $    & 0.1477   &0.1786    & \textbf{0.1317}  \\
$\alpha$ &MAPE   $\downarrow $   & 0.0999    &  0.1209  & \textbf{0.0872} \\
&Correlation   $\uparrow $ &0.7356     & 0.6023    & \textbf{0.7942}\\
&Cosine $\uparrow $  &  0.9929   &   0.9906   & \textbf{0.9946}  \\\hline

&RMSE  $\downarrow $   &0.8121 & 0.6967      & \textbf{0.6332}\\ 
&MAE  $\downarrow $    &     0.5726  &      0.6172   &    \textbf{0.5154}    \\
  $\beta$&MAPE  $\downarrow $    &    0.6332  &     0.6823    &    \textbf{0.5759}\\
    &Correlation $\uparrow $   &    0.7166  &  0.5649  & \textbf{0.7909} \\
    &Cosine  $\uparrow $  &   0.7107    &  0.5625    & \textbf{0.7793}    \\\hline

&RMSE $\downarrow $     & 2.5603  &  3.1433    & \textbf{1.9713 }\\ 
&MAE  $\downarrow $    &    2.1374  &     2.6088    &   \textbf{1.4707}   \\
$\gamma$&MAPE $\downarrow $     &     0.3301     &      0.4329    &   \textbf{0.2501} \\
   &Correlation   $\uparrow $ &     0.5963    &   0.1492  &\textbf{0.7799}\\ 
   &Cosine  $\uparrow $  &  0.9720   &  0.9588    &  \textbf{0.9829} \\\hline

&RMSE $\downarrow $     &  12.0456 &15.0112 & \textbf{10.7417} \\ 
&MAE  $\downarrow $    &    8.3995    &    10.8045   &   \textbf{7.2546} \\
  $\mu$ &MAPE $\downarrow $     &    0.5332  &      0.6142   &   \textbf{0.4265} \\
   &Correlation  $\uparrow $  &    0.8813&   0.8051& \textbf{0.9044}\\
   &Cosine  $\uparrow $  &   0.9764  &  0.9634  &  \textbf{0.9815}  \\\hline\hline

\end{tabular}
	\label{t6}
\end{table}

\section{Conclusion}
\label{Conclusion}
In this paper, we propose to describe image quality using a parameterized distribution, that it, the alpha stable model based IQSD.
 Moreover, we also propose an objective method to predict the alpha stable model based IQSD. Since the existing databases have not provided the quality scores given by specific subjects, a subjective quality assessment study has been carried out to get the subjective quality scores of all images in the LIVE database. 
After that, the alpha stable model is used to model the IQSD. In our objective prediction framework, the quality features of images are extracted based on both structural and statistical information, and four SVRs are trained to predict the parameters of alpha stable model based IQSD. 
Finally, experimental validations and analyses show that the alpha stable model can describe the IQSD well compared with other common distributions, and it can express much more information about the subjective image quality than a single MOS. 
Further, the experimental results also show that the prediction framework used in this paper is practicable and effective.


%


\ifCLASSOPTIONcaptionsoff
  \newpage
\fi



\bibliographystyle{IEEEtran}

\bibliography{IEEEabrv,refs}
%
%
%

%




\end{document}